\documentclass[a4paper,fleqn,usenatbib]{mnras}

\usepackage[T1]{fontenc}
\usepackage{ae,aecompl}
\pdfoutput=1

\usepackage{graphicx}	
\usepackage{amsmath}	
\usepackage{amssymb}	

\newcommand{\imw}{$i$--$W3$}
\newcommand{\imwf}{$i$--$W4$}
\newcommand{\rmwf}{$r$--$W4$}
\newcommand{\imwt}{$i$--$W2$}
\newcommand{\wtmwf}{$W3$--$W4$}
\newcommand{\kms}{km s$^{-1}$}

\newcommand{\lam}{$\lambda$}
\newcommand{\mum}{$\mu$m}
\newcommand{\ebv}{$E(B$$-$$V)$}
\newcommand{\heii}{\mbox{He\,{\sc ii}}}

\newcommand{\civ}{\mbox{C\,{\sc iv}}}
\newcommand{\ciii}{\mbox{C\,{\sc iii}}}

\newcommand{\nv}{\mbox{N\,{\sc v}}}

\newcommand{\ovi}{\mbox{O\,{\sc vi}}}

\newcommand{\oiii}{\mbox{O\,{\sc iii}}}

\newcommand{\oi}{\mbox{O\,{\sc i}}}
\newcommand{\feii}{\mbox{Fe\,{\sc ii}}}

\newcommand{\mgii}{\mbox{Mg\,{\sc ii}}}

\newcommand{\aliii}{\mbox{Al\,{\sc iii}}}
\newcommand{\siiv}{\mbox{Si\,{\sc iv}}}
\newcommand{\siiii}{\mbox{Si\,{\sc iii}}}
\newcommand{\siii}{\mbox{Si\,{\sc ii}}}
\newcommand{\lya}{\mbox{Ly$\alpha$}}
\newcommand{\lyb}{\mbox{Ly$\beta$}}
\newcommand{\hi}{\mbox{H\,{\sc i}}}

\title[Extremely Red Quasars]{Extremely Red Quasars in BOSS}

\author[F. Hamann et al.]{Fred Hamann$^{1,2}$\thanks{E-mail:
fhamann@ufl.edu (FH)}, Nadia L. Zakamska$^{3,4}$, Nicholas Ross$^{5}$, Isabelle Paris$^{6}$,\newauthor Rachael M. Alexandroff$^{3}$, Carolin Villforth$^{7}$, 
Gordon T. Richards$^8$, Hanna Herbst$^2$,\newauthor W. Niel Brandt$^{9,10,11}$, Ben Cook$^{12}$, Kelly D. Denney$^{13}$, Jenny E. Greene$^{14}$,\newauthor  
Donald P. Schneider$^{9,15}$, Michael A. Strauss$^{14}$
\\
$^{1}$ Department of Physics \& Astronomy, University of California, Riverside, CA 92507, USA \\
$^{2}$ Department of Astronomy, University of Florida, Gainesville, FL 32611, USA\\
$^{3}$ Department of Physics \& Astronomy, Johns Hopkins University, 3400 N. Charles St, Baltimore, MD 21218, USA\\
$^4$ Deborah Lunder \& Alan Ezekowitz Founders' Circle Member, Institute for Advanced Study, Einstein Dr., Princeton, NJ 08540, USA\\
$^{5}$ Institute for Astronomy, SUPA, University of Edinburgh, Royal Observatory, Edinburgh EH9 3HJ, UK\\
$^{6}$ INAF Osservatorio Astronomico di Trieste, Via G. B. Tiepolo 11, I-34131 Trieste, Italy\\
$^{7}$ Department of Physics, University of Bath, Claverton Down, Bath, BA2 7AY, United Kingdom \\
$^{8}$ Department of Physics, Drexel University, 3141 Chestnut Street, Philadelphia, PA 19104, USA\\
$^9$ Department of Astronomy \& Astrophysics, The Pennsylvania State University, University Park, PA 16802, USA\\
$^{10}$ Institute for Gravitation and the Cosmos, Pennsylvania State University, University Park, PA 16802, USA\\
$^{11}$ Department of Physics, Pennsylvania State University, University Park, PA 16802, USA\\
$^{12}$ Harvard-Smithsonian Center for Astrophysics, 60 Garden Street, Cambridge, MA 02138, USA\\
$^{13}$ Department of Astronomy, The Ohio State University, McPherson Laboratory, 140 West 18th Avenue Columbus OH, 43210, USA\\
$^{14}$ Department of Astrophysical Sciences, Princeton University, Princeton, NJ 08544, USA\\
$^{15}$ Institute for Gravitation \& the Cosmos, The Pennsylvania State University, University Park, PA 16802, USA\\
}

\date{Accepted XXX. Received YYY; in original form ZZZ}

\pubyear{2016}

\begin{document}
\label{firstpage}
\pagerange{\pageref{firstpage}--\pageref{lastpage}}
\maketitle

\begin{abstract}
Red quasars are candidate young objects in an early transition stage of massive galaxy evolution. Our team recently discovered a population of extremely red quasars (ERQs) in the Baryon Oscillation Spectroscopic Survey (BOSS) that has a suite of peculiar emission-line properties including large rest equivalent widths (REWs), unusual ``wingless'' line profiles, large \nv /\lya , \nv /\civ , \siiv /\civ\ and other flux ratios, and very broad and blueshifted [\oiii ] \lam 5007. Here we present a new catalog of \civ\ and \nv\ emission-line data for 216,188 BOSS quasars to characterize the ERQ line properties further. We show that they depend sharply on UV-to-mid-IR color, secondarily on REW(\civ ), and not at all on luminosity or the Baldwin Effect. We identify a ``core'' sample of 97 ERQs with nearly uniform peculiar properties selected via \imw\ $\ge 4.6$ (AB) and REW(\civ ) $\ge$ 100 \AA\ at redshifts 2.0--3.4. A broader search finds 235 more red quasars with similar unusual characteristics. The core ERQs have median luminosity $\left<\log L ({\rm ergs/s})\right> \sim 47.1$, sky density 0.010 deg$^{-2}$, surprisingly flat/blue UV spectra given their red UV-to-mid-IR colors, and common outflow signatures including BALs or BAL-like features and large \civ\ emission-line blueshifts. Their SEDs and line properties are inconsistent with normal quasars behind a dust reddening screen. We argue that the core ERQs are a unique obscured quasar population with extreme physical conditions related to powerful outflows across the line-forming regions. Patchy obscuration by small dusty clouds could produce the observed UV extinctions without substantial UV reddening.
\end{abstract}




\section{Introduction}

Quasars are signposts of rapid accretion onto supermassive black holes (SMBHs) in the centers of galaxies. The observed present-day correlation between the masses of SMBHs and their surrounding galactic spheroids suggests that SMBH accretion/growth is intimately connected to star formation and mass assembly in the host galaxies \citep{Gebhardt00, Tremaine02, Haring04, Gultekin09, Shankar09, Kormendy13}. The similar redshift peaks in the space density of quasars and the cosmic star formation rate at $z\sim 2$--3 indicate that these phenomena occurred together, perhaps in a physically-related way, at early cosmic times \citep{Boyle98,Merloni04, Marconi04, Wall05, Silverman05, Richards06, Rudnick06}. Popular models of galaxy evolution describe major episodes of SMBHs growth occurring in obscurity, deep inside dusty starbursts that appear observationally as sub-mm galaxies (SMGs) or ultra-luminous infrared galaxies \citep[ULIRGs, e.g.,][]{Sanders88,Hopkins05, Hopkins08,Veilleux09b,Simpson14}. Visibly luminous quasars are thought to appear near the end of this evolution when the SMBHs are massive enough to power quasars and a major blowout of gas and dust unveils the bright central source. ``Feedback" from quasar outflows during this evolution stage might play a role in driving the blowouts and regulating star formation in the host galaxies \citep[see also][]{DiMatteo05, Hopkins05, Hopkins10,Rupke11, Rupke13, Liu13, Wagner13}. 

Quasars that are obscured and reddened by dust can provide important tests of this evolution scheme if they appear preferentially during the brief transition phase from dusty starburst to normal blue quasar \citep[e.g.,][]{Hopkins05, Urrutia08, Glikman12, Glikman15,Wu14,Banerji15, Assef15}. However, other explanations for quasar reddening and obscuration are also possible. The Unified Model of AGN attributes the observed differences between Type 1 (broad line) and Type 2 (narrow line) AGN to orientation effects associated with an axisymmetric dusty torus that resides near the central engine of all AGN \citep{Antonucci93, Urry95,Netzer15}. In this scenario, Type 1 AGN offer direct views of the central engine and broad emission line regions while in Type 2s these regions are heavily obscured due to our nearly edge-on view of the torus/accretion disk geometry. Intermediate orientations might produce intermediate amounts of obscuration such that we observe Type 1 quasars with red colors and perhaps a wavelength-dependent mix of Type 1 and Type 2 properties \citep{Greene14}. In this context, red quasars provide valuable tests of the geometry and physical structure of quasar environments. 

Searches for red and obscured quasars have been propelled recently by wide-field surveys such as the Sloan Digital Sky Survey \citep[SDSS,][]{Zakamska03, Reyes08, Alexandroff13}, the Two Micron All Sky Survey \citep[2MASS,][]{Gregg02,Glikman07,Glikman12}, the United Kingdom Infrared Deep Sky Survey \citep[UKIDSS,][]{Glikman13}, Spitzer Space Telescope \citep{Lacy04, Lacy13, Stern05, Stern07,Hickox07,Donley12}, and the Wide-field Infrared Survey Explorer \citep[WISE,][]{Mateos12,Stern12,Assef13,Yan13}. Most of these searches combine broad-band photometry with other data such as visible-wavelength spectra or radio or X-ray fluxes to identify regions of color space populated by obscured AGN \citep[see also][and refs. therein]{Hao13}. Obscured quasars also turn up serendipitously in galaxy searches. For example, ``HotDOG'' satisfy the color selection criteria of dust obscured galaxies \citep[DOGs,][]{Dey08} even though their luminosities and especially their mid-IR emissions are believed to be dominated by hot dust powered by luminous embedded AGN \citep[][and refs. therein]{Eisenhardt12,Wu12,Tsai15,Toba16,Fan16b}. 

In \cite{Ross15}, our team discovered a unusual population of extremely red quasars (ERQs) in Data Release 10 (DR10) of the Baryon Oscillation Sky Survey \citep[BOSS,][]{Dawson13,Ross12} in the Sloan Digital Sky Survey-III \citep[SDSS-III,][]{Eisenstein11}. Starting with spectroscopically confirmed quasars in the BOSS quasar catalogs \citep[][]{Paris14, Paris16}, we combined photometry from the SDSS and WISE to select the most extreme cases with red colors similar to DOGs, e.g., with $r-W4 > 14$ and $W4 < 8.0$ (in Vega magnitudes, where $W4$ measures observed-frame $\sim$22 \mum ). This search finds 65 quasars across a wide range of redshifts ($0.28 < z_e < 4.36$) with a variety of properties. It includes a mix of Type 1 and 2 quasars, some starburst-dominated quasars, and several with broad absorption lines (BALs) that are strong and broad enough to suppress the $r$ band flux and satisfy the $r-W4$ color criterion even though the emitted spectrum is not extremely red. However, there was also a remarkable discovery that many ERQs at $z_e\ga 2$ appear to be a unique population with an ensemble of peculiar emission-line characteristics including very large rest equivalent widths (REWs), line profiles that are lacking strong Lorentzian (or logarithmic) wings characteristic of other broad-line AGN, and unusual line flux ratios that can include \nv\ \lam 1240 $>$ \lya , strong \aliii\ \lam 1860, and large ratios of \nv /\civ\ \lam 1549 and \siiv\ \lam 1400/\civ\ (see Figure~15 in \citealt{Ross15} for examples). These properties were discussed earlier by \cite{Polletta08} for an individual red quasar that is clearly in the same class as ERQs. 

Followup near-IR observations of ERQs have revealed even more remarkable properties, notably [\oiii ] \lam 5007 emission lines with the largest FWHMs and highest blueshifted wing velocities ever reported, both reaching $\sim$5000 \kms\ \citep[][Hamann et al. 2016a, in prep.]{Zakamska16}. The [OIII] lines identify powerful quasar-driven outflows in relatively low-density environments that are inferred (from photoionization arguments) to reside at least $\sim$1~kpc from the quasars. The near-IR observations also reveal that these ERQs have extreme kinematics in their broad emission-line regions, including blueshifts that can exceed 2500 \kms\ in \civ\ and other high-ionization UV lines, e.g., compared to the \hi\ Balmer lines and low-ionization permitted lines in the UV (Hamann et al. 2016a, in prep., also \S5.8 below). 

This ensemble of exotic emission-line properties is central to the physical nature of ERQs and their possible relationship to an early transition stage of quasar-galaxy evolution. We present a detailed analysis of the emission lines and line-forming regions of ERQs in Hamann et al. (2016a, in prep.). In the present paper, we combine broad-band photometry from SDSS and WISE with new measurements of the \civ\ and \nv\ emission lines in the final BOSS data release (DR12) to 1) quantify the emission-line properties of ERQs compared to the overall BOSS quasar population, 2) examine the relationships of these properties to quasar colors and luminosities, and 3) revise the selection criteria to find many more ERQ with similar exotic properties. How rare are the emission-line properties of ERQs in BOSS quasars overall? Are they closely tied to reddening and obscuration? Do they correlate with outflow signatures such as blueshifted broad absorption lines (BALs)? Are ERQs with exotic properties a unique population or just outliers in trends that occur across the larger BOSS quasar population?  

Section 2 describes the quasar samples and photometric data used in this study. \S3 and Appendix A presents our new catalog of UV line and continuum measurements. \S4 examines the relationships of ERQ emission-line properties to the quasar colors and luminosities across the BOSS quasar population. \S5 describes the selection and important characteristics of a new large sample of ERQs with exotic properties. \S6 discusses some of the implications of our results and \S7 provides a summary. Appendix B tabulates a supplemental sample of ``ERQ-like'' quasars. Throughout this paper, we adopt a cosmology with $H_o = 71$ \kms\ Mpc$^{-1}$, $\Omega_M = 0.27$ and $\Omega_{\Lambda}=0.73$. We also use magnitudes in the AB system except as noted.

\section{Quasar Samples and Datasets}

Table 1 provides a summary of the quasar samples discussed in this paper. Our starting point is the BOSS quasar catalog for Data Release 12 \citep[hereafter DR12Q,][]{Paris14, Paris16}. From this sample, we develop a new catalog of \civ\ and \nv\ emission-line data for 216,188 quasars in the redshift range $1.53\le z_e\le 5.0$. This new catalog is described in \S3 and Appendix A. Most of our analysis focusses on quasars with measured line properties in this catalog. However, we also examine all quasars with extreme red colors in DR12Q to ensure that our final ERQ samples are complete in the BOSS DR12Q database.

\begin{table}
	\centering
	\caption{Quasar samples designated by a name, selection criteria, number of quasars, and text sections where the sample is described. Samples listed above the dotted line are subsets of all samples higher in the table. The ERQ-like sample below the dotted line is selected more broadly to have emission line properties like the core ERQs. The numbers 95+2 and 228+7 listed for the last two samples refer to 95/228 quasars in our emission-line catalog plus 2/7 more identified by visual inspections and additional line fits for all BOSS DR12Q quasars with \imw\ $\ge 4.6$. }
	\begin{tabular}{lccc} 
		\hline
		Sample Name& Selection Criteria & Number & Text Ref.\\
		\hline
		DR12Q& ---& 297,301& \S2\vspace{4pt}\\
		Emission-line& $1.53 \le z_e\le 5.0$& 216,188& \S3, App. A\\
		~~~ catalog\vspace{4pt}\\
		Full Sample& $2.0<z_e<3.4$& 173,636& \S2\\
		& $i$ mag in DR12Q\\
		& well-measured \civ  \vspace{4pt}\\
		$W3$-detected& SNR($W3$) $>$ 3& 36,854& \S2\\
		& cc\_flags = 0000\vspace{4pt}\\
		ERQs& \imw\ $\ge 4.6$& 205& \S2, \S4.5\vspace{1pt}\\
		Core ERQs& \ REW(\civ ) $> 100$\AA & 95+2& \S5.1\\
		\multicolumn{4}{c}{\dotfill}\\
		ERQ-like & $2.0 < z_e < 4.0$& 228+7 & \S5.7, App. B\\
		& core ERQ-like \civ \\
		\hline
	\end{tabular}
\end{table}

We begin by defining a ``full sample'' of 173,636 quasars in the redshift range $2.0 < z_e < 3.4$ (Table 1). This limited redshift range encompasses most of the BOSS survey while ensuring that 1) \lya\ and \nv\ \lam 1240 are within the BOSS spectral coverage, 2) \civ\ \lam 1549 does not contaminate the $i$ band magnitudes, and 3) there is no significant redshift dependence in the \imw\ color (Figure 9 in \S4.3) that we use to select and study the ERQs below. We also require that the quasar has an $i$ magnitude recorded in DR12Q and that the \civ\ lines are well measured in our catalog based on signal-to-noise ratio SNR $\ge$ 4 in both REW(\civ) and FWHM(\civ), a reasonable value of $500 <$ FWHM(\civ ) $< 20,000$ \kms , and quality flag {\tt qflag = 0} indicating no significant problems with the line or continuum fits (Appendix A). The median redshift of this full sample is $\left<z_e\right> \approx 2.5$. 

We color select using broad-band photometry from the SDSS \citep{York00,Alam15}, WISE \citep{Wright10,Yan13} and UKIDSS \citep{Casali07}, as provided in the BOSS DR12Q quasar catalog \citep[see][]{Paris14, Paris16}. SDSS photometry in the filters $ugriz$ is available for every quasar \citep[e.g.,][and refs. therein]{Aihara11}. We correct these magnitudes for Galactic extinction using offsets from \cite{Schlafly11}. WISE provides fluxes in four bands $W1$, $W2$, $W3$ and $W4$ centered at 3.4, 4.6, 12 and 22 \mum , respectively. The WISE data are from the most recent ``AllWISE" release \citep{Cutri13}, which yields deeper photometry in the two shorter wavelength bands and better results overall than the earlier data release ``AllSky" \citep[see also][for more discussion]{Ross15}. We convert WISE magnitudes in the Vega system to AB using offsets provided by \cite{Cutri11}. UKIDSS provides photometry in the filters $YJHK$, which are already converted to fluxes in DR12Q \citep[][]{Paris14, Paris16}. 

Note that the broad-band rest-frame UV properties of quasars in our study are constrained by the target selection criteria used in BOSS \citep{Bovy11,Ross12}. This is important to keep in mind because it skews our searches for ERQs toward a particular unusual SED that is relatively flat across the rest-frame UV while being extremely red from the UV to mid-IR \citep[see \S5.5, also][]{Ross15}. 

WISE $W3$ and the specifically the \imw\ color are critical for the selection and analysis of ERQs at the redshifts of our study. The subset of our full sample with good $W3$ measurements based on WISE contamination and confusion flag {\tt cc\_flags} = 0000, which indicates no significant problems, and SNR $\ge$ 3 in the $W3$ flux (as listed in DR12Q) includes 36,854 quasars. 

Within this $W3$-detected sample, we find 205 ERQs defined by \imw\ $\ge 4.6$ and a subset of 95 ``core'' ERQs that have both \imw\ $\ge$ 4.6 and REW(\civ ) $>$ 100 \AA . These parameter thresholds are based on our analyses in \S4. We also identify 228 ``ERQ-like'' quasars that span wider ranges in redshift and color with emission-line properties like the core ERQs (\S5.7). We supplement these core ERQ and ERQ-like samples selected from our emission-line catalog by performing visual inspections and additional line fits to all quasars that satisfy \imw\ $>$ 4.6 in DR12Q. This additional search finds only 2 more core ERQs and 7 more ERQ-like quasars in DR12Q that are not in our emission-line catalog (see S3 and Appendix A for details). Thus the total numbers of quasars in the core ERQ and ERQ-like samples are 95+2=97 and 228+7= 235, respectively.

Finally, for some of our discussions, we consider a crude division between Type 1 and Type 2 sources based on FWHM(\civ ) $\geq$ 2000 \kms\ or $<$ 2000 \kms , respectively \citep[following previous studies by][]{Alexandroff13,Ross15}. We add to this the requirement that Type 2 quasars cannot have BALs based on the visual inspection flag in DR12Q. This distinction between Type 1 and 2 sources is just a guideline to help describe and compare the different samples. The ambiguities in Type 1 versus Type 2 classifications based on UV spectra are discussed further in \S5.4 \citep[also][]{Zakamska03, Zakamska16, Reyes08, Alexandroff13, Greene14}.

\section{Line \& Continuum Measurements}

Emission line measurements are an important part of our study. We developed simple robust procedures to fit the UV continuum and the \civ\ \lam1548,1551 (hereafter \lam 1549) and \nv\  \lam 1238,1242 (hereafter \lam 1240) emission lines in 216,188 quasars in BOSS DR12Q. The quasar redshift are limited to the range $1.53 \le z_e \le 5.0$ so that \civ\ and the adjacent continuum are covered by BOSS. Our measurements provide line profile information and \nv /\civ\ flux ratios not available in DR12Q, and they can yield better results for quasars like the ERQs that have unusual line properties and/or faint rest-frame UV continua that lead to noisy spectra in BOSS. The only BOSS quasars in this redshift range excluded from our measurements have strong \civ\ BALs that are reported in DR12Q to be at velocities that might interfere with our fits to the CIV emission line profiles or to the continuum beneath CIV. Appendix A describes the fitting procedures and the final catalog of results. Here we provide a brief summary. 

Throughout this paper we adopt the best available emission-line redshifts ($z_e$) from DR12Q, e.g., derived from spectral fitting except in rare cases where those are not available we use the visual inspection redshifts. In either case, precise redshifts are not important for our study. We use separate power laws to fit the continuum beneath the \civ\ and \nv\ emission lines, and then another single power law across the wavelength range 1360-2230 \AA\ (rest) to measure the overall UV continuum slope, $\alpha_{\lambda}$ (for $f_{\lambda}\propto \lambda^{\alpha_{\lambda}}$). The values of $\alpha_{\lambda}$ that we use below are derived from the BOSS spectra after applying wavelength-dependent flux corrections \citep[from][see Appendix A]{Harris15}. 

We fit the \civ\ emission line profiles with two Gaussian components and then use these fits to measure basic line parameters including the REW, FWHM, and a kurtosis index we call $kt_{80}$ that characterizes the profile shape in terms of the width of the line core relative to the wings. Specifically, this index measures the velocity width at 80\% of the peak height divided by the width at 20\%, e.g., $kt_{80}\equiv \Delta {\rm v}(80\%)/\Delta {\rm v}(20\%)$. For comparison, a single Gaussian has $kt_{80} = 0.372$ while most quasars have substantial logarithmic or Lorentzian line wings that yield $kt_{80} \sim 0.15 - 0.3$ (\S4.2 below). The $kt_{80}$ kurtosis index is a variation on the index used by \cite{Marziani96} and \cite{Zamfir10} that provides a slightly greater dynamic range between different profile shapes. 

The \nv\ emission line presents unique problems because it is usually blended with the \lya\ emission line, which, in turn, can be distorted by absorption in the \lya\ forest at rest wavelengths $\la$1216 \AA . We avoid these problems by using the fitted \civ\ profile as a template that we shift and scale to match the data at the \nv\ wavelengths. Visual inspections of several thousand spectra show that the \civ\ and \nv\ fits are generally excellent. We note, however, that the \nv\ line strength can be overestimated in cases where \lya\ is unusually broad and strong leading to substantial flux relative to \nv\ across the \nv\ wavelengths. This is not a factor for any of the ERQs. 

\section{Analysis}

Here we investigate the relationships of the emission-line properties of ERQs to the colors and luminosities of quasars in BOSS DR12Q. Figures 1 and 2 plot the measured REW(\civ ) and FWHM(\civ ) distributions for all quasars in our full sample, the $W3$-detected subsample, and two ERQ samples defined in Table 1 (\S2). 

\begin{figure}
\begin{center}
 \includegraphics[scale=0.45,angle=0.0]{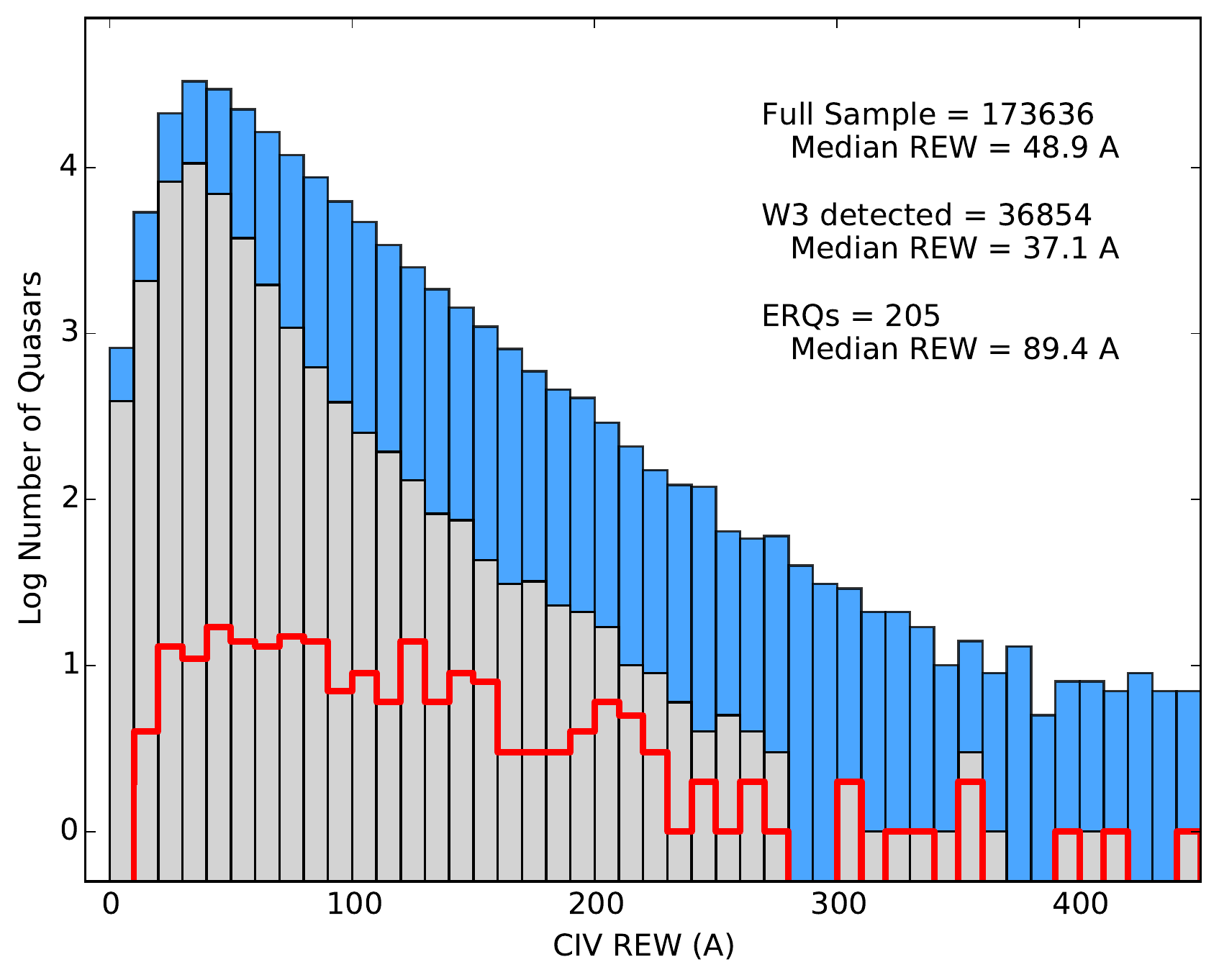}
 \vspace{-8pt}
 \caption{REW(\civ ) distributions for all quasars in our full sample (blue histogram), the WISE $W3$-detected sample (gray histogram), and extremely red quasars (ERQs) with \imw\ $\ge 4.6$ (red histogram). The log numbers of quasars on the left axis are per 10 \AA\ bin in REW(\civ ). See \S2.}
 \end{center}
\end{figure}

\begin{figure}
\begin{center}
 \includegraphics[scale=0.45,angle=0.0]{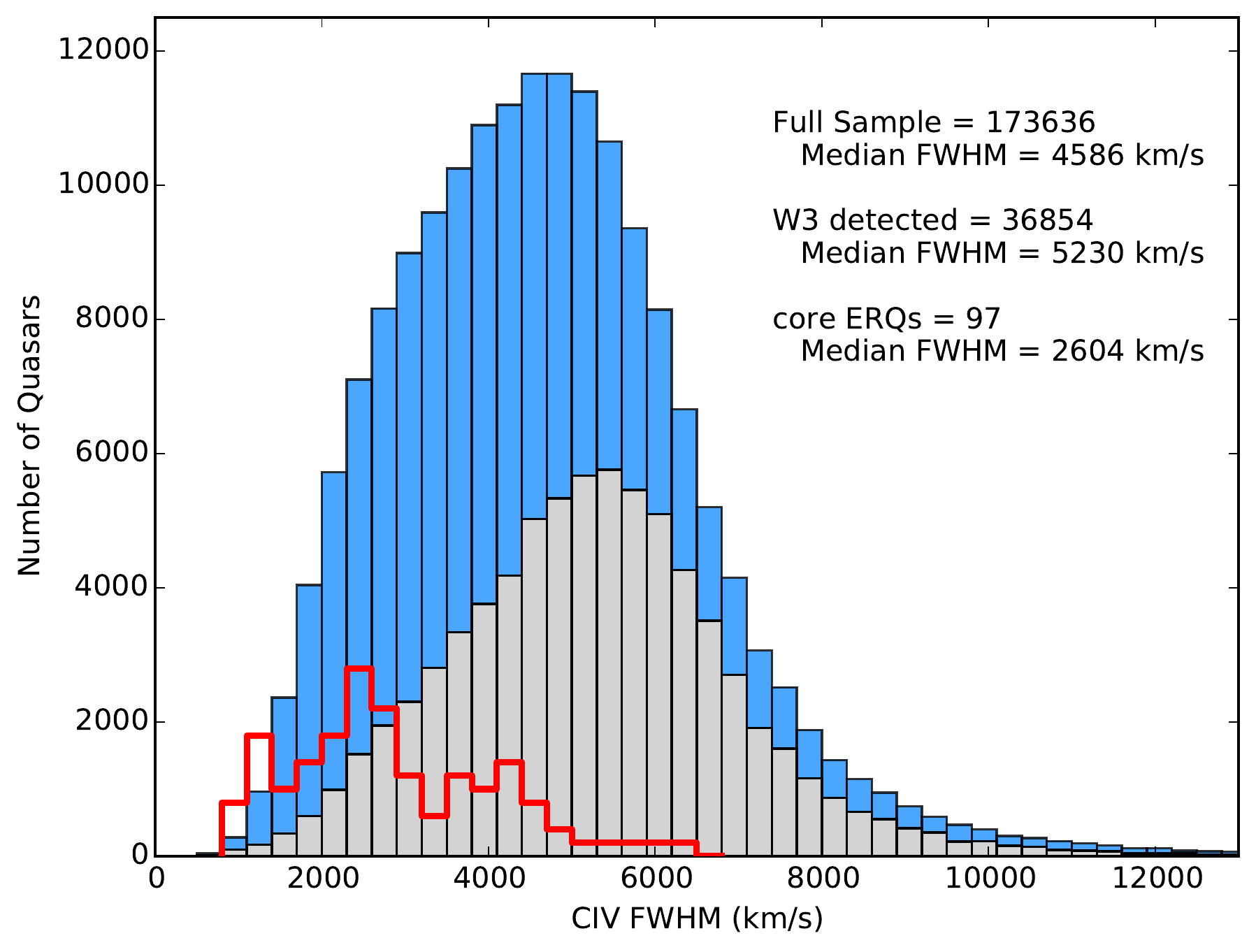}
 \vspace{-8pt}
 \caption{FWHM(\civ ) distributions for all quasars in our full sample (blue histogram), the WISE $W3$-detected sample (gray histogram), and ERQs in the ``core" sample with \imw\ $\ge 4.6$ and REW(\civ ) $\ge$ 100 \AA\ (red histogram). The numbers of quasars on the left axis are per 300 km/s bin in FWHM(\civ ). The $W3$-detected and core ERQ distributions are multiplied by 2 and 200, respectively. See \S2.}
 \end{center}
\end{figure}

We retain the term ``ERQ'' from \cite{Ross15} even though we use it to indicate a less stringent color constraint of $i-W3 \ge 4.6$ compared to $r-W4> 7.5$ (corresponding to $r-W4 > 14$ Vega magnitudes) in the Ross et al. study. There are 205 ERQs defined by $i-W3 \ge 4.6$ in our $W3$-detected sample. Note that the $i$ and $W3$ filters measure the quasar fluxes at rest wavelengths of $\sim$0.2 \mum\ and $\sim$3.4 \mum , respectively, at the median redshift $\left<z_e\right> \sim 2.5$ of our samples. We use $i$ instead of $r$ to define ERQs because $r$ can be severely contaminated by \civ\ in emission or absorption at redshifts $\ga$2.7. The advantages of $W3$ over $W4$ for ERQ selection are described in \S4.4. Also note that there is no significant redshift dependence in the observed \imw\ colors in our samples (Figure 9 below).

\subsection{Color versus Luminosity Dependence of REW(\civ )}

One important characteristic of ERQs is a tendency for large emission-line REWs (Figure~1). \cite{Ross15} noted that 45\% of Type 1 non-BAL ERQs have REW(\civ ) $>$ 150 \AA\ compared to only 1.3\% for BOSS quasars overall. This is a remarkable result, but there are thousands of BOSS quasars with REW(\civ ) $>$ 150 \AA\ (Figure~1) and most of them are {\it not} red. This raises the question of whether the large REWs in ERQs are related to their extreme red colors or perhaps some other property of the quasars, such as their luminosities. 

The Baldwin Effect \citep{Baldwin77} is an empirical inverse correlation between emission-line REWs and luminosity in Type 1 quasars \citep[see][for a review]{Shields07}. The REW distributions shown in Figure~1 are strongly affected by this correlation. In particular, the $W3$-detected sample is skewed toward smaller REWs than the full sample because the limited sensitivity of $W3$ excludes many low-luminosity quasars with large REWs. 

Figure~3 illustrates the \civ\ Baldwin Effect more directly for all non-BAL Type 1 quasars in our full and $W3$-detected samples. BAL quasars are excluded from this plot based on the visual inspection flag  {\tt bal\_flag\_vi = 0} in DR12Q. The continuum luminosities at $\lambda = 1450$ \AA\ in the rest frame, $\lambda L_{\lambda}(1450)$, are estimated by extrapolation from the observed broad-band $r$ or $i$ fluxes (for redshifts $z_e < 2.7$ or $\ge 2.7$, respectively) using the UV spectral slopes $\alpha_{\lambda}$ determined from our spectral fits (Appendix A). The results shown are in good agreement with previous studies. In particular, the slope of the log-log distribution for the full sample, $\beta\approx -0.23$, is similar to the value $\beta\approx -0.20$ derived by \cite{Dietrich02} for quasars with similar luminosities. 

\begin{figure}
\includegraphics[scale=0.48,angle=0.0]{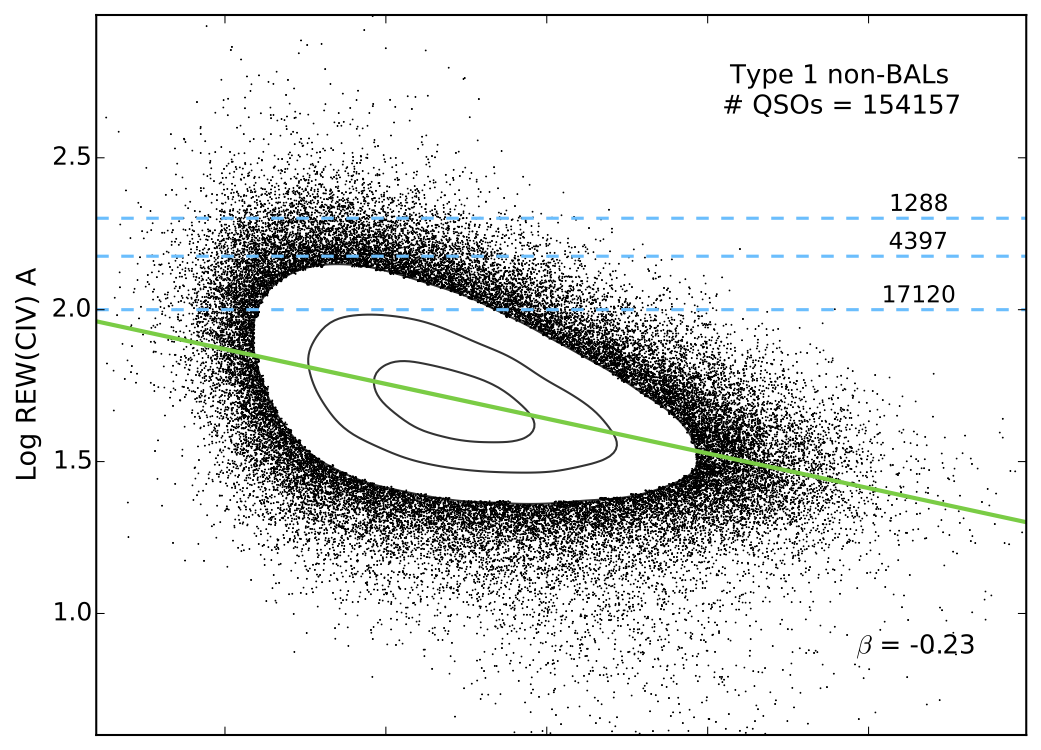}
\includegraphics[scale=0.48,angle=0.0]{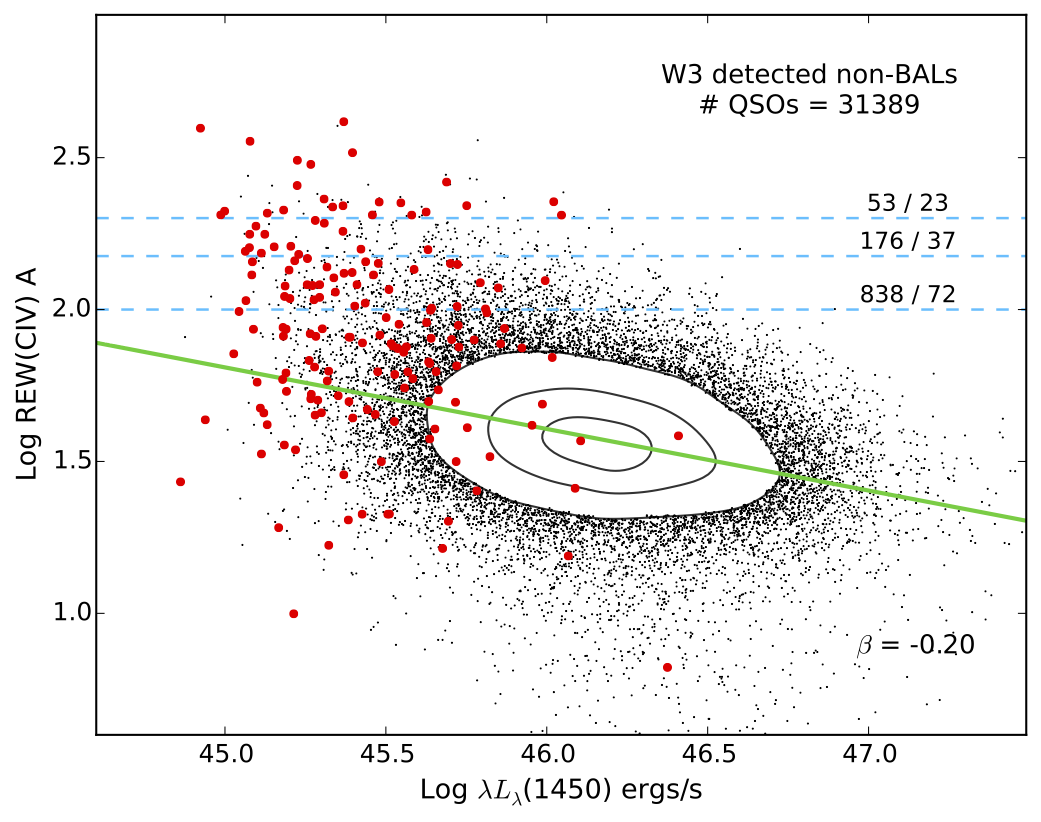}
  \vspace{-10pt}
 \caption{\civ\ Baldwin Effect for Type 1 non-BAL quasars in our full sample (top panel) and the $W3$-detected subsample (bottom). The green lines are linear fits to the log-log distributions with slopes given by $\beta$ in the lower right. The red dots in the bottom panel mark ERQs with \imw\ $\ge 4.6$. The dashed blue lines at REW(\civ ) = 100, 150 and 200 \AA\ are labeled by the numbers of quasars above these thresholds. In the bottom panel, the two numbers indicate all quasars plotted / ERQs only. The contours shown here and in all subsequent figures mark the quasar point densities at 85\%, 55\% and 25\% of the maximum in the plot, where the maximum is measured in a box whose size is 1/50th of the full $x$ and $y$ axis dimensions.}
\end{figure}

The REW thresholds marked by the dashed blue lines in Figure~3 show that there are 4397 Type 1 non-BAL quasars with REW(\civ ) $>$ 150 \AA\ in our full sample. As noted above, Type 1 non-BAL quasars with $W3$ detections (bottom panel in Figure~3) tend to have smaller REWs because this sample favors large $\lambda L_{\lambda}(1450)$. The ERQs (marked by red dots) strongly favor large REWs at {\it apparent} low luminosities. This seems consistent with the Baldwin Effect. However, the faint $i$ magnitudes of ERQs are caused by  typically $\sim$3 magnitudes of UV obscuration (\S5.5, \S6.1), their actual luminosities are typically $\sim$1.2 dex larger than depicted in this plot (\S5.1) and the Baldwin Effect does {\it not} explain their large REWs. 

Figure~4 shows that quasars with REW(\civ ) $>$ 150 \AA\ strongly favor red \imw\ colors. This figure plots the \imw\ distribution for Type 1 quasars in our $W3$-detected sample (grey histogram), the fractions of these quasars with REW(\civ ) $>$ 150 \AA\ (blue histogram), and the distribution of Type 1 $W3$-detected quasars with REW(\civ ) $>$ 150 \AA\ (red histogram). The median\footnote{Throughout this paper, we provide sample medians with ``standard errors'' that are derived by a bootstrapping technique (resampling with replacement). For our samples, these errors are equivalent to standard deviations. The uncertainties contributed by individual measurement errors are insignificant compared to the standard errors.} color in the $W3$-detected sample is $\left<i-W3\right> = 2.50\pm 0.56$. The fraction of quasars with REW(\civ ) $>$ 150 \AA\ at this median color is only about 0.4\%. The red histogram shows that the color distribution of quasars with REW(\civ ) $>$ 150 \AA\ is dramatically offset toward the red with a strong extension and secondary peak at \imw\ $\ga 4.6$. This redward extension/peak is the ERQs defined by \imw\ $\ga 4.6$, for which the fraction with REW(\civ ) $>$ 150 \AA\  reaches $\sim$50\% at the red extreme.
 
\begin{figure}
\begin{center}
 \includegraphics[scale=0.43,angle=0.0]{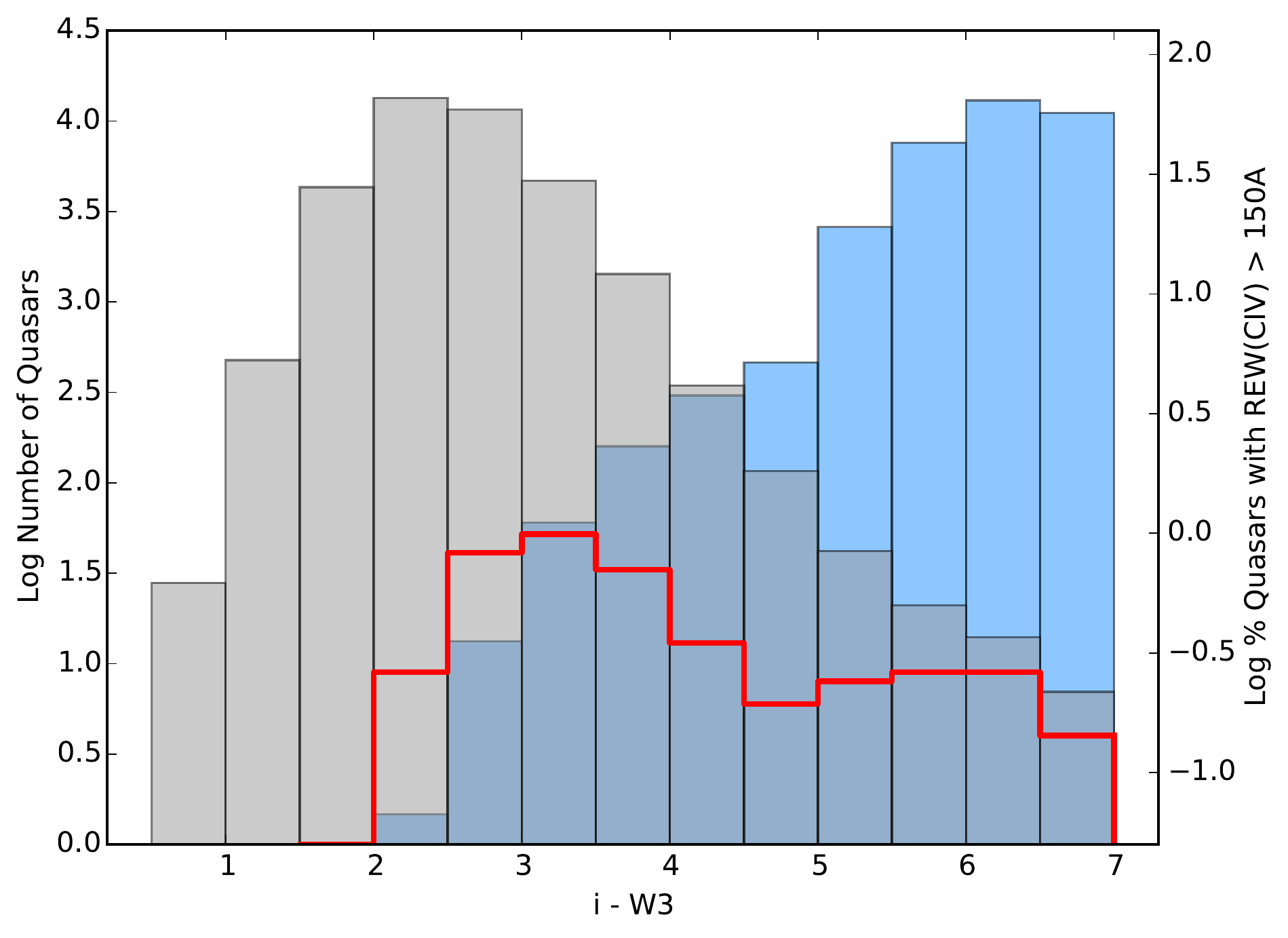}
  \vspace{-8pt}
 \caption{Distributions in \imw\ color for Type 1 quasars in the $W3$-detected sample (gray histogram) and the subset with REW(\civ ) $>$ 150 \AA\ (red histogram). The blue histogram shows the percentages of all Type 1 $W3$-detected quasars with REW(\civ ) $>$ 150 \AA\ at each \imw\ color, given on the right-hand vertical scale.}
 \end{center}
\end{figure}

Unfortunately, Figure~4 also includes selection effects that couple red \imw\ colors to faint $i$ magnitudes and thus larger REWs via the Baldwin Effect. Figure~5 shows these effects for each of the colors \imwt\, \imw\ and \imwf . In particular, quasars that are faint in $i$ in our $W3$-detected sample are necessarily red in \imw . Faint blue quasars are excluded because they are below the sensitivity limits of the WISE photometry (below the green curves in Figure~5). This makes the dependence of large REWs on red colors versus faint $i$ magnitudes difficult to disentangle. However, it is clear from the upward extensions of red/orange points representing quasars with REW(\civ ) $>$ 150 \AA\ in Figure~5, and particularly from the red/orange squares representing large REWs with unusual wingless profiles ($kt_{80} > 0.33$), that red colors are important to find rare quasars with both of these peculiar emission-line properties. 

\begin{figure}~
\begin{center}
\includegraphics[scale=0.48,angle=0.0]{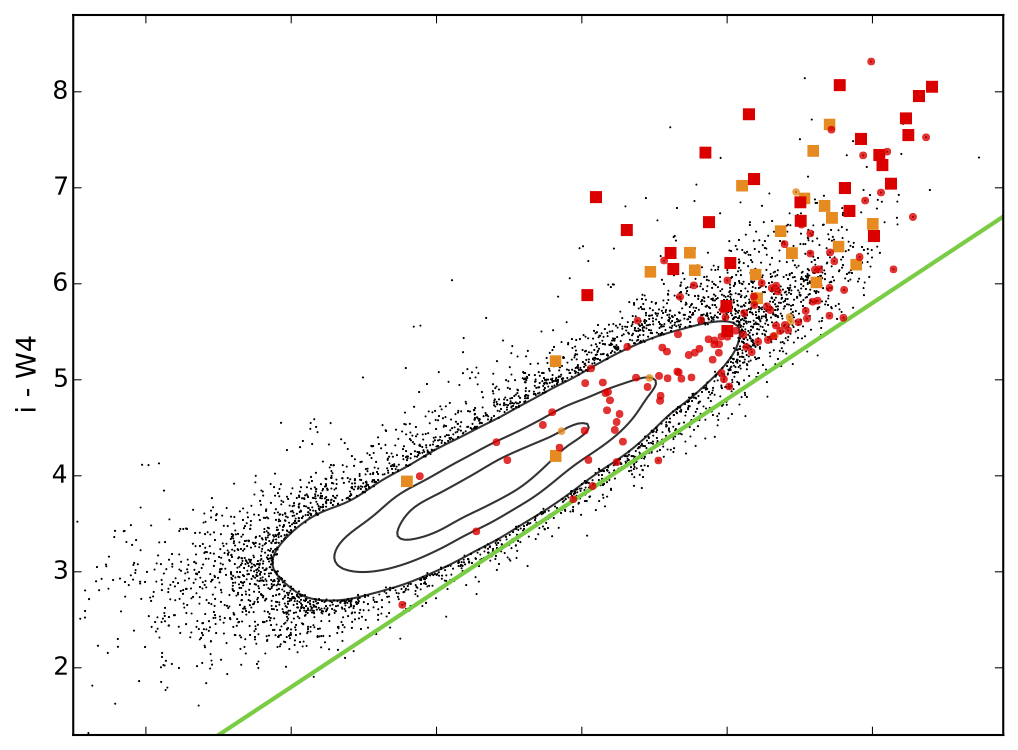}  
\includegraphics[scale=0.48,angle=0.0]{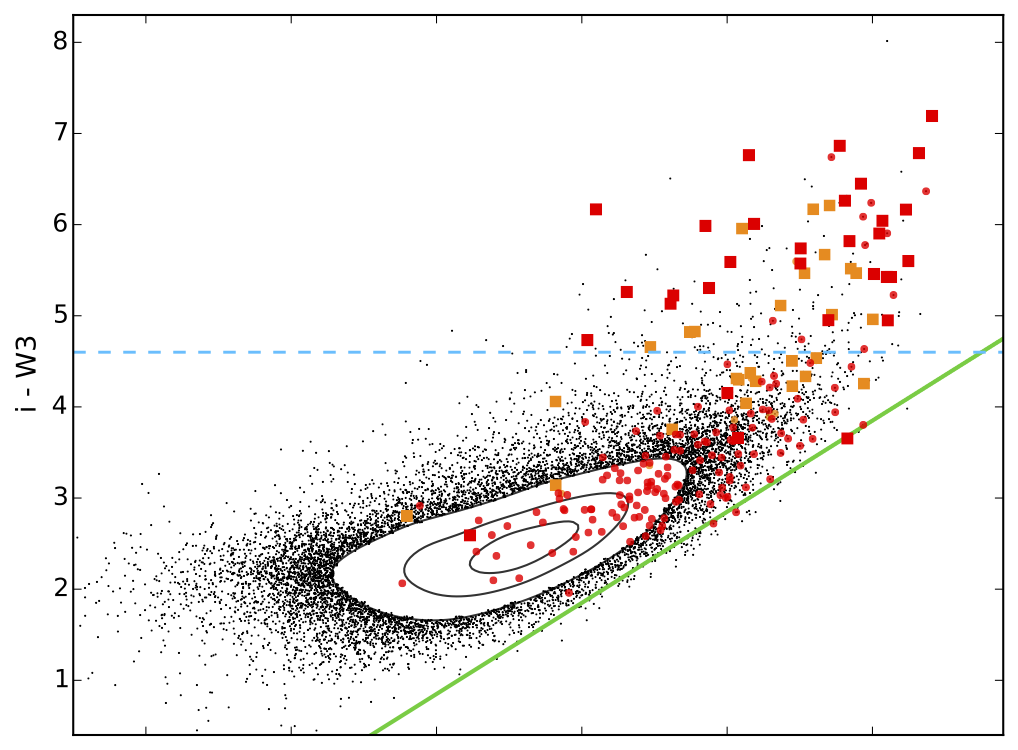}  
\includegraphics[scale=0.48,angle=0.0]{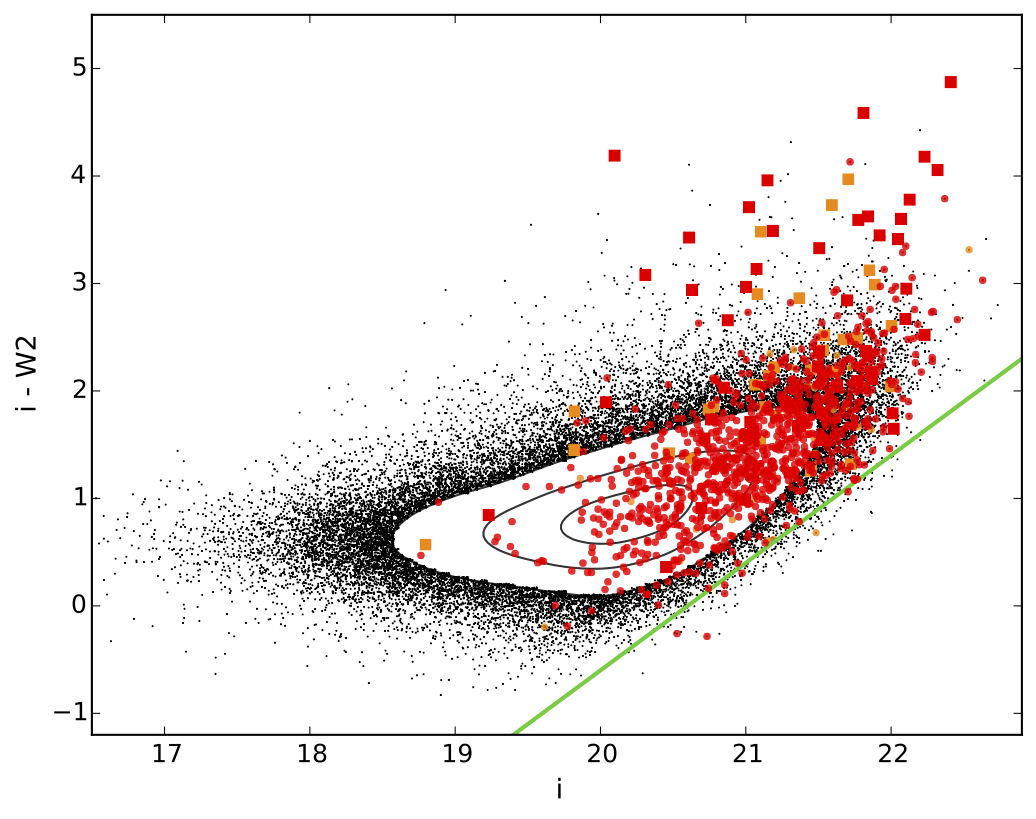}  
\end{center}
  \vspace{-8pt}
 \caption{SDSS $i$ magnitude versus \imwf\ (top panel), \imw\ (middle) and \imwt\ (bottom) for all quasar in our full sample with good $W4$, $W3$ or $W2$ detections (SNR $>$ 3 and cc\_flag = 0000), respectively. The green lines show approximate sensitivity limits of the WISE filters. The red and orange dots indicate Type 1 and 2 quasars, respectively, with REW(\civ ) $>$ 150 \AA . The squares indicate REW(\civ ) $>$ 150 \AA\ {\it and} ``wingless" profiles with $kt_{80} > 0.33$. The dashed blue line in the middle panel marks the ERQ threshold at \imw\ = 4.6.}
\end{figure}

Figure~6 attempts to isolate the $i$ magnitude dependence by plotting the $i$ distributions and the fractions of quasars with REW(\civ ) $>$ 150 \AA\ only for blue quasars with $i-W3< 3.5$. Many quasars in this plot are not detected in $W3$, e.g., below the green line in the lower right corner of Figure~5. For these $W3$ non-detections, we still require good WISE measurement in the more sensitive $W1$ or $W2$ bands (with {\tt cc\_flags = 0000}) to ensure that the $W3$ non-detections are not due to measurement problems. We see that faint blue quasars also have a high incidence of REW(\civ ) $>$ 150 \AA , as expected from the Baldwin Effect. However, at a magnitude equal to the median for ERQs, $\left<i\right>\sim 21.4$ (Figure~5 middle panel), the percentage of blue quasars with REW(\civ ) $>$ 150 \AA\ is only $\sim$5\% compared to 25\% for the ERQs overall and $\sim$50\% for the reddest ERQs with \imw\ $\ga 5.5$ (e.g., Figures 1 and 4). This shows that extreme large REWs are more closely related to red \imw\ colors than to faint $i$ magnitudes. 

\begin{figure}
 \includegraphics[scale=0.43,angle=0.0]{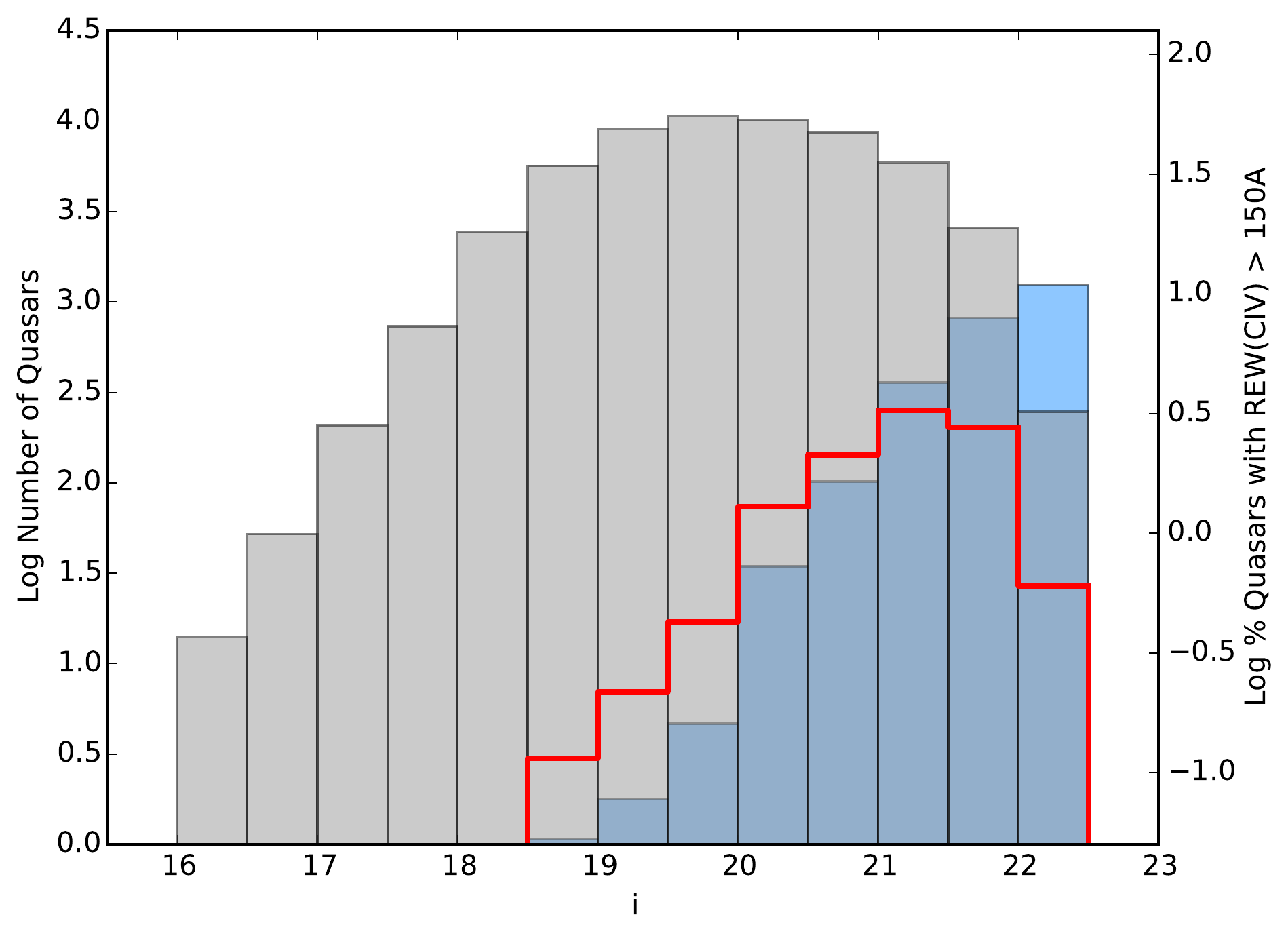}
  \vspace{-8pt}
 \caption{Distributions in $i$ magnitude for blue Type 1 quasars with \imw\ $<$ 3.5 (gray histogram) and the subset of them with REW(\civ ) $>$ 150 \AA\ (red histogram). The blue histogram shows the percentages of blue quasars with REW(\civ ) $>$ 150 \AA\ at each $i$ magnitude, given on the right-hand vertical scale. See \S4.1 for more on the samples plotted.}
\end{figure}

Moreover, if the red colors and faint $i$ magnitudes of ERQs are caused by UV obscuration (as expected, \S6.1), then ERQs are luminous and their actual peers in the Baldwin Effect are luminous blue quasars with similar $W3$ magnitudes (not similar $i$) that strongly favor {\it small} REWs in the Baldwin Effect (Figure~3). The median $W3$ magnitude of ERQs with $i-W3 > 4.6$ is $\left< W3\right> \approx 16.1\pm 0.7$. Out of 6119 blue quasars in our emission-line catalog with similar $W3$ magnitudes (in the range $15.3 < W3 < 16.5$ to yield a median $\left< W3\right> \approx 16.2\pm 0.3$), only 2 (0.03\%) have REW(\civ ) $>$ 150 \AA\ and 25 (0.4\%) have REW(\civ ) $\ge 100$ \AA . The median REW(\civ ) in this $W3$-matched blue sample is also only $31.8\pm 14.2$ \AA\ compared to $89.4\pm 81.4$ \AA\ for the ERQs (where the large standard error for the ERQs reflects the distribution reaching very large REW(\civ ), Figure~1). If the Baldwin Effect is operating at all in the ERQs, it would push them toward {\it smaller} REWs like these blue quasars, not larger ones as observed. We conclude that the Baldwin Effect plays no role whatsoever in the extreme large REWs of ERQs.

\subsection{Exotic Line Properties tied to \imw\ Color}

Another important feature of ERQs is that the large REWs are accompanied by peculiar wingless line profiles and often by exotic line ratios like \nv\ $>$ \civ\ and \nv\ $>$ \lya  . Figure~5 above indicates that \imw\ $\ge$ 4.6 (middle panel) is particularly effective at separating quasars with large REWs and wingless profiles from the rest of the quasar population. Figure~7 shows more explicitly how these ERQ line properties are related to each other and strongly correlated with red \imw\ color. The top row (panels A and B) shows that ERQs defined by \imw\ $\ge$ 4.6 tend to be bright in $W3$ compared to other quasars in the $W3$-detected sample. If $W3$ is a reasonable surrogate for unobscured luminosity (\S4.1), then these plots show that ERQs tend to be about a magnitude more luminous than other quasars in our $W3$-detected sample (see also \S5.1 and \citealt{Ross15}). Panels C and E show that ERQs strongly favor both wingless line profiles and large REWs. They cluster in the upper right corner of both plots exhibiting a strong preference for REW(\civ ) $\ga$ 100 \AA\ and $kt_{80}\ga 0.33$. Panels D and F show that ERQs span a wide range in \nv /\civ\ ratios, but they favor large \nv /\civ\ much more than blue quasars (panel F) and they uniquely have large \nv /\civ\ accompanied by {\it large} REWs (panel D). 

\begin{figure*}
\begin{center}
\includegraphics[scale=0.48,angle=0.0]{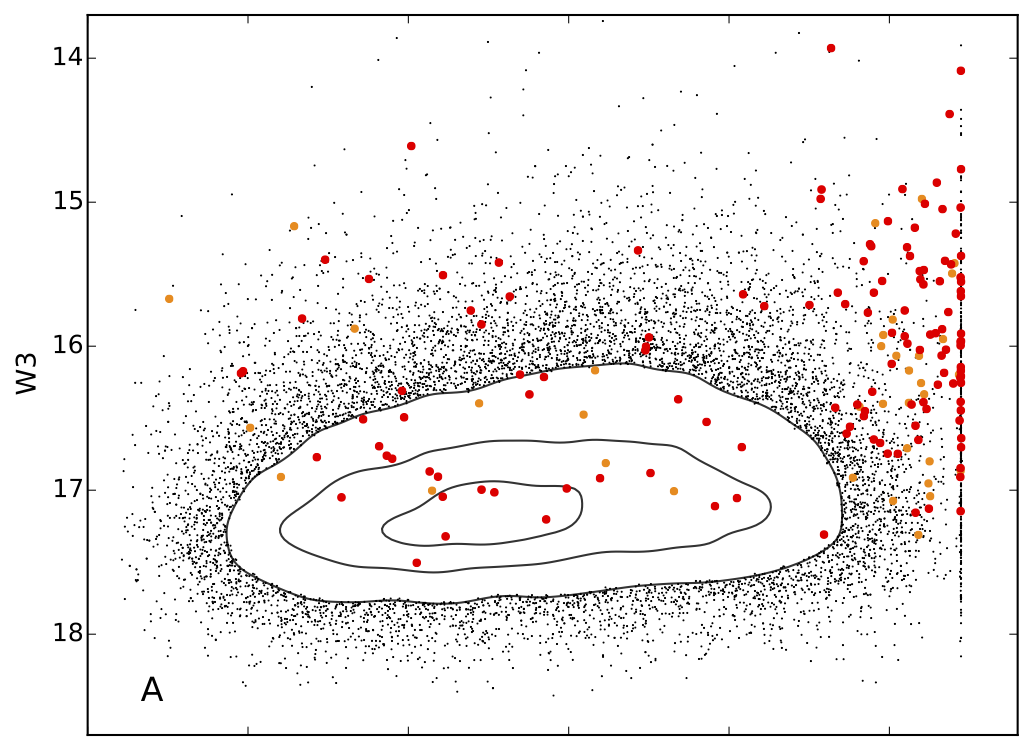}  
\includegraphics[scale=0.48,angle=0.0]{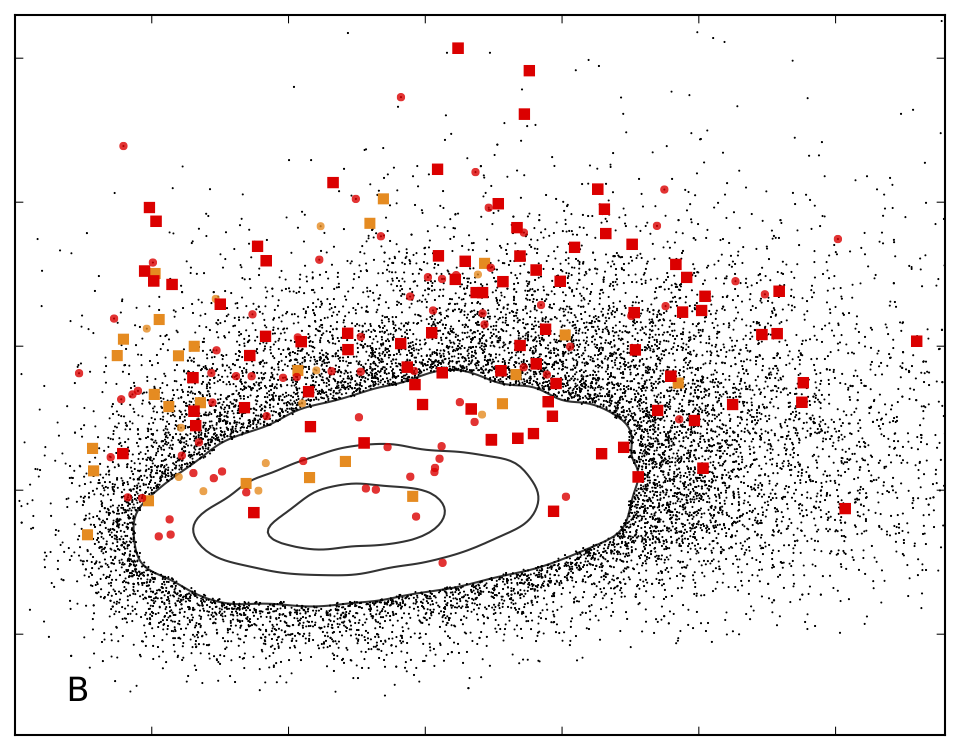}  
\includegraphics[scale=0.48,angle=0.0]{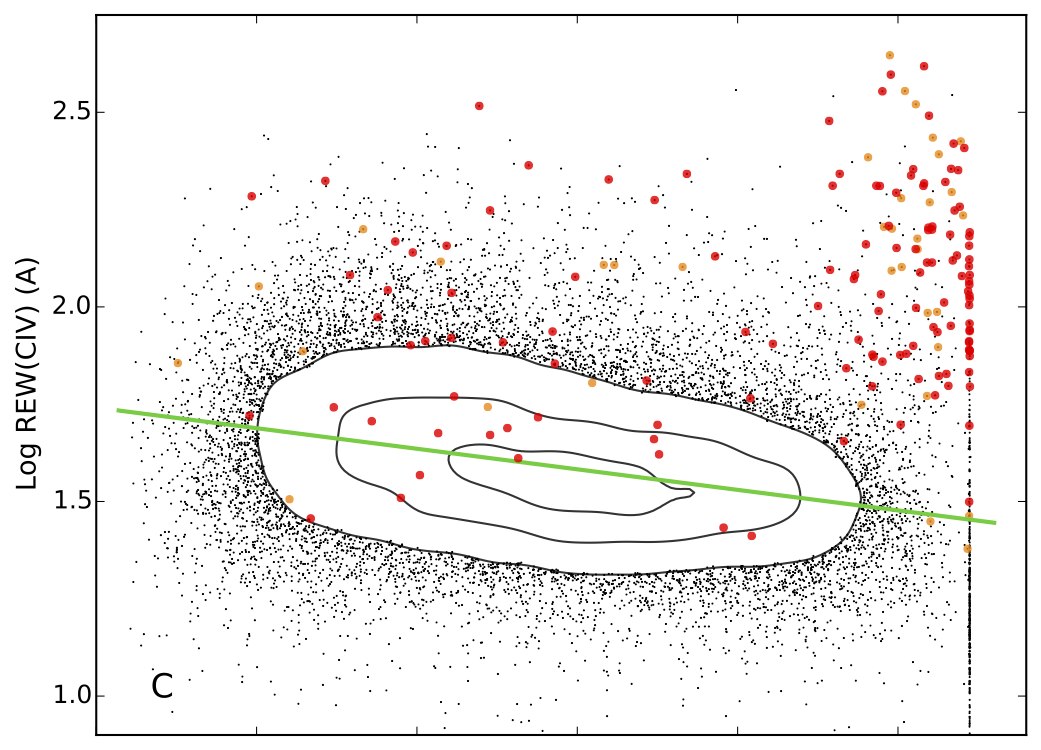}  
\includegraphics[scale=0.48,angle=0.0]{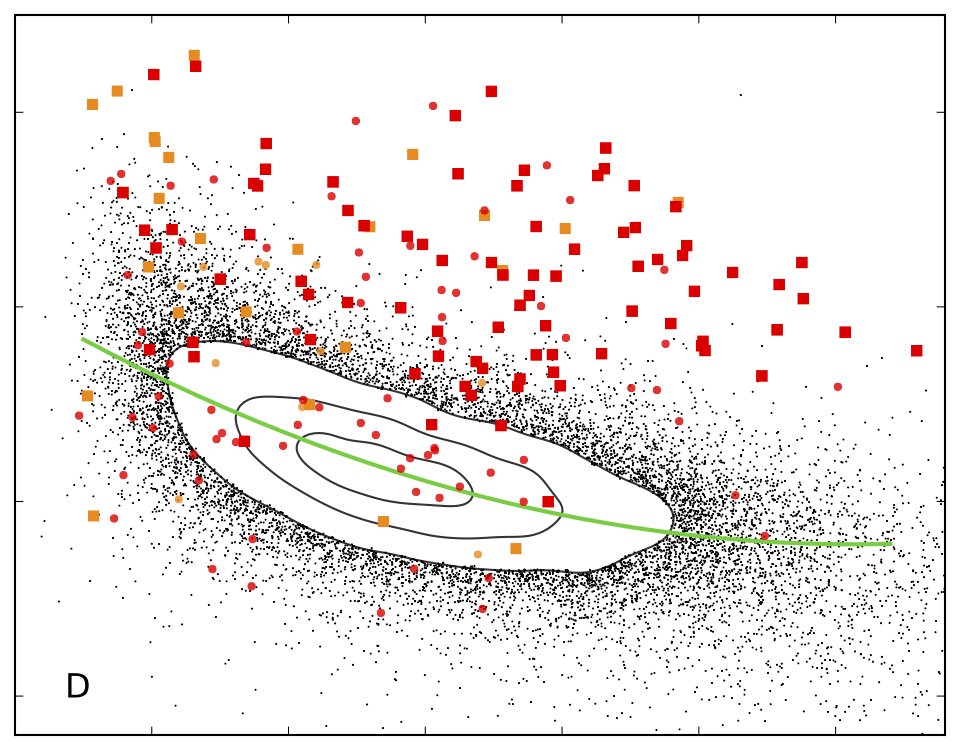}  
\includegraphics[scale=0.48,angle=0.0]{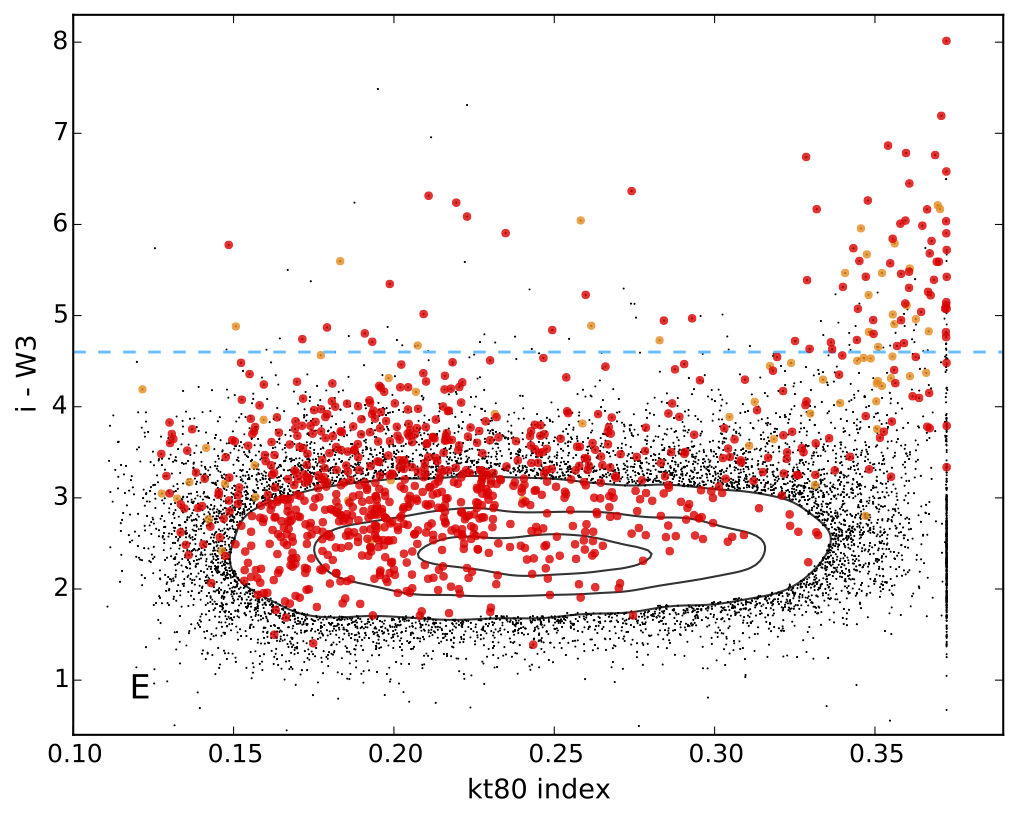}  
\includegraphics[scale=0.48,angle=0.0]{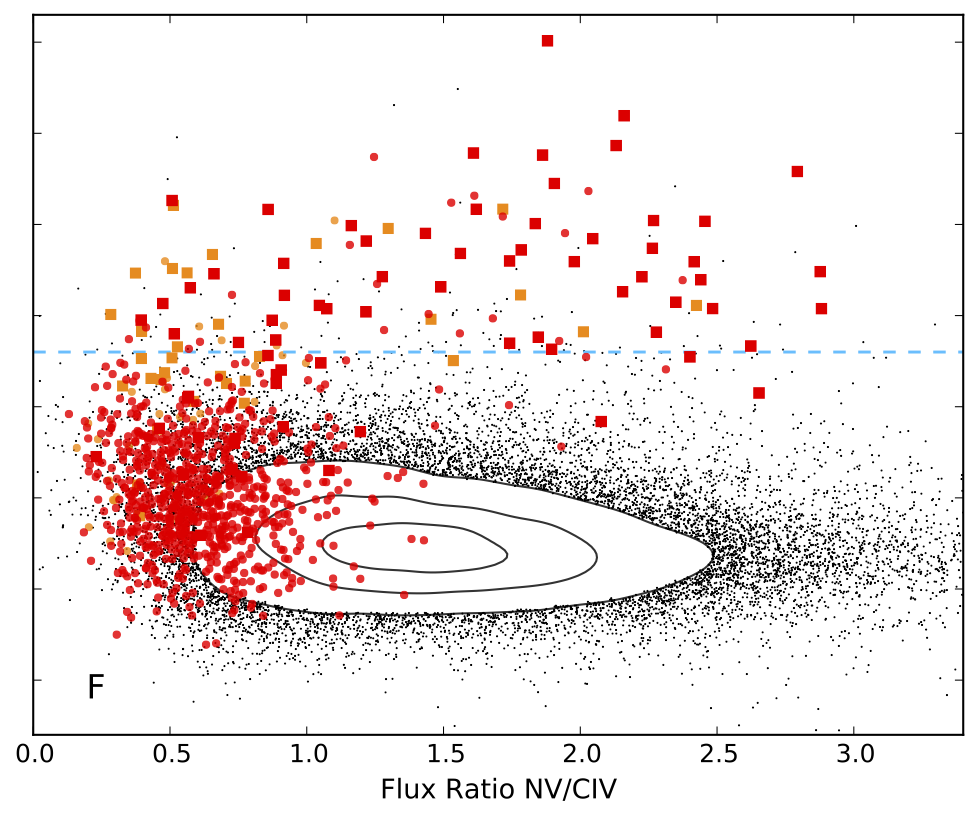}  
  \vspace{-4pt}
 \caption{Kurtosis index $kt_{80}$ for \civ\ (left panels) and line flux ratio \nv /\civ\ (right panels) versus $W3$ magnitude (top), $\log\,$REW(\civ ) in \AA\ (middle), and \imw\ color (bottom) for quasars in the $W3$-detected sample. In the top two rows (panels A-D), the red and orange symbols mark Type 1 and 2 ERQs, respectively defined by $i-W3\ge 4.6$. In the bottom row (panels E-F), the red and orange symbols mark all Type 1 and 2 quasars, respectively, with REW(\civ ) $\ge 100$ \AA. The dashed blue line marks the ERQ threshold at \imw\ = 4.6. In the right-hand panels, the square red and orange symbols additionally indicate $kt_{80} > 0.33$. Quasars with weak lines bunched up at $kt_{80} = 0.37$ are artifacts of the fitting procedure, which defaults to a single Gaussian if the fit is not improved by a second component. The green curves in panels C and D are fits to the point distributions intended only to guide the eye.}
 \end{center}
\end{figure*}

Figure~8 shows dramatically how the median line properties of ERQs differ from normal blue quasars matched to the ERQs in $i$ or $W3$ magnitude. This figure plots median BOSS spectra of non-BAL Type 1 quasars normalized to unity in the continuum to facilitate comparisons between the line profiles and REWs. The black curve represents all 57 non-BAL Type 1s with REW(\civ ) $\ge 100$ \AA\ in the core ERQ sample that we define in \S2 and discuss extensively in \S5. The blue quasars in this plot have $i-W3 < 3.5$ color. The red spectrum represents low-luminosity blue quasars that are matched to the ERQs in $i$ magnitude and also required to have REW(\civ ) $>$ 100 \AA . Faint blue quasars often have large REWs in the Baldwin Effect, but Figure~8 shows that their line profiles, FWHMs, peak heights relative to the continuum, and flux ratios such as \nv /\civ , \nv /\lya , \siiv /\civ , and \ovi /\civ , are all very different from the core ERQs. 

\begin{figure*}
\begin{center}
\includegraphics[scale=0.6,angle=0.0]{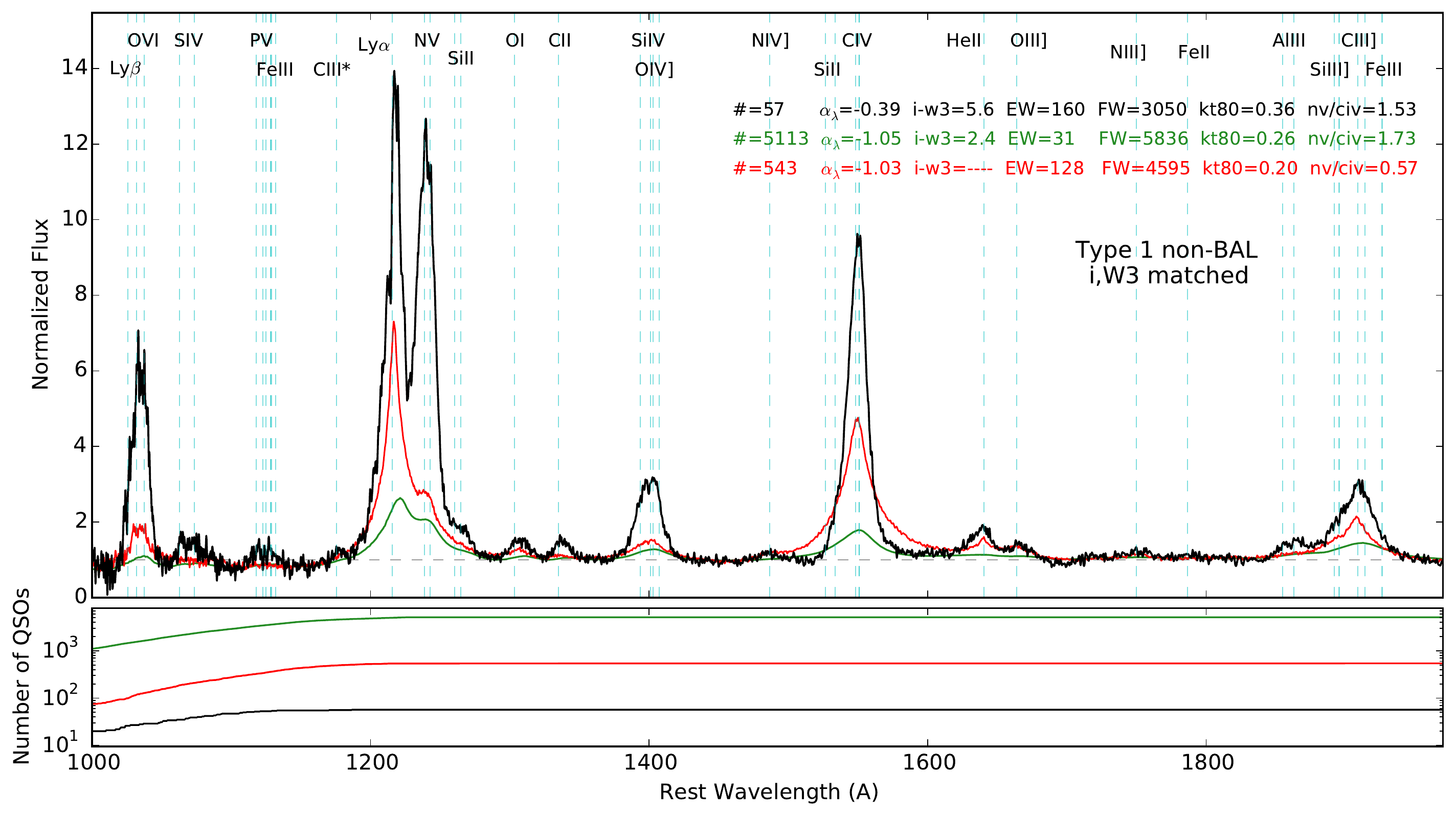}
  \vspace{-8pt}
 \caption{{\it Top panel:} Normalized median BOSS spectra of Type 1 non-BAL quasars in three samples: core ERQs with $i-W3 \ge 4.6$ and REW(\civ ) $\ge 100$ \AA\ (black curve, \S5.1), faint blue quasars with REW(\civ ) $\ge 100$ \AA\ matched to the core ERQs in $i$ magnitude (red curve), and luminous blue quasars matched to the core ERQs in $W3$ magnitude as a surrogate for luminosity (green curve, \S4.1). The spectra are shifted to the \civ\ frame (using {\tt wciv0} from our emission-line catalog, Appendix A) before calculating the medians. Prominent emission lines are labeled in this frame across the top. The numbers of quasars and the median values of properties measured in the individual quasars are given for each composite in colored text matching the spectra, where EW is REW(\civ ) in \AA , FW is FWHM(\civ ) in \kms , kt80 is the \civ\ kurtosis index $kt_{80}$, and \nv /\civ\ is the line flux ratio. {\it Bottom panel:} Numbers of quasars contributing to the median spectra at each wavelength. See \S4.2.}
 \end{center}
\end{figure*}

The green spectrum in Figure~8 represents blue quasars matched to the ERQs in $W3$ magnitude as a surrogate for luminosity (\S4.1). This sample has no REW(\civ ) constraint, so the spectrum represents a typical luminous blue quasar with small REWs in the Baldwin Effect. It is interesting that ERQs do not have anomalously large \nv /\civ\ flux ratios compared to these luminous blue quasars with {\it small} REWs (see also Figure~7 panels D and F). Large \nv /\civ\ and \siiv /\civ\ flux ratios can be attributed to higher metallicities and they are known to correlate generally with large quasar luminosities \citep{Hamann93, Hamann99, Dietrich02, Warner03, Nagao06}. The underlying cause of this relationship might be that more luminous quasars reside in more massive host galaxies, which naturally have higher metallicities in the well-known galactic mass-metallicity relation  \citep{Ferland96, Hamann99, Hamann02}. The ERQs are also luminous and might be particularly metal rich \citep[\S6.2,][Hamann et al. 2016a, in prep.]{Polletta08}.

\subsection{Redshift Distribution}

Figure~9 plots \imw\ color versus redshift for all quasars in our $W3$-detected sample. There is a very weak trend for bluer \imw\ colors (by 0.1-0.2 magnitudes) at higher redshifts in this sample. This trend is negligible for our purposes and we do not discuss it further. There are 205 ERQs above the upper dashed line that marks the threshold $i-W3 \ge 4.6$. We see again that a very high fraction of ERQs have large \civ\ REWs and unusual wingless line profiles compared to the normal blue quasar population. 

\begin{figure}
 \begin{center}
\includegraphics[scale=0.48,angle=0.0]{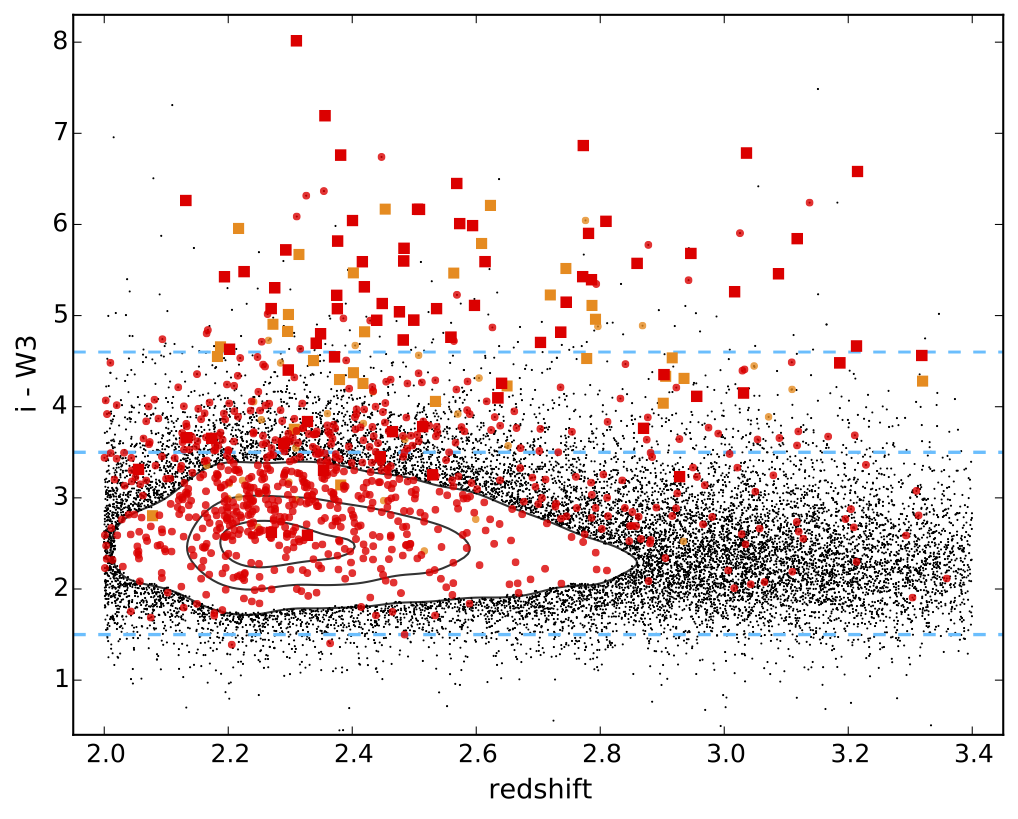}
\end{center}
\vspace{-8pt}
 \caption{\imw\ color versus redshift for the $W3$-detected sample. The red/orange symbols mark type 1/2 quasars with REW(\civ ) $\ge 100$ \AA , while the squares identify subsets of these quasars with wingless profiles, e.g., $kt_{80} > 0.33$. The dashed blue lines mark specific colors $i-W3 = 1.5$, 3.5 and 4.6 used to construct composites for Figure~11 below (\S4.5). }
\end{figure} 

\subsection{\wtmwf\ and the advantages of \imw }

There are two reasons why \imw\ is better for selecting ERQs than \imwf . One is the greater sensitivity of $W3$ compared to $W4$ in the WISE survey \citep{Yan13}. For example, only 39.5\% of quasars in our $W3$-detected sample also have $W4$ detections at SNR $>$ 3. The second reason is illustrated by Figure~10. At the median redshift of our samples, $z_e\sim 2.5$, the filters $i$, $W3$ and $W4$ measure rest wavelengths of roughly 0.2 \mum , 3.4 \mum\ and 6.4 \mum , respectively. Figure~10 plots \imw\ and \imwf\ versus \wtmwf\ color for quasars that are securely detected in all three bands. We can see from the quasar distributions in these plots that \imw\ is better for isolating ERQs with unusual line properties. In particular, in the top panel, quasars with REW(\civ ) $\ge 100$ \AA\ and unusual wingless profiles (red/orange squares) are strongly offset toward red \imw\ colors regardless of their color in \wtmwf . Similarly, quasars with REW(\civ ) $<$ 100 \AA\ and normal profiles (black dots) are almost exclusively blue in \imw\ in spite of the range they exhibit in \wtmwf  . However, in the bottom panel, the quasars with weak lines and normal profiles overlap with the ERQs in \imwf\ color because of the way \imwf\ depends on \wtmwf . 

\begin{figure}
\includegraphics[scale=0.48,angle=0.0]{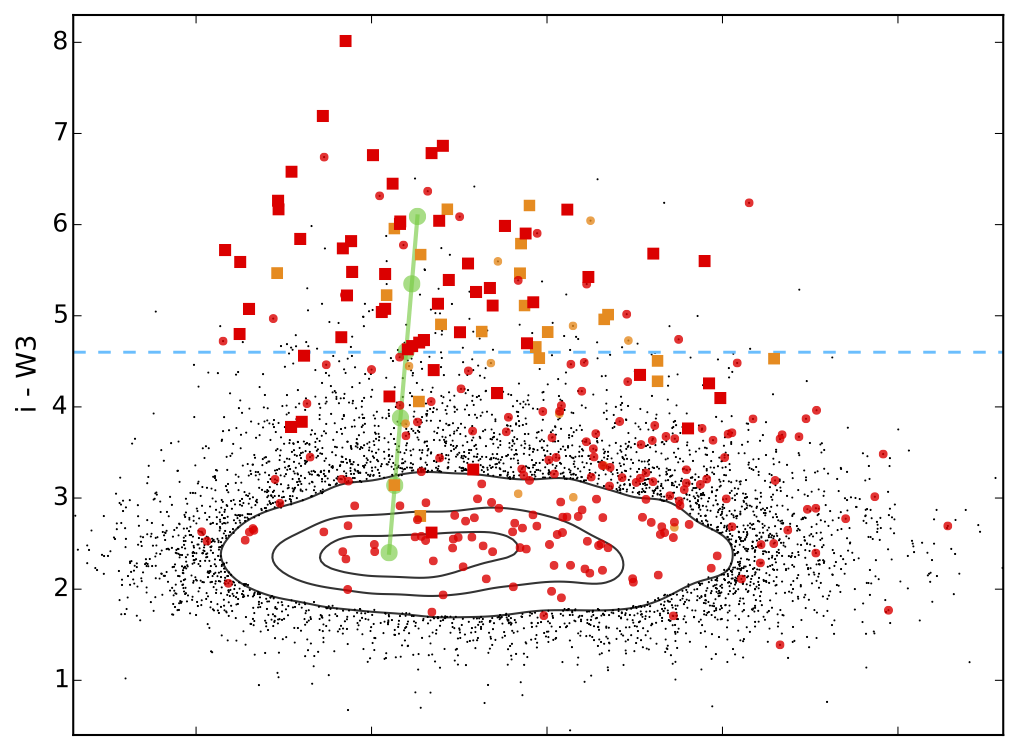}
\includegraphics[scale=0.48,angle=0.0]{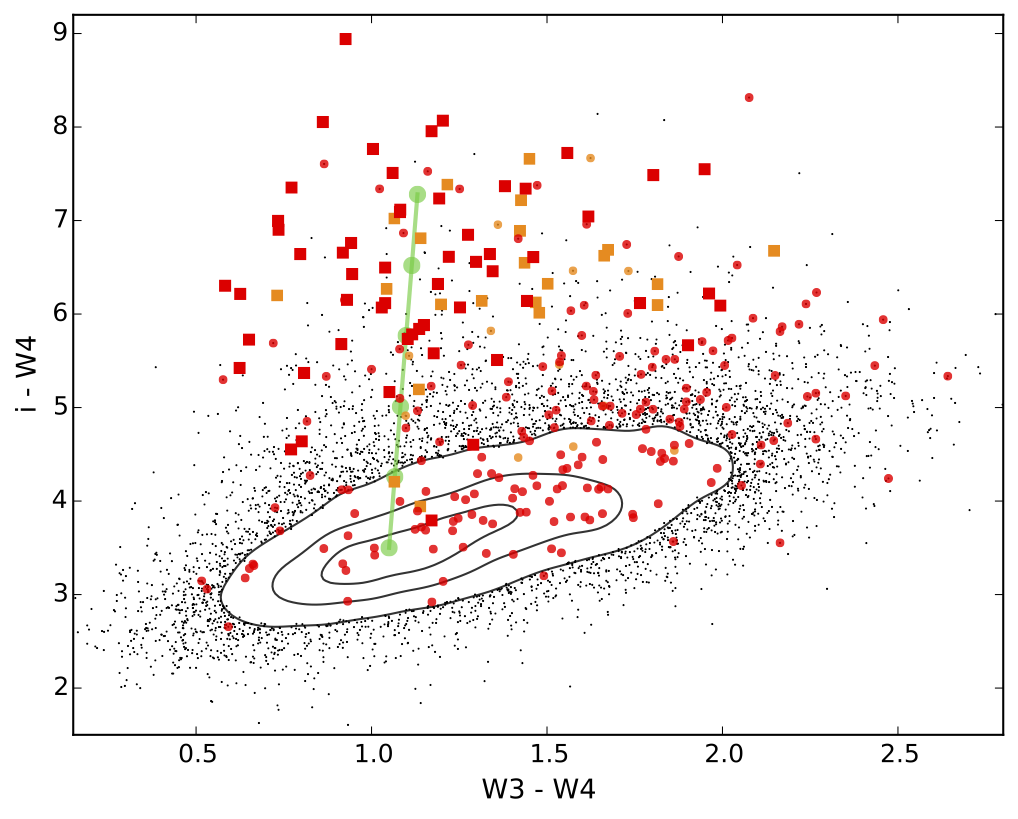}
  \vspace{-8pt}
 \caption{\imw\ (top panel) and \imwf\ (bottom) versus \wtmwf\ for quasars detected in all three filters. The red/orange symbols have the same meaning as in Figure~9. The green lines are reddening vectors with dots marking $E(B-V) = 0.0$, 0.1, 0.2, 0.3, 0.4 and 0.5 from bottom to top (based on the quasar reddening curve in Hamann et al. 2016b, in prep.). The dashed blue line in the top panel marks the ERQ threshold at \imw\ $\ge$ 4.6.}
\end{figure}

Median BOSS spectra constructed for quasars in different regions of Figure~10 confirm that the UV emission-line properties have no relationship to \wtmwf\ color but they depend strongly on \imw . Also note that the observed range in \wtmwf\ color runs orthogonal to the reddening vector in the top panel of Figure~10, implying that redder \wtmwf\ colors are not related to the obscuration and reddening measured by \imw . Thus it appears that the \wtmwf\ colors at these redshifts are regulated by something that is not UV obscuration and not related to the ERQ phenomenon. We speculate that if the mid-IR fluxes at $\sim$3.4 and $\sim$6.4 \mum\ arise from hot dust near the quasars, then the observed range in \wtmwf\ colors might be controlled by optical depth and viewing angle effects within a warm dusty torus \citep{Efstathiou95,Nenkova08,Mor09} or by different dust spatial geometries leading to a range of dust temperatures (or temperature distributions) across the quasar sample.

\subsection{Specific Thresholds in \imw\ \& REW(\civ )}

Panels E and F in Figure~7 above show that there is a surprisingly sharp boundary near \imw\ $\ga$ 4.6 where the exotic line properties of ERQs start to appear in the majority of quasars. Panels C and D in this figure indicate that there is also a strong dependence on REW(\civ ). Figure~11 defines these dependences further by plotting normalized median BOSS spectra (left panels) and corresponding median spectral energy distributions (SEDs, right panels) for quasars in different intervals of \imw\ and REW(\civ ). The top row plots results for Type 1 non-BAL quasars with REW(\civ ) $\ge 100$ \AA\ in three color bins defined by the dashed blue lines in Figure~9. We see that the median line properties of ERQs with REW(\civ ) $\ge 100$ \AA\ differ markedly from their counterparts with blue and intermediate colors. Experiments with other \imw\ color cuts show that these dramatic changes in the median properties occur across a narrow color range at approximately $i-W3 = 4.6\pm 0.2$. The dramatic shift to peculiar line properties across this boundary illustrates the distinct nature of the core ERQs defined by both \imw\ $>$ 4.6 and REW(\civ ) $>$ 100 \AA\ (\S5.1). In particular, they are not simply an extension of trends with \imw\ color that exists across the BOSS quasar population. 

\begin{figure*}
\begin{center}
\includegraphics[scale=0.413,angle=0.0]{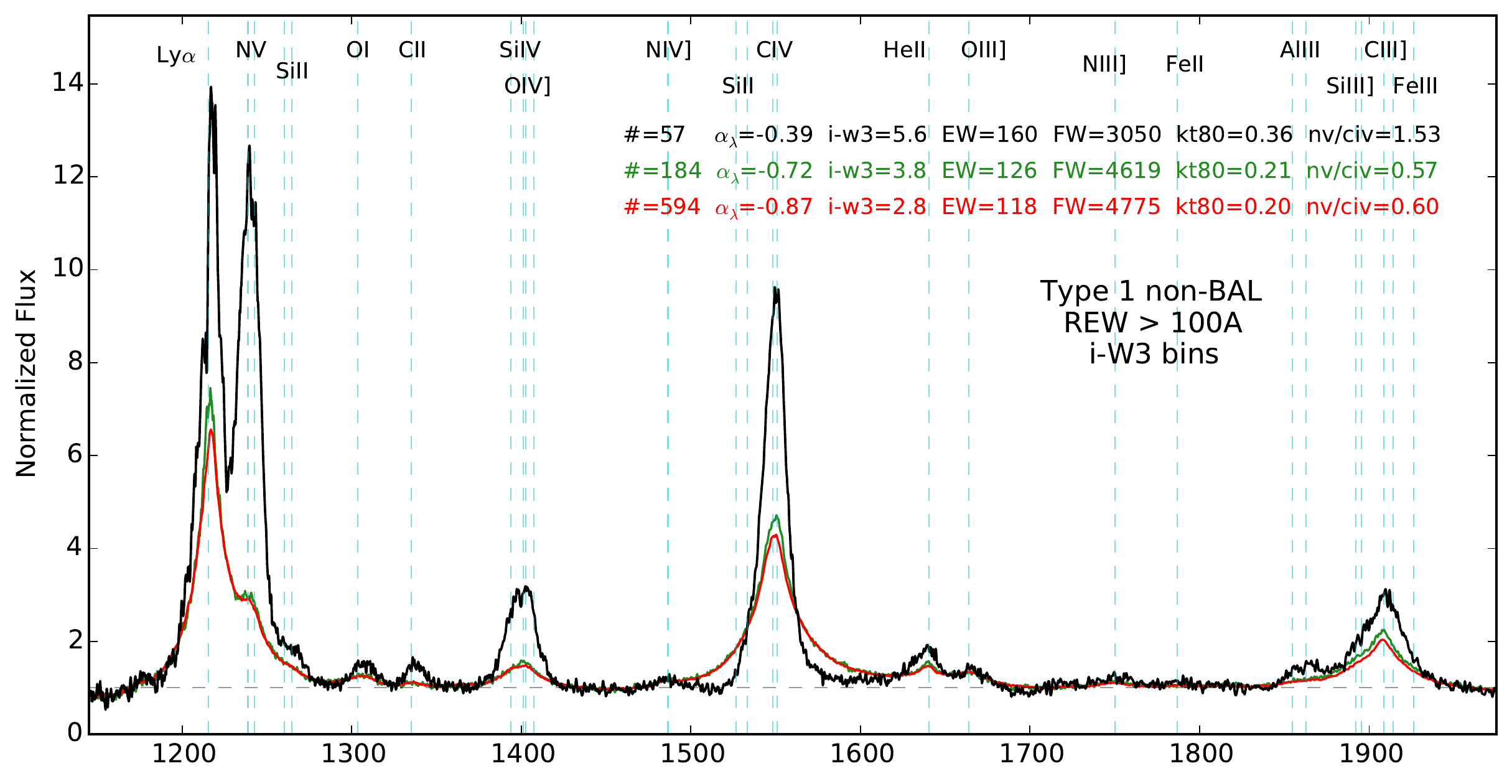}
\includegraphics[scale=0.4,angle=0.0]{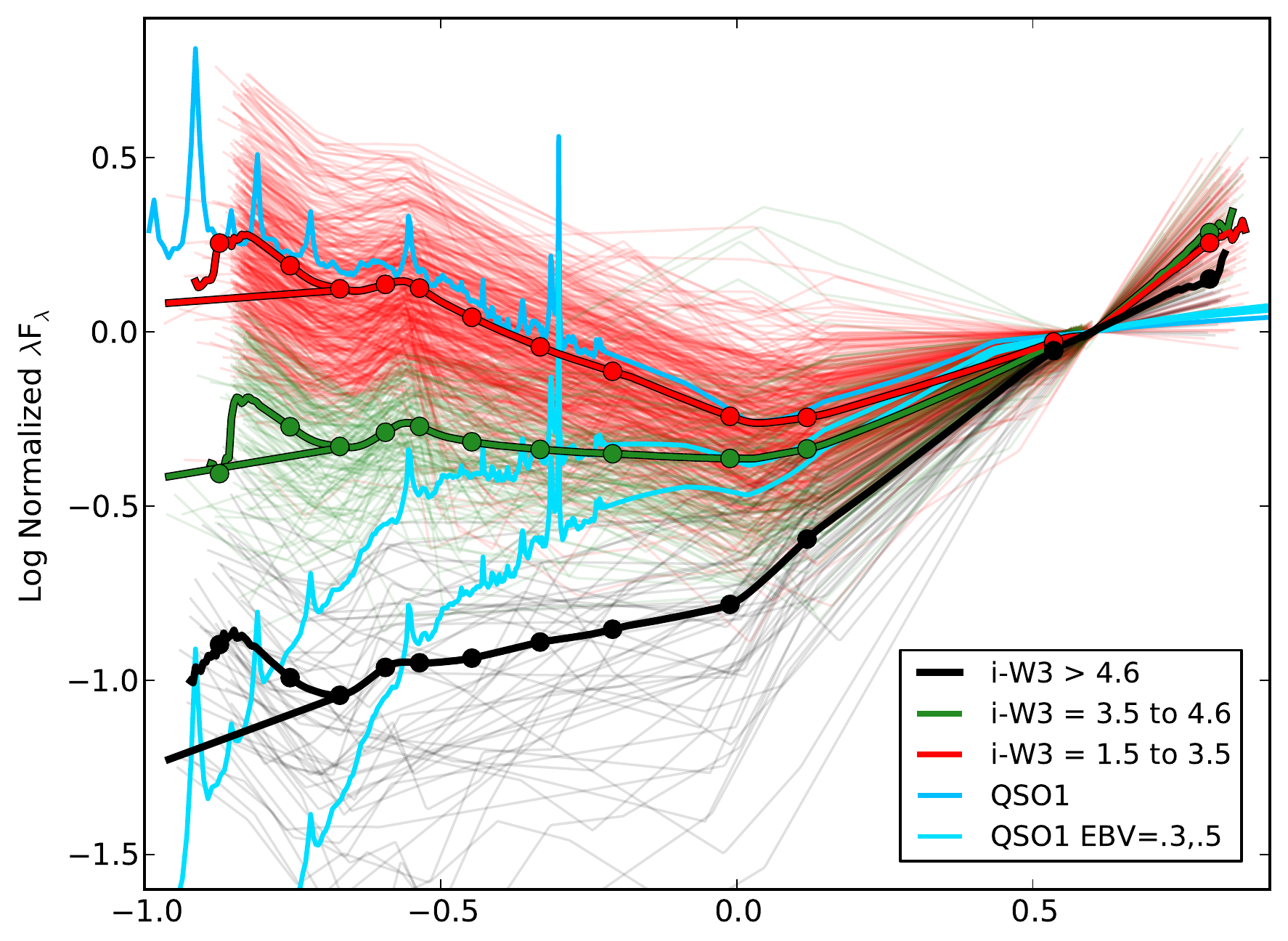}
\includegraphics[scale=0.413,angle=0.0]{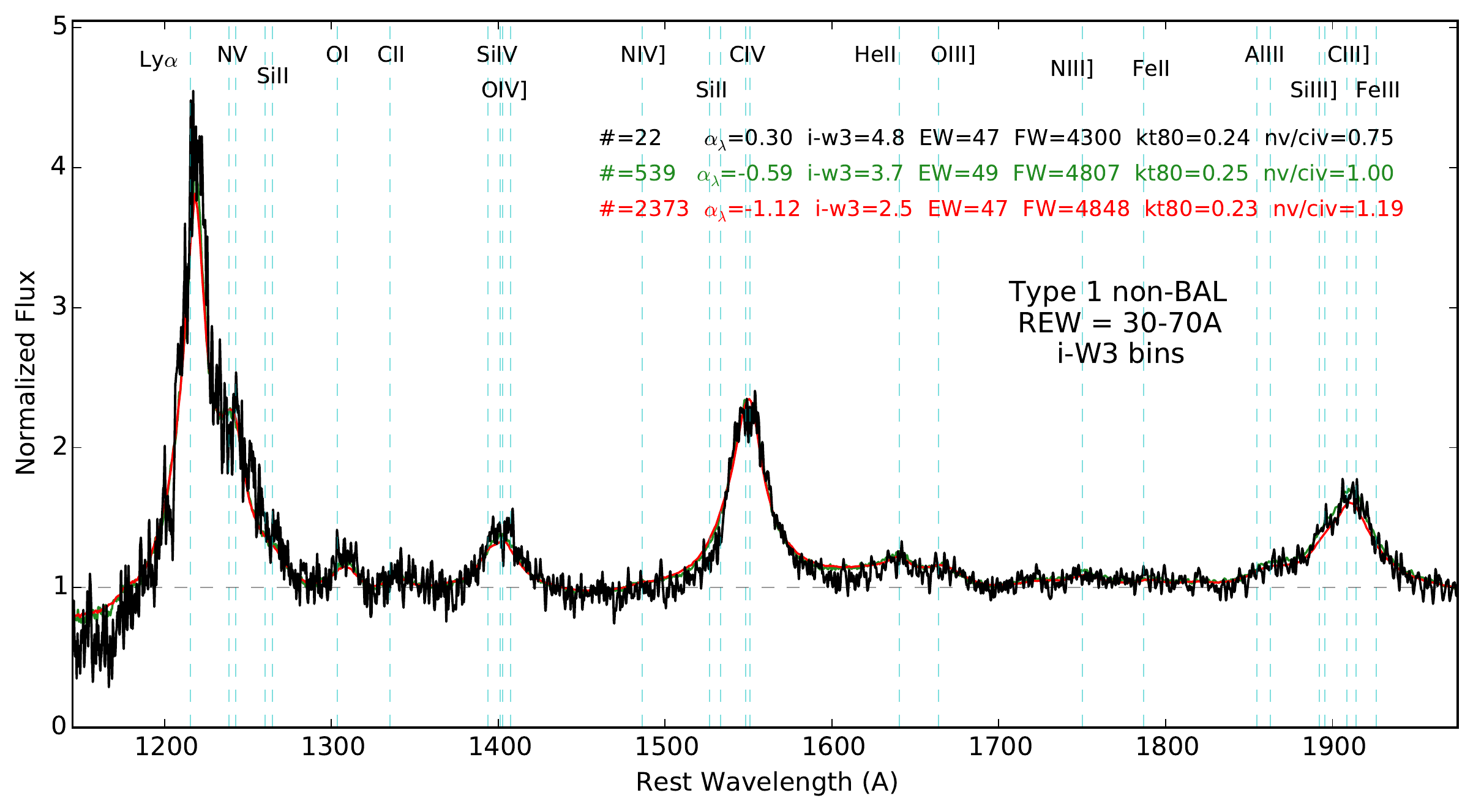}
\includegraphics[scale=0.4,angle=0.0]{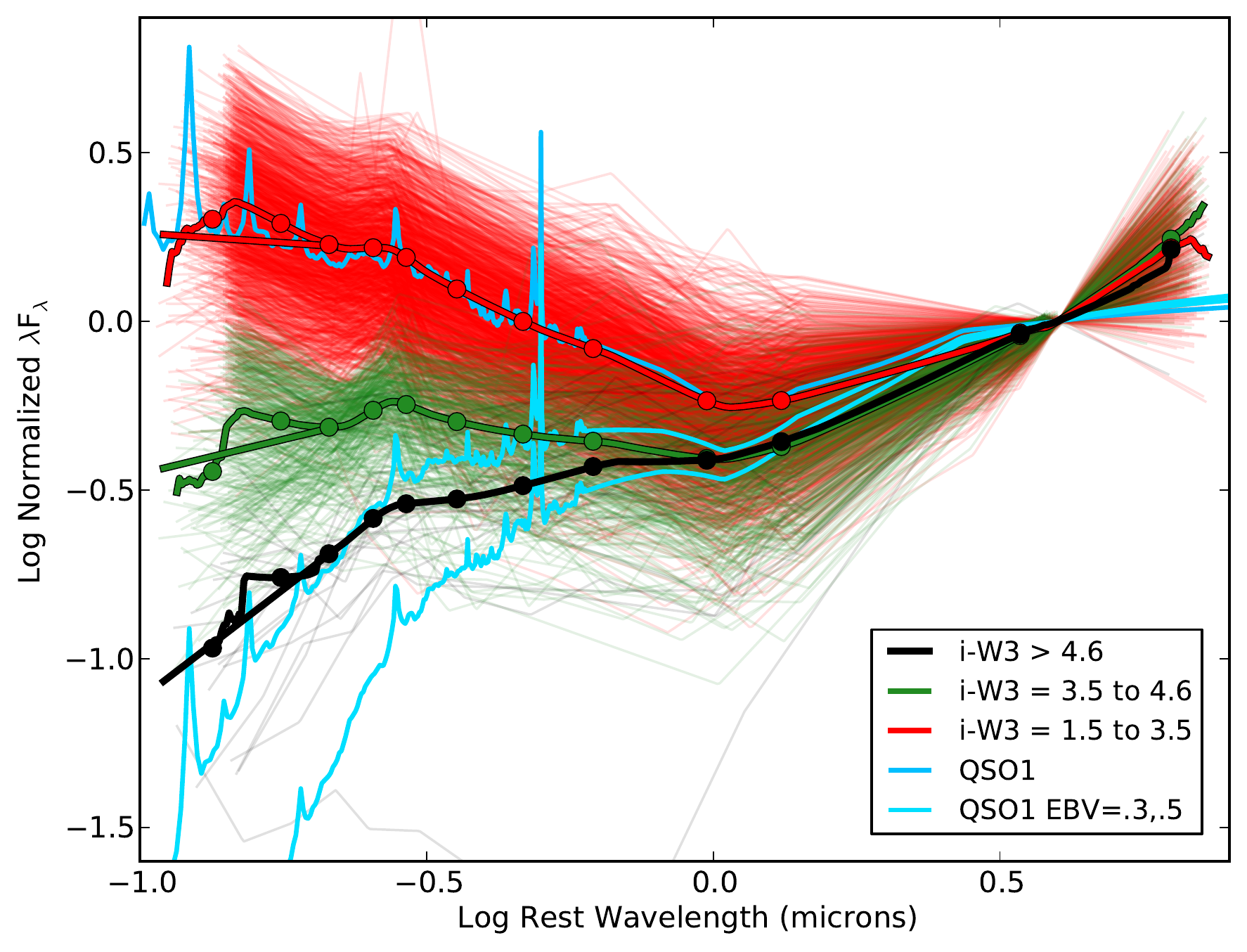}
  \vspace{-8pt}
 \caption{{\it Top Row:} Normalized median BOSS spectra (left) and SEDs (right) for Type 1 non-BAL $W3$-detected quasars with REW(\civ ) $\ge 100$ \AA\ in the three color bins shown in Figure~9: ERQs with \imw\ $\geq 4.6$ (black curves), intermediate colors $4.6>$ \imw\ $\geq  3.5$ (green curves), and blue quasars with $3.5 >$ \imw\ $> 1.5$ (red curves). The left-hand panels provide median values of some measured parameters for these quasar samples, as in Figure~8 above. The SEDs are scaled to unity at 4 \mum\ and plotted with the same color scheme as the BOSS spectra. The thin faint curves are individual quasar SEDs while the bold curves are the sample medians. The dots on the bold median SEDs mark wavelengths of the broad-band filters at a typical redshift $z_e = 2.5$, from left to right: $griz$, $YJHK$, and $W1$, $W2$, $W3$ and $W4$. The median SEDs split at short wavelengths to show results directly from the photometry (upper curves with filter dots) and for a median power law derived from fits to the UV spectra of individual quasars that are more reliable because they avoid spectral line contamination (lower curves without dots). The slopes of these median UV power laws are listed in the left panels. The light blue SEDs show the Type 1 quasar template QSO1 from Polletta et al. (2007) without and with reddening for $E(B-V) = 0.3$ and 0.5 using an SMC-like reddening curve (from Hamann et al. 2016b). {\it Bottom Row:} Same as above but for quasars with $30 <$ REW(\civ ) $< 70$ \AA . See \S4.5.}
 \end{center}
\end{figure*}

The bottom row in Figure~11 shows the importance of REW(\civ ) to the overall properties of ERQs. These panels plot median spectra and SEDs for the same three color bins as the top panels, but now with line strengths in a range more typical of BOSS quasars, namely $30 \le$ REW(\civ ) $\le 70$ \AA\ (see Figure~1). There are only 22 ERQs with REW(\civ ) in this range but many thousands of blue quasars. To simplify the plots, we randomly exclude blue quasars to limit their numbers and impose stricter REW(\civ ) constraints to force their median REW(\civ ) to match the ERQs. The result is that all three color samples have remarkably similar line profiles and relative line strengths. Unlike the core ERQs in top-left panel, these ERQs with normal line strengths have median line properties that are virtually identical to normal blue quasars. This suggests that ERQs with normal REWs generally do not belong to the same unique population as the core ERQs; they appear to be just normal quasars behind a dust-reddening screen.

This conclusion is supported by the median SEDs plotted in right-hand panels of Figure~11. First note that the blue quasars in these plots (red curves) have SEDs very similar to the unobscured Type 1 quasar template QSO1 from \cite{Polletta07}. The ERQs with normal line strengths (black curve, bottom right panel) have SEDs roughly consistent with a standard reddening curve applied to this template (Hamann et al. 2016b, in prep.). In particular, these ERQs exhibit a sharp decline in the near-UV with only moderately red colors across the near-IR, similar to QSO1 reddened by selective extinction $E(B-V) \sim 0.3$. This behavior again indicates that ERQs with normal REWs tend to be normal quasars reddened by dust. 

In contrast, the core ERQs have SEDs (black curve, top right panel) much flatter across the rest-frame UV in spite of their red \imw\ colors. These unusual SEDs are another important characteristic of the core ERQs that helps to define them as a unique red quasar population. 

\section{A Unique Red Quasar Population}

Our analysis in \S4 shows that many ERQs defined by \imw\ $>$ 4.6 have a suite of peculiar emission-line properties (Figures~7 and 8) accompanied by SEDs that are surprisingly flat across rest-UV given their red \imw\ colors \citep[Figure 11, see also][]{Ross15}. This ensemble of properties starts to appear in a majority of quasars across a surprisingly abrupt color boundary near \imw\ $\ga$ 4.6. However, ERQs defined only by \imw\ $>$ 4.6 include interlopers that look like normal quasars reddened by dust (\S4.5). These interlopers tend have normal \civ\ line strengths while the ERQs with peculiar properties tend to have REW(\civ) $\ga$ 100 \AA . 

\subsection{The Core ERQ Sample}

Here we combine the selection criteria \imw\ $\ge$ 4.6 and REW(\civ ) $\ge 100$ \AA\ to define a ``core'' sample of ERQs that excludes most of the interlopers and strongly favors the exotic properties that identify a unique new red quasar population. We find 95 core ERQs satisfying these criteria in our $W3$-detected sample. We also searched the entire DR12Q catalog for more core ERQs that might be missing from emission-line catalog, e.g., because they have BALs at wavelengths that might affect our emission-line measurements (\S2, Appendix A). We do this by visually inspecting the BOSS spectra and performing additional line fits for all DR12Q quasars with \imw\ $>$ 4.6. This search yields only 2 more sources for a total sample of 97 core ERQs that is complete among $W3$-detected quasars in DR12Q at the redshifts of our study. 

Thirteen of these 97 core ERQs are in the original ERQ sample of \cite{Ross15}. The larger number here is due partly to our use of the final BOSS data release DR12 instead of DR10, but mostly to our less stringent color constraint using a more sensitive WISE filter, e.g., \imw\ $\ge 4.6$ instead of \rmwf\ $> 7.5$. The median color of the core ERQs is $\left< i-W3\right> \approx 5.31\pm 0.65$ compared to $\left<i-W3\right> \approx 2.50\pm 0.57$ for quasars in the $W3$-detected sample overall (\S4.1, also Figures 4 and 9). The median redshift of the core ERQs is $\left<z_e\right> = 2.50\pm 0.27$. 

The median line properties and SEDs of Type 1 non-BAL ERQs in the core sample were shown already in Figures~8 and 11 above. Seventy four (76\%) of the 97 core ERQs are Type 1s by our definition in \S2, although closer examination suggests that some of them with FWHM(\civ ) $<$ 2000 \kms\ are actually Type 1s. This is issue discussed further in \S5.4. 

Bolometric luminosities are difficult to estimate for ERQs because we have limited wavelength coverage and the amounts of obscuration in the rest-frame visible/UV are uncertain. However, if we assume that their $W3$ fluxes (rest frame $\sim$3.4 \mum ) are relatively unaffected by dust extinction and their emitted/intrinsic SEDs are like other luminous quasars \citep[as described in][]{Hamann13}, then the median magnitude $\left<W3\right> \sim 16.1\pm 0.7$ of the core ERQs corresponds to a median bolometric luminosity of $\log L ({\rm ergs/s}) \sim 47.1\pm 0.3$. Given the range of SEDs observed in blue/unobscured quasars \citep[Figures 9 and 11,][]{Elvis94,Richards06}, the uncertainties in their bolometric correction factors \citep[also][]{Richards11}, and the possibility that the core ERQs might have intrinsically peculiar SEDs in the unobserved far-UV (Hamann et al. 2016a, in prep.), this median luminosity should be considered only a crude estimate uncertain by at least a factor of two.

Figure~12 plots BOSS spectra for some individual Type 1s in the core ERQ sample to illustrate both the range and similarity of properties across the sample. This figure includes all 10 Type 1 core ERQs from \cite{Ross15} plus 20 new ones discovered here. All of them have unusual wingless \civ\ profiles and many have some type of blueward extension on \civ\  -- either BALs or BAL-like absorption (e.g., J004713+264024, J102130+214438, J103146+290324, J131047+322518, J135608+073017, plus others) or extended blue emission wings (not matched on the red side, e.g., J005233-055653, J113834+473250, J134417+445459, J150117+231730, J160431+563354) that can lead to a kurtosis index smaller than the majority of core ERQs with $kt_{80}>0.33$ (Figure~7). 

\begin{figure*}
\centering
\includegraphics[scale=0.53,angle=0.0]{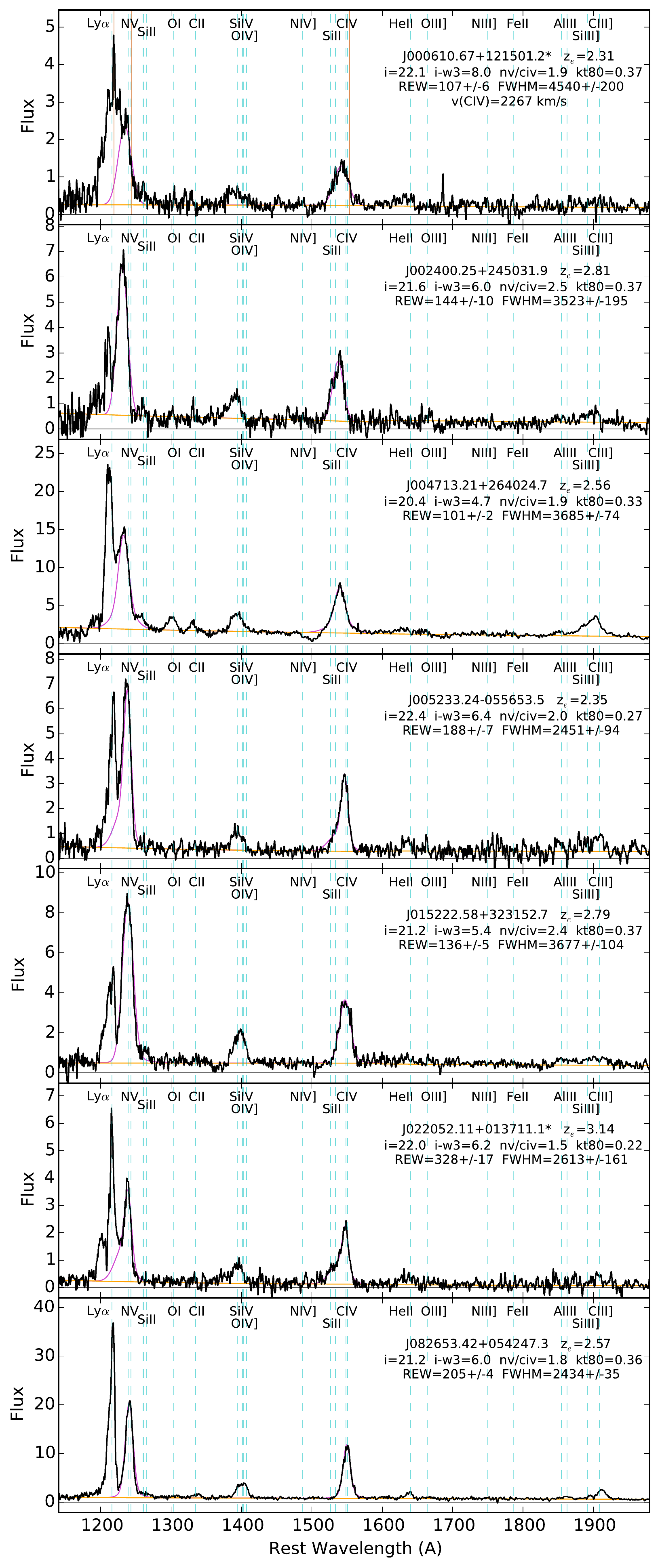}
\includegraphics[scale=0.53,angle=0.0]{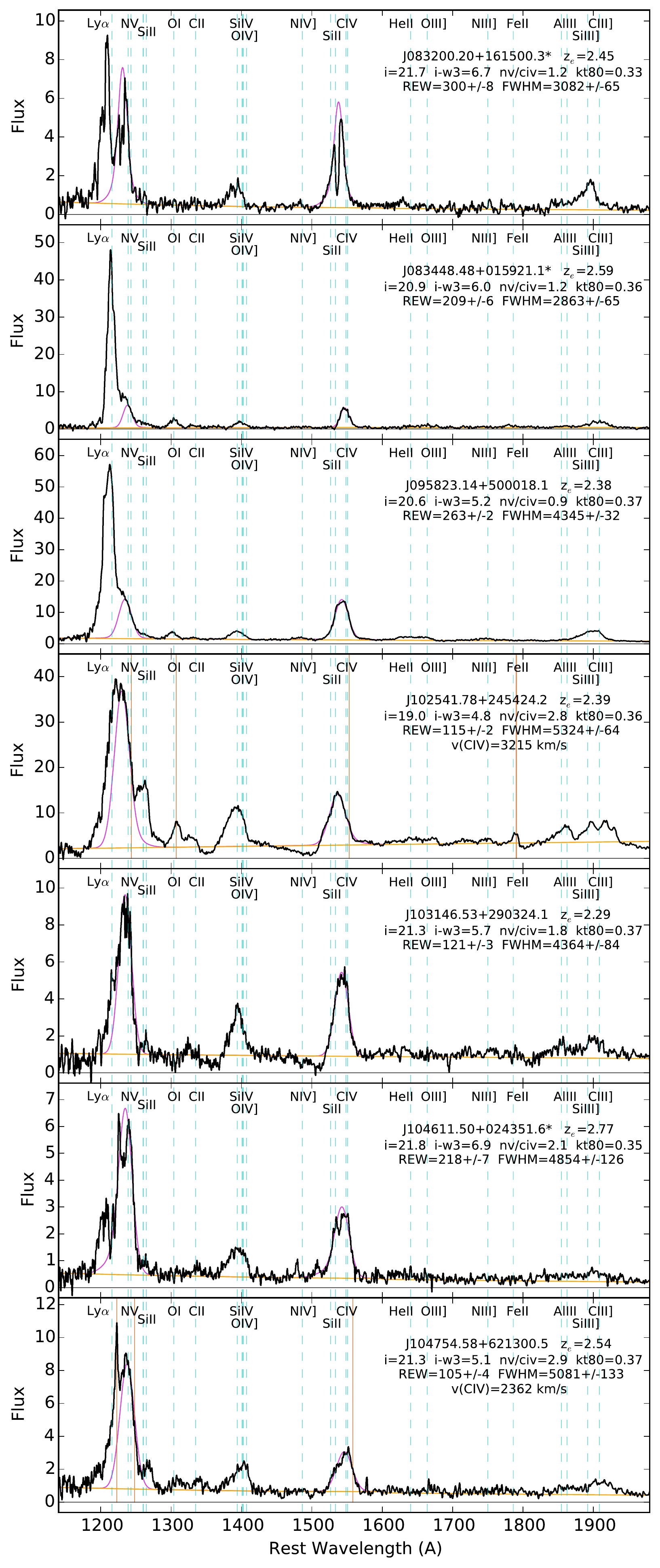}
\vspace{-8pt}
 \caption{BOSS spectra of some representative Type 1 core ERQs (\S5.1) plotted at rest wavelengths using the best available redshifts from DR12Q (\S3). Common emission lines are labeled across the top at positions marked by dashed blue lines. The orange and magenta curves show our fits to the continuum and the \civ\ and \nv\ emission lines, respectively. The quasar names, redshifts and other measured properties are given in each panel (see also Table 2). Names marked by `*' are in the Ross et al. (2015) sample. Vertical brown lines in some panels mark estimated systemic line wavelengths based on distinct narrow \lya\ emission spikes or \oi\ and \feii\ emission lines (in J102541+245424 only). These markings can reveal large blueshifts in the observed broad \civ\ and \nv\ emission lines, which are listed as v(\civ ) in those panels (see \S5.8). The flux units are 10$^{-17}$ ergs s$^{-1}$ cm$^{-2}$ \AA$^{-1}$. }
\end{figure*} 
\setcounter{figure}{11}
\begin{figure*}
\centering
\includegraphics[scale=0.53,angle=0.0]{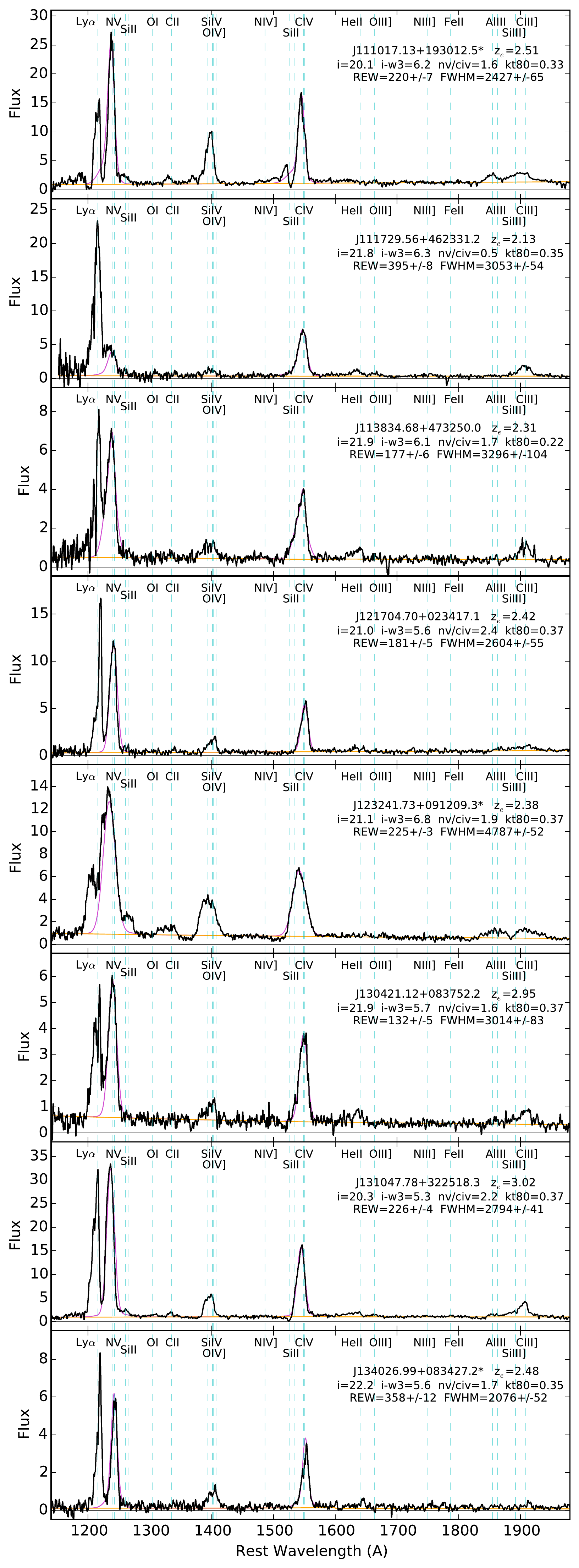}
\includegraphics[scale=0.53,angle=0.0]{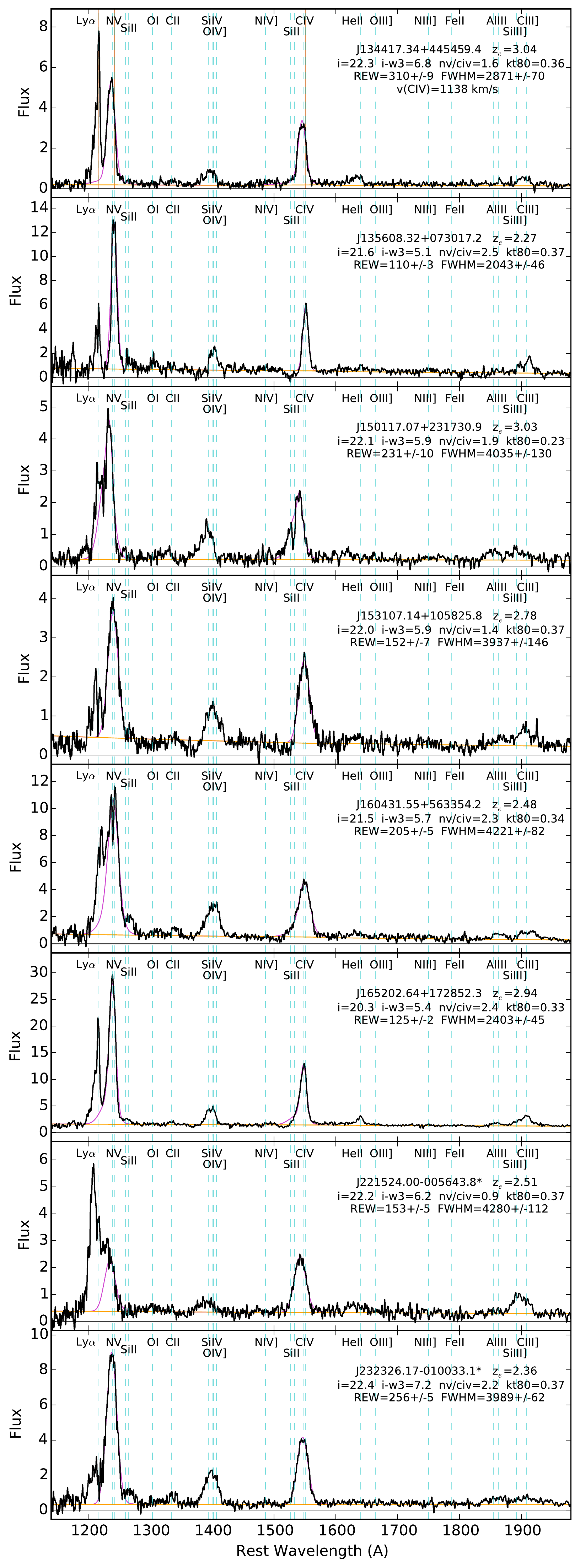}
\vspace{-10pt}
\caption{(Continued)}
\end{figure*} 

We also see in Figure~12 the unusual line flux ratios described above, most notably \nv\ $>$ \lya\ and large \nv / \civ , but there are some dramatic exceptions with strong \lya\ and weak/normal \nv\ (e.g., J083448+015921, J095823+500018, J111729+462331). It is interesting, but possibly coincidental, that these exceptions stand out for {\it not} having significant blue-red asymmetries or BAL-like absorption like the other core ERQs (\S5.3). The core ERQs also tend to have unusually large \siiv /\civ\ ratios and in some cases strong \aliii\ \lam1860 compared to the intercombination lines with similar ionization, namely \ciii ] \lam 1909 and \siiii ] \lam 1892. These line properties carry information about the gas metallicities, physical conditions, and the locations of the emitting regions, as well as the shape and intensity of the incident ionizing spectra. We discuss these issues further in Hamann et al. (2016a, in prep., see also \citealt{Polletta08}). 

Table 1 lists some basic data for the 97 core ERQs. The selective extinctions in this table, \ebv , are based on $r,i,z-W1$ colors corrected for Galactic extinction (from Hamann et al. 2016b, in prep.). They are available only for quasars at $2.1 < z_e<3.4$ with detections in both $i$ and $W1$ at SNR $>$ 5. They provide a conventional measure of the reddening for comparison to other studies. However, these \ebv\ values underestimate the true amounts of obscuration in the core ERQs because they derive from spectral slopes only across the rest-frame visible to UV (see \S5.5 for more discussion). 

\begin{table*}
 \begin{center}
 \begin{minipage}{176mm}
  \caption{Core ERQ properties: $z_e$ is the best available emission-line redshift from DR12Q (\S3). $i$ magnitude and \imw\ color corrected for Galactic extinction. REW, FWHM and $kt_{80}$ for the \civ\ emission lines, \nv /\civ\ is the line flux ratio, and $\alpha_{\lambda}$ is the UV continuum slope all from our emission line catalog (Appendix A). BAL is the visual inspection flag ({\tt bal\_flag\_vi}) from DR12Q where 1 indicates that a BAL is present. \ebv\ is the selective extinction derived from $r,i,z-W1$ colors by Hamann et al. (2016b, in prep.). FIRST is the 20 cm radio flux from FIRST where no entry means the source was not covered by FIRST, 0.0 indicates a non-detection with 5$\sigma$ upper limit $\sim$1 mJy \citep{Becker95, Helfand15}, non-zero entries are measurements with SNR $>$ 3 as recorded in DR12Q.}
  \begin{tabular}{@{}lccccccccccc@{}}
  \hline
 Quasar Name& $z_{e}$& $i$ & \imw & REW & FWHM& $kt_{80}$& \nv /\civ& BAL & $\alpha_{\lambda}$& \ebv & FIRST\\
 & & (mag) & (mag) & (\AA ) & (km/s)& & & & & & (mJy)\\
\hline
J000610.67+121501.2$^a$&  2.31&  22.1&   8.0&   107$\pm$6&   4540$\pm$200&   0.37&   1.88&   0&   $-$0.20&   0.66$\pm$0.01&  0.0\cr
J000746.19+122223.9&  2.43&  21.3&   4.9&   220$\pm$6&   2433$\pm$71&   0.28&   0.39&   0&   $-$0.17&   0.19$\pm$0.01&  0.0\cr
J002400.25+245031.9&  2.81&  21.6&   6.0&   144$\pm$10&   3523$\pm$195&   0.37&   2.46&   0&   $-$1.14&   0.20$\pm$0.02&  ---\cr
J004713.21+264024.7&  2.56&  20.4&   4.7&   101$\pm$2&   3685$\pm$74&   0.33&   1.92&   1&   $-$1.08&   0.15$\pm$0.01&  ---\cr
J005044.95$-$021217.6&  2.25&  21.5&   4.7&   147$\pm$9&   4343$\pm$487&   0.19&   0.61&   0&   $-$1.73&   0.14$\pm$0.02&  0.0\cr
J005233.24$-$055653.5&  2.35&  22.4&   6.4&   188$\pm$7&   2451$\pm$94&   0.27&   2.03&   0&   1.33&   ---&  0.0\cr
J014111.13$-$031852.5&  2.56&  20.8&   4.8&   101$\pm$2&   2986$\pm$45&   0.37&   1.85&   0&   $-$0.29&   0.15$\pm$0.01&  0.0\cr
J015222.58+323152.7&  2.79&  21.2&   5.4&   136$\pm$5&   3677$\pm$104&   0.37&   2.44&   0&   $-$0.27&   0.37$\pm$0.01&  ---\cr
J020932.15+312202.7&  2.38&  21.4&   5.1&   108$\pm$6&   2180$\pm$97&   0.34&   1.07&   0&   0.61&   0.23$\pm$0.02&  ---\cr
J022052.11+013711.1$^a$&  3.14&  22.0&   6.2&   328$\pm$17&   2613$\pm$161&   0.22&   1.53&   0&   $-$1.78&   ---&  0.0\cr
J080425.75+470159.0&  2.77&  22.1&   5.4&   162$\pm$5&   2371$\pm$64&   0.35&   1.28&   0&   $-$0.61&   0.25$\pm$0.03&  0.0\cr
J080547.66+454159.0&  2.33&  21.8&   6.3&   109$\pm$4&   2667$\pm$107&   0.21&   1.61&   0&   $-$0.43&   0.31$\pm$0.01&  0.0\cr
J082536.31+200040.3&  2.09&  21.5&   4.7&   211$\pm$9&   3265$\pm$209&   0.17&   0.35&   0&   $-$1.35&   ---&  0.0\cr
J082649.30+163945.2&  2.32&  21.9&   4.6&   205$\pm$6&   5633$\pm$150&   0.33&   0.57&   0&   $-$0.60&   0.33$\pm$0.02&  0.0\cr
J082653.42+054247.3&  2.57&  21.2&   6.0&   205$\pm$4&   2434$\pm$35&   0.36&   1.84&   0&   $-$0.62&   0.28$\pm$0.01&  1.1\cr
J083200.20+161500.3$^a$&  2.45&  21.7&   6.7&   300$\pm$8&   3082$\pm$65&   0.33&   1.25&   0&   $-$1.24&   0.30$\pm$0.01&  1.0\cr
J083448.48+015921.1$^a$&  2.59&  20.9&   6.0&   209$\pm$6&   2863$\pm$65&   0.36&   1.16&   0&   1.15&   0.09$\pm$0.01&  0.0\cr
J084447.66+462338.7$^a$&  2.22&  21.1&   6.0&   161$\pm$4&   1656$\pm$36&   0.35&   1.30&   0&   $-$1.06&   0.31$\pm$0.01&  0.0\cr
J085451.11+173009.1&  2.61&  21.9&   5.6&   120$\pm$9&   4199$\pm$207&   0.37&   1.98&   0&   $-$1.40&   0.21$\pm$0.02&  0.0\cr
J090306.18+234909.8&  2.26&  21.9&   5.0&   144$\pm$5&   2481$\pm$98&   0.21&   1.45&   0&   $-$0.79&   0.26$\pm$0.02&  0.0\cr
J091303.90+234435.2&  2.42&  21.7&   5.3&   145$\pm$5&   2190$\pm$65&   0.34&   1.49&   0&   $-$0.23&   0.20$\pm$0.02&  0.0\cr
J091508.45+561316.0&  2.86&  21.5&   5.6&   226$\pm$9&   2867$\pm$102&   0.35&   0.92&   0&   $-$2.12&   0.11$\pm$0.03&  1.2\cr
J092049.59+282200.9&  2.30&  20.8&   4.8&   197$\pm$4&   1048$\pm$17&   0.37&   0.40&   0&   $-$1.70&   ---&  0.0\cr
J093226.93+461442.8&  2.31&  21.7&   5.7&   443$\pm$13&   1960$\pm$53&   0.35&   0.66&   0&   1.99&   0.26$\pm$0.02&  0.0\cr
J093506.96$-$024137.7&  2.17&  21.8&   4.8&   119$\pm$5&   2404$\pm$106&   0.25&   1.28&   1&   $-$0.60&   0.25$\pm$0.02&  0.0\cr
J093638.41+101930.3$^a$&  2.45&  21.6&   6.2&   172$\pm$3&   1271$\pm$19&   0.37&   1.72&   0&   $-$0.60&   0.36$\pm$0.01&  0.0\cr
J095033.51+211729.1&  2.74&  21.9&   5.5&   272$\pm$7&   1387$\pm$29&   0.36&   0.51&   0&   $-$0.34&   0.35$\pm$0.02&  1.5\cr
J095823.14+500018.1&  2.38&  20.6&   5.2&   263$\pm$2&   4345$\pm$32&   0.37&   0.92&   0&   $-$0.98&   0.20$\pm$0.01&  10.3\cr
J101324.53+342702.6&  2.48&  20.0&   4.7&   205$\pm$3&   4157$\pm$51&   0.34&   0.89&   0&   $-$0.67&   0.08$\pm$0.01&  0.0\cr
J101533.65+631752.6&  2.23&  21.9&   5.5&   130$\pm$5&   2012$\pm$59&   0.36&   2.88&   0&   $-$1.31&   0.30$\pm$0.02&  0.0\cr
J102130.74+214438.4&  2.19&  22.1&   5.4&   155$\pm$6&   3567$\pm$92&   0.37&   2.23&   1&   $-$1.39&   0.32$\pm$0.01&  0.0\cr
J102353.44+580004.9&  2.60&  21.1&   5.1&   116$\pm$5&   2107$\pm$73&   0.37&   1.05&   0&   $-$0.47&   0.24$\pm$0.01&  0.0\cr
J102447.32$-$013633.8&  2.88&  21.9&   5.8&   192$\pm$15&   2843$\pm$396&   0.15&   1.16&   0&   $-$0.49&   0.23$\pm$0.04&  9.3\cr
J102541.78+245424.2$^b$& 2.34& 19.0& 4.8& 115$\pm$2& 5324$\pm$64& 0.36& 2.82& 1& 1.24& 0.20$\pm$0.00& 0.0\cr
J103146.53+290324.1&  2.29&  21.3&   5.7&   121$\pm$3&   4364$\pm$84&   0.37&   1.78&   1&   $-$0.16&   0.33$\pm$0.01&  0.0\cr
J103456.95+143012.5&  2.96&  21.0&   4.8&   102$\pm$8&   5949$\pm$367&   0.37&   1.26&   1&   1.83&   0.18$\pm$0.01&  0.0\cr
J104611.50+024351.6$^a$&  2.77&  21.8&   6.9&   218$\pm$7&   4854$\pm$126&   0.35&   2.13&   0&   $-$1.10&   0.29$\pm$0.03&  0.0\cr
J104718.35+484433.8&  2.28&  20.9&   5.3&   158$\pm$4&   2521$\pm$51&   0.36&   0.57&   0&   $-$0.47&   0.19$\pm$0.01&  0.0\cr
J104754.58+621300.5&  2.54&  21.3&   5.1&   105$\pm$4&   5081$\pm$133&   0.37&   2.88&   0&   $-$0.92&   0.20$\pm$0.01&  0.0\cr
J110202.68$-$000752.7&  2.63&  21.9&   4.9&   121$\pm$6&   3767$\pm$282&   0.18&   0.41&   0&   $-$1.73&   0.29$\pm$0.01&  0.0\cr
J111017.13+193012.5$^a$&  2.51&  20.1&   6.2&   220$\pm$7&   2427$\pm$65&   0.33&   1.62&   1&   1.30&   0.48$\pm$0.00&  0.0\cr
J111346.10+185451.9&  2.52&  21.7&   4.6&   127$\pm$3&   986$\pm$26&   0.35&   0.49&   0&   $-$1.09&   0.18$\pm$0.02&  0.0\cr
J111355.72+451452.6&  2.19&  20.5&   4.7&   190$\pm$7&   1243$\pm$38&   0.35&   0.53&   0&   $-$0.59&   0.16$\pm$0.01&  0.0\cr
J111516.33+194950.4&  2.79&  22.0&   5.0&   247$\pm$13&   1739$\pm$75&   0.36&   1.45&   0&   0.43&   0.29$\pm$0.03&  0.0\cr
J111729.56+462331.2&  2.13&  21.8&   6.3&   395$\pm$8&   3053$\pm$54&   0.35&   0.51&   0&   0.27&   0.55$\pm$0.02&  0.0\cr
J113349.71+634740.0&  2.20&  21.2&   4.6&   121$\pm$4&   2081$\pm$69&   0.34&   1.90&   1&   $-$1.38&   0.26$\pm$0.01&  0.0\cr
J113834.68+473250.0&  2.31&  21.9&   6.1&   177$\pm$6&   3296$\pm$104&   0.22&   1.72&   0&   $-$0.17&   0.30$\pm$0.01&  0.0\cr
J121253.47+595801.2$^b$& 2.58& 20.8& 4.9& 107$\pm$3& 1402$\pm$41& 0.34& 2.3& 1& 0.13& 0.26$\pm$0.01& 0.0\cr
J121704.70+023417.1&  2.42&  21.0&   5.6&   181$\pm$5&   2604$\pm$55&   0.37&   2.42&   0&   2.01&   0.40$\pm$0.01&  0.0\cr
J122000.68+064045.3&  2.80&  21.4&   4.9&   113$\pm$6&   1047$\pm$57&   0.15&   0.61&   0&   0.04&   0.23$\pm$0.02&  0.0\cr
J123241.73+091209.3$^a$&  2.38&  21.1&   6.8&   225$\pm$3&   4787$\pm$52&   0.37&   1.86&   0&   $-$0.74&   0.28$\pm$0.01&  0.0\cr
J124106.97+295220.8&  2.79&  21.8&   5.3&   138$\pm$7&   2600$\pm$133&   0.20&   1.26&   0&   0.12&   ---&  0.0\cr
J124738.40+501517.7&  2.39&  21.5&   5.0&   135$\pm$5&   3268$\pm$118&   0.29&   1.68&   0&   $-$0.35&   0.30$\pm$0.01&  0.0\cr
J125019.46+630638.6&  2.40&  21.9&   5.5&   242$\pm$4&   1881$\pm$28&   0.34&   0.56&   0&   $-$1.20&   0.20$\pm$0.01&  0.0\cr
J125449.50+210448.4&  3.12&  21.2&   5.8&   141$\pm$3&   2482$\pm$48&   0.36&   2.05&   0&   $-$0.08&   0.41$\pm$0.01&  0.0\cr
J125811.25+212359.6&  2.61&  21.5&   5.6&   158$\pm$5&   1599$\pm$53&   0.18&   0.48&   0&   $-$0.42&   ---&  0.0\cr
J130114.46+131207.4&  2.79&  21.4&   5.1&   186$\pm$4&   1877$\pm$35&   0.36&   2.42&   0&   $-$0.33&   0.27$\pm$0.02&  0.0\cr
\hline
\end{tabular}
\end{minipage}
\end{center}
\end{table*}
\setcounter{table}{1}
\begin{table*}
\begin{center}
\begin{minipage}{176mm}
\caption{Core ERQ Properties (Continued)}
  \begin{tabular}{@{}lccccccccccc@{}}
  \hline
 Quasar Name& $z_{e}$& $i$ & \imw & REW & FWHM& $kt_{80}$& \nv /\civ& BAL & $\alpha_{\lambda}$& $E(B$$-$$V)$& FIRST\\
 & & (mag) & (mag) & (\AA ) & (km/s)& & & & & & (mJy)\\
\hline
J130421.12+083752.2&  2.95&  21.9&   5.7&   132$\pm$5&   3014$\pm$83&   0.37&   1.56&   0&   $-$0.91&   0.34$\pm$0.02&  0.0\cr
J130630.66+584734.7&  2.30&  21.7&   5.0&   331$\pm$6&   1133$\pm$18&   0.36&   0.28&   0&   $-$1.24&   ---&  0.0\cr
J130654.76+132704.8&  2.50&  21.7&   5.0&   196$\pm$7&   2511$\pm$71&   0.35&   0.39&   0&   $-$0.17&   0.22$\pm$0.02&  0.0\cr
J130936.14+560111.3&  2.57&  21.9&   6.4&   161$\pm$6&   3630$\pm$114&   0.36&   1.91&   0&   $-$0.56&   0.26$\pm$0.01&  0.0\cr
J131047.78+322518.3&  3.02&  20.3&   5.3&   226$\pm$4&   2794$\pm$41&   0.37&   2.15&   1&   0.18&   0.31$\pm$0.01&  0.0\cr
J131722.85+322207.5&  2.40&  22.1&   6.0&   160$\pm$5&   2311$\pm$65&   0.36&   2.27&   0&   $-$0.36&   0.31$\pm$0.01&  0.0\cr
J131833.76+261746.9&  2.27&  21.3&   4.9&   150$\pm$4&   1280$\pm$29&   0.36&   0.68&   0&   0.26&   0.21$\pm$0.01&  0.0\cr
J134001.90+322155.9&  2.40&  21.7&   4.7&   131$\pm$7&   1823$\pm$125&   0.21&   0.89&   0&   0.38&   0.41$\pm$0.01&  0.0\cr
J134026.99+083427.2$^a$&  2.48&  22.2&   5.6&   358$\pm$12&   2076$\pm$52&   0.35&   1.74&   0&   $-$0.71&   ---&  0.0\cr
J134417.34+445459.4&  3.04&  22.3&   6.8&   310$\pm$9&   2871$\pm$70&   0.36&   1.61&   0&   $-$0.31&   0.46$\pm$0.02&  0.0\cr
J134450.51+140139.2&  2.75&  21.6&   5.1&   132$\pm$7&   4487$\pm$191&   0.37&   2.35&   0&   0.27&   0.32$\pm$0.02&  0.0\cr
J134535.66+600028.4&  2.94&  21.6&   5.1&   142$\pm$21&   6460$\pm$742&   0.37&   0.92&   0&   $-$1.38&   0.13$\pm$0.02&  0.0\cr
J135557.60+144733.1&  2.70&  20.4&   4.7&   118$\pm$3&   2958$\pm$60&   0.34&   0.75&   0&   $-$0.10&   0.16$\pm$0.01&  0.0\cr
J135608.32+073017.2&  2.27&  21.6&   5.1&   110$\pm$3&   2043$\pm$46&   0.37&   2.48&   1&   $-$1.29&   0.31$\pm$0.01&  0.0\cr
J140506.80+543227.3&  3.21&  21.1&   4.7&   123$\pm$3&   2640$\pm$59&   0.36&   2.62&   1&   $-$1.64&   0.09$\pm$0.02&  0.0\cr
J141350.76+214307.7&  2.44&  22.1&   4.9&   208$\pm$6&   2356$\pm$57&   0.36&   0.87&   1&   $-$0.75&   0.19$\pm$0.02&  0.0\cr
J143159.76+173032.6&  2.38&  21.8&   5.8&   177$\pm$3&   2084$\pm$29&   0.37&   1.22&   0&   $-$0.45&   0.31$\pm$0.01&  0.0\cr
J145354.70+190343.9&  2.35&  19.9&   4.8&   142$\pm$5&   2046$\pm$60&   0.35&   0.52&   1&   1.08&   0.28$\pm$0.00&  0.0\cr
J145623.35+214516.2&  2.48&  21.0&   5.0&   103$\pm$5&   4422$\pm$148&   0.36&   1.22&   0&   $-$0.32&   0.32$\pm$0.01&  0.0\cr
J150117.07+231730.9&  3.03&  22.1&   5.9&   231$\pm$10&   4035$\pm$130&   0.23&   1.94&   0&   0.00&   0.30$\pm$0.02&  0.0\cr
J152941.01+464517.6&  2.42&  20.7&   4.8&   159$\pm$4&   1896$\pm$42&   0.35&   2.01&   0&   $-$1.16&   0.12$\pm$0.01&  0.0\cr
J153107.14+105825.8&  2.78&  22.0&   5.9&   152$\pm$7&   3937$\pm$146&   0.37&   1.43&   0&   $-$1.13&   0.25$\pm$0.02&  0.0\cr
J153108.10+213725.1&  2.57&  22.1&   5.2&   213$\pm$11&   2767$\pm$143&   0.26&   0.73&   0&   1.29&   0.20$\pm$0.02&  0.0\cr
J153446.26+515933.8&  2.26&  21.7&   4.7&   127$\pm$5&   1156$\pm$57&   0.28&   0.69&   0&   $-$0.72&   0.32$\pm$0.01&  0.0\cr
J154243.87+102001.5&  3.21&  22.2&   6.6&   114$\pm$11&   3901$\pm$286&   0.37&   2.79&   0&   $-$0.82&   0.48$\pm$0.02&  0.0\cr
J154743.78+615431.1&  2.87&  21.7&   4.9&   128$\pm$10&   1177$\pm$111&   0.26&   0.92&   0&   $-$1.15&   0.33$\pm$0.02&  0.0\cr
J154831.92+311951.4&  2.74&  21.7&   4.8&   127$\pm$6&   3050$\pm$104&   0.37&   2.28&   0&   $-$0.71&   0.31$\pm$0.02&  0.0\cr
J160431.55+563354.2&  2.48&  21.5&   5.7&   205$\pm$5&   4221$\pm$82&   0.34&   2.26&   0&   $-$1.66&   0.20$\pm$0.01&  0.0\cr
J164725.72+522948.6&  2.72&  21.6&   5.2&   124$\pm$4&   1905$\pm$52&   0.35&   1.78&   0&   $-$1.05&   0.11$\pm$0.02&  1.2\cr
J165202.64+172852.3&  2.94&  20.3&   5.4&   125$\pm$2&   2403$\pm$45&   0.33&   2.37&   0&   $-$0.41&   0.31$\pm$0.01&  1.6\cr
J170558.64+273624.7&  2.45&  20.6&   5.1&   157$\pm$3&   1301$\pm$22&   0.36&   0.47&   1&   0.32&   0.36$\pm$0.00&  0.0\cr
J171420.38+414815.7&  2.34&  21.3&   4.7&   130$\pm$5&   3816$\pm$109&   0.36&   1.74&   1&   $-$0.74&   0.20$\pm$0.01&  0.0\cr
J220337.79+121955.3$^a$&  2.62&  21.7&   6.2&   266$\pm$3&   1070$\pm$9&   0.37&   0.51&   0&   0.23&   0.40$\pm$0.01&  0.0\cr
J221524.00$-$005643.8$^a$&  2.51&  22.2&   6.2&   153$\pm$5&   4280$\pm$112&   0.37&   0.86&   0&   0.06&   0.36$\pm$0.02&  0.0\cr
J222421.63+174041.2&  2.17&  21.5&   4.8&   110$\pm$4&   2749$\pm$102&   0.19&   1.56&   1&   $-$0.93&   0.24$\pm$0.01&  ---\cr
J223754.52+065026.6&  2.61&  22.0&   5.8&   141$\pm$5&   1391$\pm$44&   0.36&   1.03&   0&   $-$1.20&   0.24$\pm$0.02&  0.0\cr
J225438.30+232714.5&  3.09&  22.0&   5.5&   415$\pm$20&   4412$\pm$146&   0.36&   0.66&   0&   $-$0.38&   ---&  ---\cr
J232326.17$-$010033.1$^a$&  2.36&  22.4&   7.2&   256$\pm$5&   3989$\pm$62&   0.37&   2.16&   0&   0.91&   0.40$\pm$0.01&  0.0\cr
J232828.47+044346.8&  2.56&  21.5&   5.5&   359$\pm$5&   1584$\pm$20&   0.35&   0.37&   0&   $-$0.28&   0.13$\pm$0.02&  0.0\cr
J233636.99+065231.0&  2.78&  22.2&   6.0&   128$\pm$6&   1484$\pm$83&   0.26&   1.10&   0&   $-$0.19&   ---&  0.0\cr
\hline
\multicolumn{12}{l}{$^a$These quasars are in the \cite{Ross15} ERQ sample.}\cr
\multicolumn{12}{l}{$^b$These quasars are in DR12Q but not in our emission line catalog.}
\end{tabular}
\end{minipage}
\end{center}
\end{table*}

\subsection{Radio Properties}

Out of the 91 core ERQs covered by FIRST, 8 are detected at $\ge$1 mJy. (None of the 6 objects without FIRST coverage are detected by the NVSS at $>$2.5 mJy). At the median redshift of the radio detections, $z_e = 2.67$, the 1 mJy flux limit of the FIRST survey corresponds to a k-corrected luminosity $\nu L_{\nu}$(1.4GHz) $\sim 4\times 10^{41}$ ergs s$^{-1}$. While radio spectral indices of this population are unknown, we assume $\alpha=-0.7$ in our calculation. As discussed by \citet{Ross15}, it is unlikely that much of the radio luminosity in the FIRST-detected sources is due to star formation, as star formation rates in excess of 8000 $M_{\odot}$/year would be required to produce such luminosity, as per calibrations by \citep{Bell03}. Thus, we suspect that radio emission at this level must be due to the quasar -- either from the jets or as a bi-product of radiatively driven winds \citep{Zakamska14}.

The luminosity that corresponds to the flux limit of the FIRST survey is close to the traditional cutoff between radio-quiet and radio-loud objects at these redshifts \citep{Chun99, Richards11}. Two of the eight ERQs with FIRST detections were targeted by BOSS exclusively due to their radio detections, and therefore the fraction of FIRST detections in a sample selected only by their optical and infrared properties is 6/89=6.7\%, similar to other luminous quasar populations \citep{Zakamska04, Jiang07}. For a direct comparison of this detection fraction between ERQs and the overall quasar population, we take all $\sim$75,000 quasars at $2.4<z<2.9$ in the DR12 quasar catalog \citep{Paris16} and use their FIRST\_MATCHED flag to calculate the fraction of quasars with matches in the FIRST survey in the same way as we do for ERQs. While the overall FIRST detection fraction is only 3.4\% for quasars in this redshift range, it rises steadily as we consider more and more luminous objects until it reaches the same percentage as ERQs for the $\sim$8000 quasars brighter than $M_{2500}=-26.6$ mag. Therefore, the FIRST detection rates of ERQs are similar to those of the brightest Type 1 quasars at the same redshift. 

Mean and median stacks \citep{White07} of the 81 non-detected core ERQs with FIRST coverage yield a measured flux of 125 $\mu$Jy/beam in the mean and 107 $\mu$Jy/beam in the median. These fluxes are likely underestimated because of the poorly understood CLEAN bias which is introduced during the non-linear radio image reconstruction from the incompletely sampled Fourier space \citep{White07}. Assuming that most of the flux is due to point sources and correcting for the CLEAN bias using estimates from \citet{White07}, we estimate that the true mean flux of the radio-quiet core ERQs is 180 $\mu$Jy. The main sources of uncertainty in this estimate are the dispersion of the radio fluxes within the population and the CLEAN bias. By resampling (with return) the images that contribute to the stacks we estimate the error in the mean flux to be 25 $\mu$Jy and the uncertainly in the CLEAN bias is on the same order. At the median redshift of the stacked sample, $z_e = 2.48$, the estimated true mean flux of 180 $\mu$Jy corresponds to k-corrected luminosity  $\nu L_{\nu}$(1.4GHz) $\sim 8\times 10^{40}$ ergs s$^{-1}$. \citet{Alexandroff16} discuss comparisons of the average radio properties of ERQs to high-redshift Type 2 quasar candidates and conclude that ERQs are brighter in the radio than optically selected Type 2 quasar candidates at the same redshift.

\subsection{Broad Outflow Absorption Lines}

Fourteen ($\sim$20\%) of the 71 core ERQs with FWHM(\civ ) $>$ 2000 \kms\ in our emission-line catalog have BALs identified by visual inspection in DR12Q. This BAL fraction is nearly 1.5 times larger than the 14\% we find for Type 1s in our $W3$-detected sample overall. We note that BALs identified by visual inspection do not always meet the rigorous definition of ``balnicity index'' BI $>$ 0 \citep[][]{Weymann91, Paris14, Paris16}. They can be narrower or at lower velocities than BALs defined by BI to have FWHM $\ga$ 2000 \kms\ and v~$ > 3000$ \kms . However, visual inspection can be more effective at identifying BALs than automated BI determinations for quasars with noisy/weak continua or with absorption in the wing of very strong broad emission lines (which tends to be the case for the ERQs). Thus we proceed with the understanding that the BALs discussed here are not always of the classic variety described by \cite{Weymann91}, but they are nonetheless broad and indicative of high-speed quasar-driven outflows. 

We also note that the BAL fractions in our study are underestimates because our emission-line catalog excludes quasars with BALs at wavelengths that might affect our emission line fits (\S3). The BAL fraction in our $W3$-detected sample is, nonetheless, similar to other BAL quasar studies based on SDSS-I/II  \citep[e.g.,][]{Gibson09, Knigge08, Trump06}. For the core ERQs, our visually inspections of all BOSS spectra of ERQs in DR12Q finds only 1 more quasar with FWHM(\civ ) $>$ 2000 \kms , which has strong BALs, bringing the final BAL fraction to 15/72 = 21\% for all core ERQs with FWHM(\civ ) $>$ 2000 \kms\ in DR12Q. 

However, our visual inspections also reveal that BALs and BAL-like features are missed preferentially in core ERQs by the visual inspection flag {\tt bal\_flag\_vi} in DR12Q. We attribute this to two factors: 1) ERQs are typically faint in $i$ ($\sim$2 magnitudes fainter than $W3$-detected quasars overall, Figure~5), which leads to noisier BOSS spectra and more difficult BAL detections, and 2) the much larger REWs in the core ERQs can mask the appearance of weak/moderate BAL features in the BOSS spectra. Some examples of core ERQs with definite BALs or BAL-like absorption not flagged in DR12Q are J123241+091209, J130421+083752, J153107+105825 and J1605252+172852 in Figure~12. Figure~13 shows these features more clearly on an expanded vertical scale. They appear in both \civ\ and \siiv\ with substantial depths below the continuum and widths ranging from FWHM $\sim$ 1900 \kms\ in J130421+083752 to $\sim$6800 \kms\ in J153107+105825. Results like this from our own visual inspections indicate that the actual BAL fraction in the core ERQs is $\ga$30\%. 

\begin{figure}
\includegraphics[scale=0.6,angle=0.0]{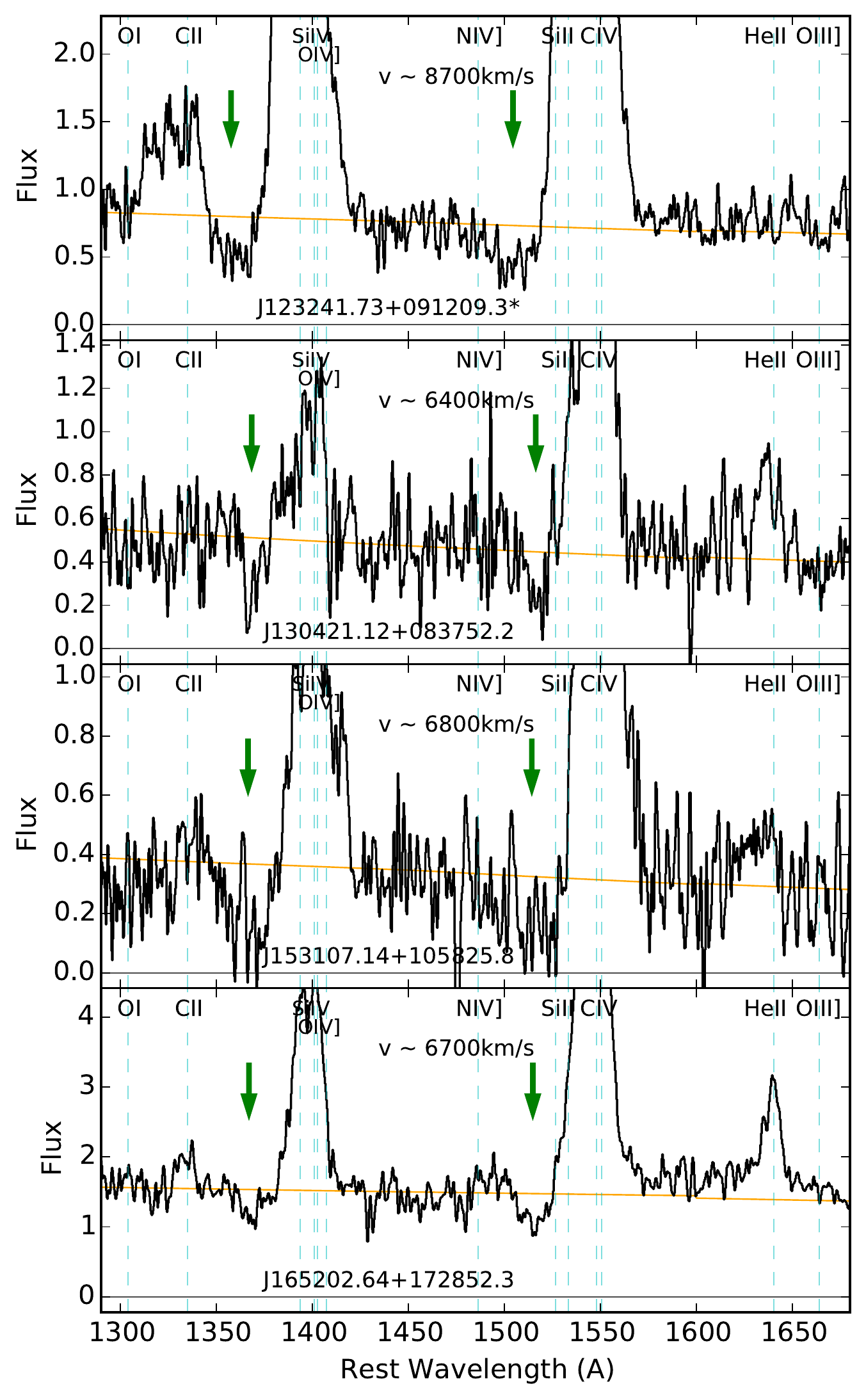}
\vspace{-8pt}
 \caption{BOSS spectra of four core ERQs from Figure 12 that have BALs or BAL-like features in \civ\ and \siiv\ but were not flagged as BALs in DR12Q. The BAL features are marked by green arrows at velocity shifts (relative to the best available DR12Q redshift) indicated above the arrows at \civ . See \S5.3 and the Figure~12 caption.}
\end{figure} 

\subsection{Emission Line Widths \& Type 1 versus 2}

Figure 2 above shows that the core ERQs have generally narrower \civ\ emission lines than other BOSS quasars. If we discard the Type 2s and consider only Type 1 non-BAL quasars (as in Figure~8), then the core ERQs have median $\left<{\rm FWHM(\civ )}\right> = 3050\pm 990$ \kms\ compared to $\left<{\rm FWHM(\civ )}\right> = 5836\pm 1576$ \kms\ for blue quasars that are matched to the core ERQs in $W3$.

Many core ERQs have FWHMs near our adopted Type 1/Type 2 boundary at FWHM(\civ ) $= 2000$ \kms . In standard Unified Models \citep{Antonucci93,Urry95,Netzer15}, the Type 1 versus Type 2 classification is useful to constrain the location of the emission-line regions and determine if we have a direct view of the quasar central engines. Broad permitted lines (Type 1s) are attributed to dense sub-parsec environments moving at large virial speeds near the accretion disk, while narrow (and forbidden) lines (Type 2s) arise from a low-density environment hundreds or thousands of parsecs farther out. Unfortunately, Type 1 versus 2 classifications for the core ERQs can be ambiguous, not only because of the small FWHMs, but also because they have characteristically strong lines with wingless profiles sitting atop weak continua resembling Type 2s (Figures~8 and 12). 

Figure~14 shows BOSS spectra of several core ERQs with narrow lines indicative of a Type 2 classification (FWHM(\civ ) $<$ 2000 \kms ) but other line properties that resemble Type 1s in our ERQ sample. In particular, these quasars have strong low-ionization permitted lines such as \oi\ \lam 1304, \siii\ \lam 1260, and \feii\ \lam 1787 (plus the blend \feii\ 2300-2750 \AA\ not shown), two of them have line flux ratios \nv /\civ\ $\ge$ 2, and J152941+464517 has weak BAL-like absorption in \civ\ that was not flagged by visual inspection in DR12Q. Low-ionization permitted lines are suggestive of Type 1 origin, perhaps like narrow-line Type 1 sources \citep{Laor97,Constantin03}, but they have also been measured in Type 2s including the prototypical Seyfert 2 galaxy NGC 1068 \citep{Kraemer00}. We are left without definitive guidelines, but a visual inspection of all 24 core ERQs with FWHM(\civ ) $<$ 2000 \kms\ suggests that up to half of them have ambiguous Type 2 versus 1 classifications based on BAL-like features or strong low-ionization emission lines.

\begin{figure*}
\includegraphics[scale=0.53,angle=0.0]{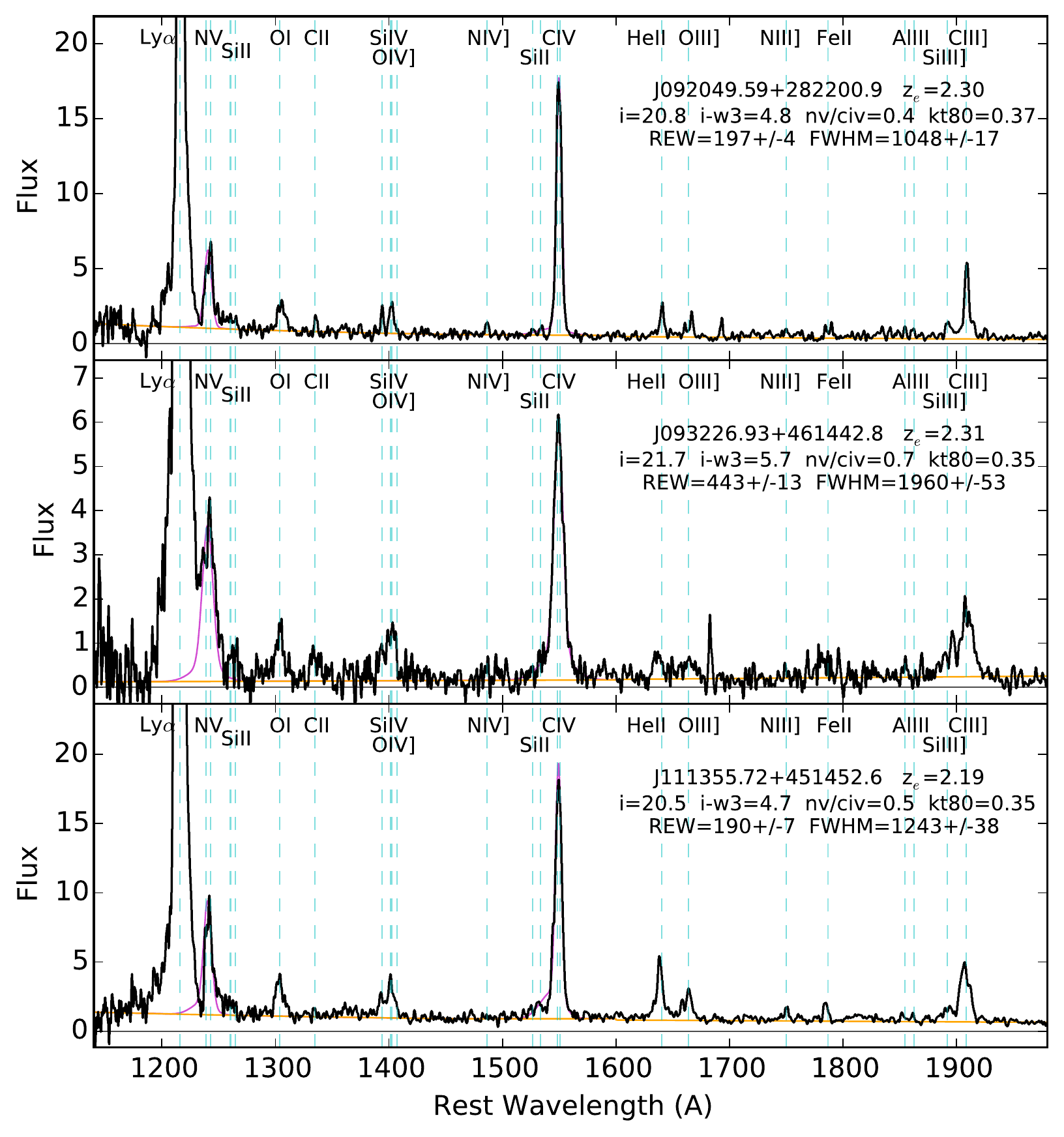}
\includegraphics[scale=0.53,angle=0.0]{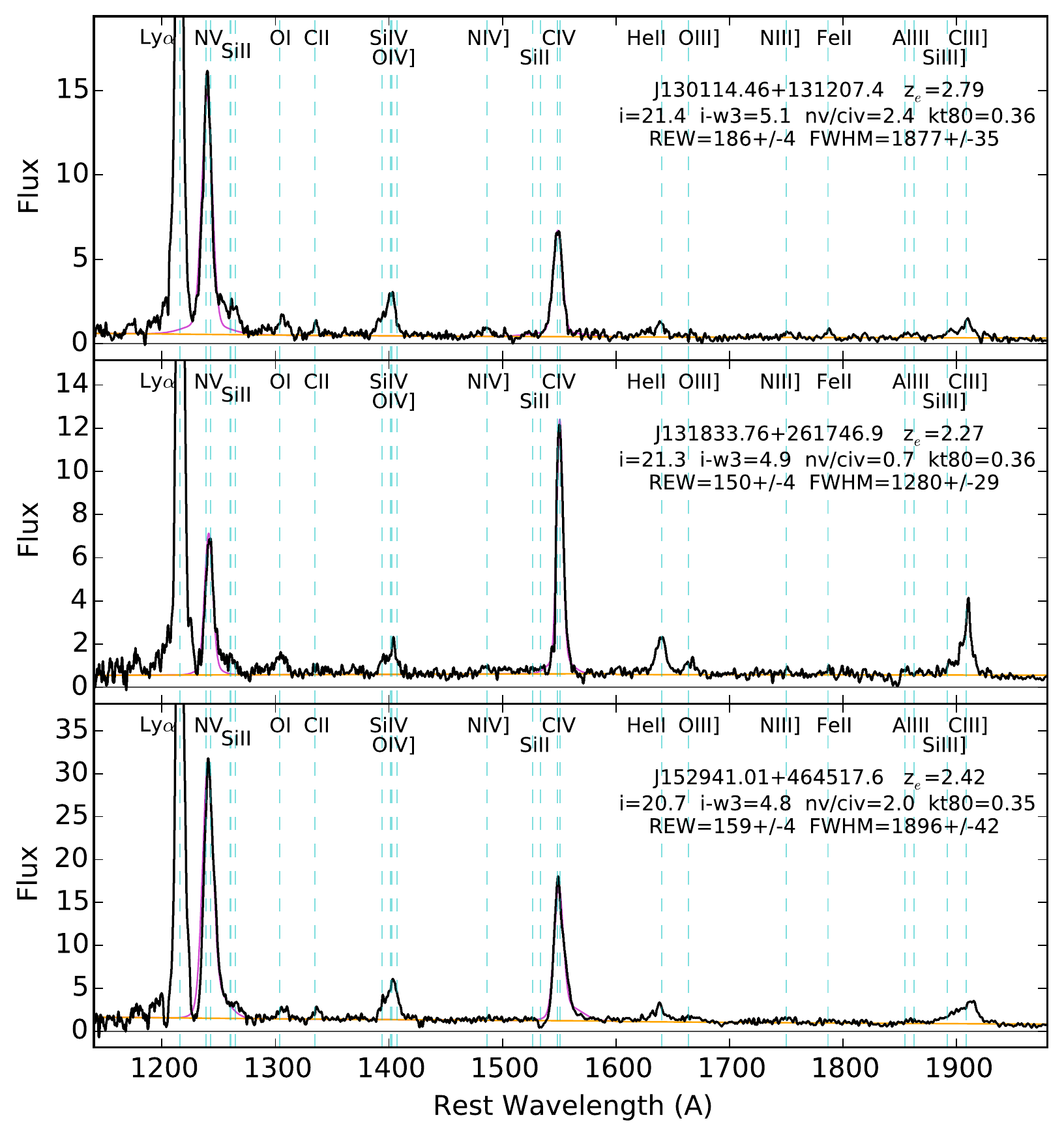}
\vspace{-8pt}
 \caption{BOSS spectra of core ERQs that resemble Type 1s in spite of narrow line widths FWHM(\civ ) $<$ 2000 \kms . All of these quasars have strong emission in \oi\ \lam 1304, \siii\ \lam 1260, and \feii\ \lam 1787, two of them have \nv /\civ\ $\ge$ 2, and J152941+464517 has BAL-like absorption in \civ . See \S5.4 and the Figure~12 caption.}
\end{figure*} 

This ambiguity raises a more general question about the meaning of ``broad" versus ``narrow" line regions in the core ERQs.  We have shown here that the permitted \civ\ lines tend to be narrow unusually narrow in the core ERQs, with half of the nominal Type 1s having FWHM(\civ ) $<$ 3050 \kms\ (also Figures 2 and 8), while other observations show that the forbidden [OIII] \lam 5007 lines can be exceptionally broad, with FWHM $\sim 2600$--5000 \kms\ \citep[in 4 core ERQs measured so far,][]{Zakamska16}. We speculate in \S6.2 that these unusual line kinematics in the core ERQs are caused by outflows and spatially-extended broad line regions that might connect to their low-density forbidden line regions farther out. 

\subsection{SEDs Inconsistent with Simple Reddening}

The median SEDs in the right-hand panels of Figure~11 reveal another important characteristic of the core ERQs. We have already noted (\S4.5) that ERQs with normal line strengths, $30 \le$ REW(\civ ) $\le$ 70 \AA\ (black curve, bottom right panel) have a median SED consistent with normal quasars reddened by a standard reddening curve. In contrast to this, the top-right panel in Figure~11 shows the core ERQs have a median SED that is {\it not} consistent with a simple reddening picture. Their SEDs are characteristically much redder than the reddened QSO1 template across the mid-to-near-IR, but much bluer (flatter) than the reddened template across the UV. 

Figure~15 shows these SED behaviors in another way for individual quasars. It plots \imw\ versus $r$$-$$z$ (top panel) and the UV power law index from our continuum fits $\alpha_{\lambda}$ (bottom panel, see Appendix A). The $r$$-$$z$ plot omits quasars with redshifts $z_e>2.7$ to avoid contamination by \civ\ emission in the $r$ band. Most of the core ERQs reside above the reddening vector shown by the light green line. The ERQs are slightly redder across the rest UV than the typical/blue BOSS quasars, but they are not nearly as red as expected from a simple reddening law. The bottom panel in Figure~15 shows similarly that the core ERQs, in spite of their extreme red \imw\ color, have only slightly redder UV continuum slopes than the overall quasar population.

\begin{figure}
\includegraphics[scale=0.48,angle=0.0]{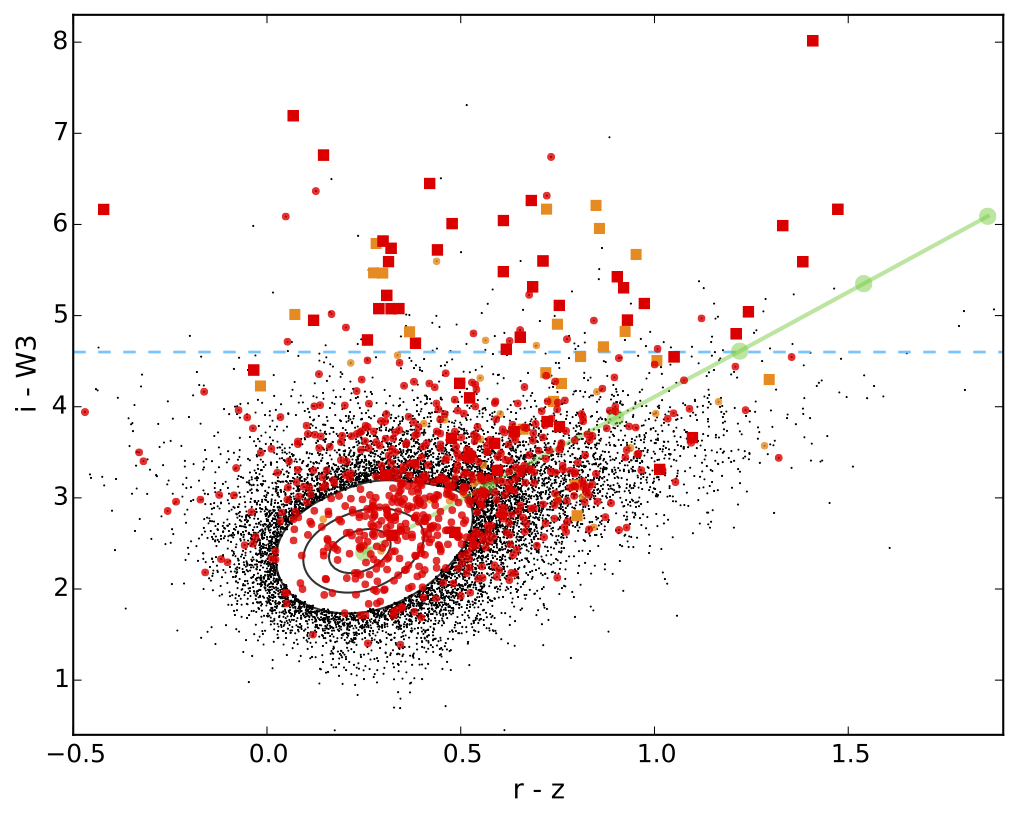}
\includegraphics[scale=0.48,angle=0.0]{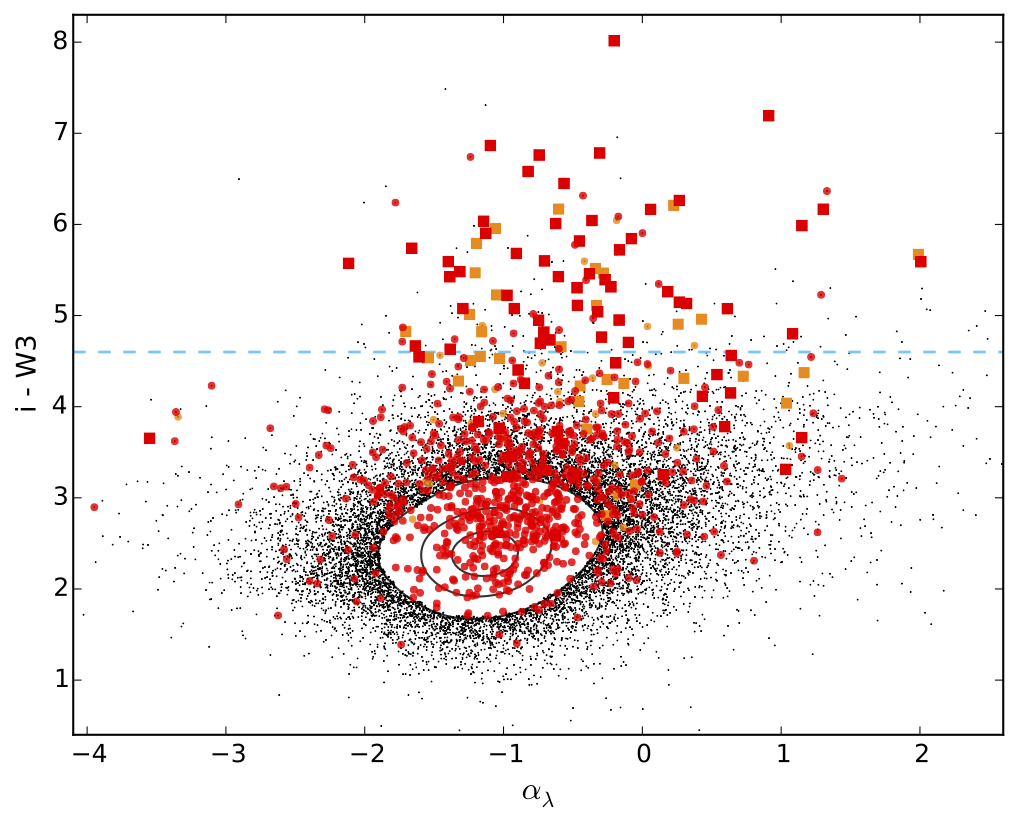}
  \vspace{-8pt}
 \caption{{\it Top panel:} \imw\ versus $r-z$ color for $W3$-detected quasars at redshifts $z_e \leq 2.7$. The symbols and line markings are the same as in Figures 9 and 10. The light green reddening vector extends from the sample centroid to the upper right with dots marking $E(B-V) = 0$, 0.1, 0.2, 0.3, 0.4 and 0.5. {\it Bottom panel:} \imw\ versus UV spectral index $\alpha_{\lambda}$ for the entire $W3$-detected sample.}
\end{figure}

The peculiar SED shapes in the core ERQs make it difficult to characterize the amount of reddening and obscuration. The median \imw\ color of the core ERQs compared to the rest of the $W3$-detected sample (Figure 11, \S5.1) indicates that the core ERQs have nominally almost 3 magnitudes of extinction in the rest-frame UV near 0.2 \mum . The \ebv\ values listed in Table 1 describe the reddening across rest wavelengths from roughly 0.2 to 1.0 \mum\ (for $z_e \sim 2.5$). This works well for the vast majority of BOSS quasars at these redshifts (Hamann et al. 2016b, in prep.). The core ERQs are clearly red across this wavelength range with median $\left<E(B-V)\right> =0.27$. However, this amount of reddening is less than expected from the red \imw\ colors. If we use the same reddening curve and color standard for $E(B-V)=0$ as Hamann et al. (2016b, in prep.), we find that the flux ratio between 0.2 \mum\ and 4.0 \mum\ in the median SED of the core ERQs is better represented by $\left<E(B-V)\right> = 0.48$. However, this characterization of the obscuration is misleading and probably incorrect because large \ebv\ values should be accompanied by steep slopes across the rest UV. Something more complex than a reddening curve is needed to explain the extreme red \imw\ colors with relatively flat UV slopes in the core ERQs. We discuss possible physical explanations in \S6.1 \citep[see also][]{Zakamska16,Alexandroff13}.

It is important to remember here that selection biases are at work. The rest-frame UV colors of all of the quasars in our study are defined ultimately by BOSS target selection \citep{Bovy11,Ross12}. This can include quasars with atypical UV colors if they were targeted in an ancillary program or they were observed early in the BOSS program when radio fluxes where used to help identify quasars \citep[see][]{Ross15}. However, overall, BOSS quasars cannot deviate greatly from the colors of ``normal" quasars and, in particular, they cannot be very red across the rest-frame UV. Our study, therefore, favors ERQ SEDs that are relatively flat across the UV in spite of extremely red \imw\ colors. However, this selection bias does not deny the reality of these unusual SEDs and it cannot explain why the core ERQs have flatter UV slopes than the other ERQs we call interlopers with normal line strengths (Figure 11). These other ERQs not in the core sample demonstrate that redder UV slopes than we find typically in the core ERQs {\it are} allowed by BOSS quasar selection but the core ERQs prefer flatter/bluer UV slopes. Thus these unusual SEDs appear to be a real property of the core ERQs and not just an artifact of selection biases.

\subsection{Comparisons to HotDOGs \& Highly-Reddened Quasars}

Figure~16 compares the median SEDs of Type 1 non-BAL quasars in the core ERQ sample (as in the upper right panel of Figure~11) to other obscured quasar samples from the literature, including a typical Type 2 quasar \cite[QSO2, from][]{Mateos13}, a typical HotDOG \citep[from][]{Tsai15}, and a typical luminous highly reddened Type 1 quasar \citep[HR1, from][]{Banerji13}. These comparison SEDs were constructed by us using tracing software to draw approximate average SEDs in the figures shown by \cite{Tsai15}, \cite{Banerji13} and \cite{Mateos13}. 

\begin{figure}
\begin{center}
\includegraphics[scale=0.48,angle=0.0]{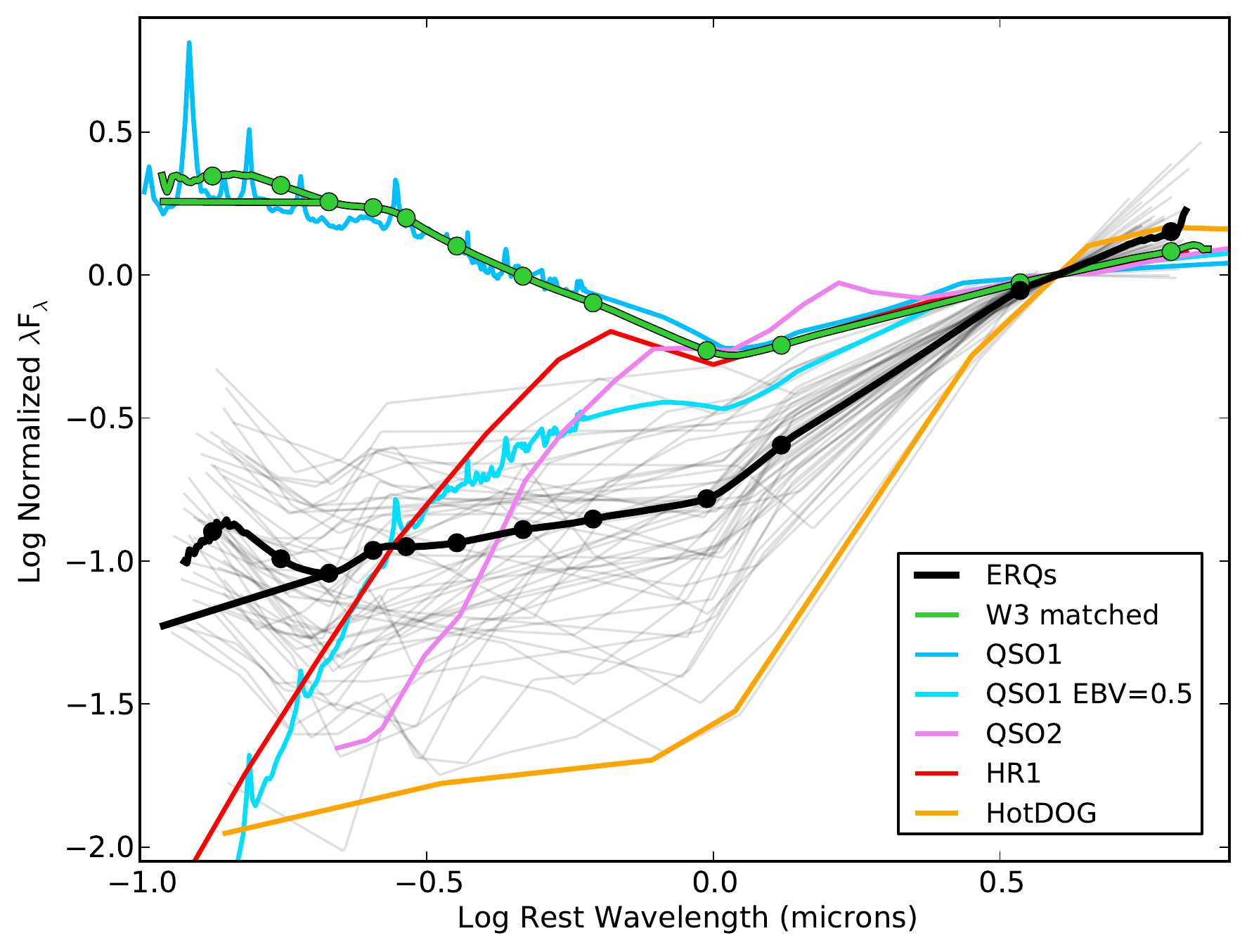}
  \vspace{-8pt}
 \caption{Normalized median SEDs for Type 1 non-BALs in the core ERQ sample (black curve) plus blue quasars matched to the core ERQs in $W3$ magnitude (green curve) as in Figure~8, the Type 1 quasar template QSO1 with and without reddening equal to $E(B-V) = 0.5$ (light blue, from Polletta et al. 2007), a typical Type 2 quasar (QSO2) from Mateos et al. (2013, purple),  a typical heavily reddened Type 1 quasar (HR1) from Banerji et al. (2013), and a typical HotDOG from Tsai et al. (2015). The light gray curves are SEDs of individual core ERQs. See \S5.5, \S5.6 and the Figure~11 caption for additional notes.}
 \end{center}
\end{figure}

The Type 2s and HR1s both exhibit steep declines across the rest-frame UV and visible. This is very different from the core ERQs but  consistent with normal quasars viewed behind a dust reddening screen. HR1s are a heavily-obscured high-redshift quasar population with luminosities similar to the ERQs. They have typical redshifts $z_e\sim 2$--3 and a median bolometric luminosity of $\log L ({\rm ergs/s}) \sim 47.1$ \citep{Banerji13, Banerji15}. However, unlike the ERQs in our study \citep[also][]{Ross15}, the HR1s are selected to be very red across the observed-frame visible through near-IR \citep{Banerji13, Banerji15}. Their SEDs, emission-line properties and estimated black hole masses lead \cite{Banerji15} to conclude that, apart from their large dust extinctions, HR1s ``appear to be similar to luminous unobscured quasars." This is the same conclusion we reached for ERQs with normal line strengths (\S4.5 and \S5.5), which are quite different from the core ERQs. 

The core ERQs have SED shapes more similar to HotDOGs. The main difference is that our modest requirement for $i-W3 > 4.6$ leads to a median color of only $\left<i-W3\right> \sim 5.6$ for the core ERQs compared to $\sim$7.6 for HotDOGs (measured by us from the SED in Figure~16). HotDOGs are an obscured quasar-dominated population with typical redshifts in the range 1.8--3.4 \citep[][and refs. therein]{Assef15,Tsai15,Fan16b}. Recent studies using WISE photometry select HotDOGs to be bright in the long-wavelength bands $W3$ and/or $W4$ but faint or undetected in $W1$ and $W2$ \citep{Eisenhardt12,Wu12,Bridge13,Wu14}. They have extremely red $r-W4$ colors like dust obscured galaxies (DOGs, e.g., \citealt{Dey08} and refs. therein) but their large mid-IR fluxes in $W3$ and $W4$ identify hot dust emissions powered by obscured AGN. The AGN contributions to HotDOG luminosities are estimated in the range $\log L ({\rm ergs/s}) \sim 47$--48 \citep{Assef15,Tsai15,Fan16b}. The core ERQs appear to be typically near the bottom of this range (\S5.1).

The relationship of ERQ and HotDOG colors is illustrated further by Figure~17, which mimics Figure~2 in Tsai et al. (2015, also \citealt{Eisenhardt12,Yan13,Ross15}). This figure plots $W1$$-$$W2$ versus $W2$$-$$W3$ for all 205 ERQs in our study plus the luminous blue quasars matched to the ERQs in $W3$ magnitude (as in Figures 8 and 16). The dashed vertical line at $W2$$-$$W3 = 3.08$ marks a minimum color used to define HotDOGs \citep{Wu12,Bridge13,Tsai15}. 

\begin{figure}
\begin{center}
\includegraphics[scale=0.48,angle=0.0]{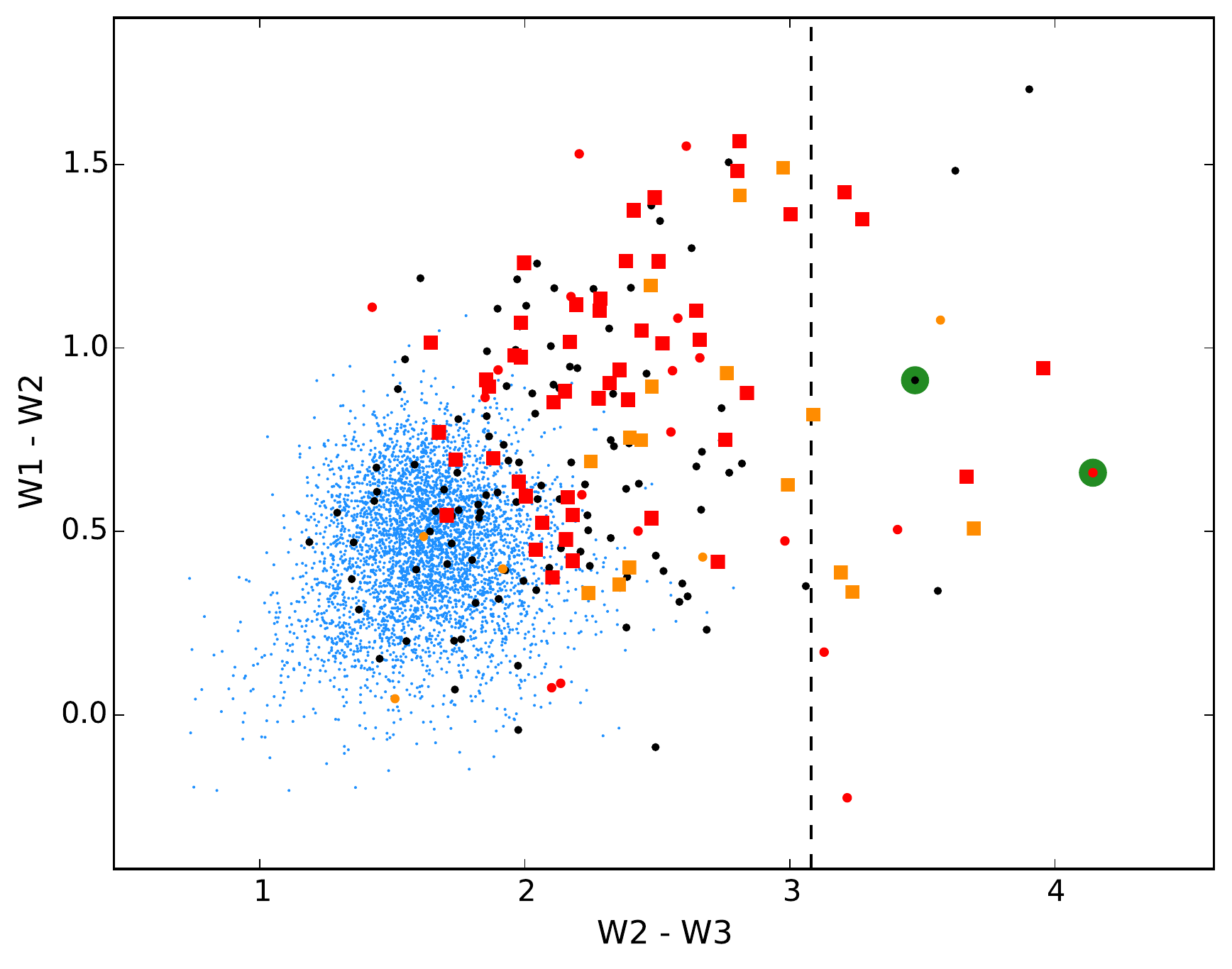}
 \vspace{-12pt}
 \end{center}
 \caption{Color-color diagram for red and blue quasars in the $W3$-detected sample. The black, red and orange symbols mark ERQs with $i-W3\ge 4.6$. The black dots indicate REW(\civ ) $<$ 100 \AA , while the red/orange symbols mark Type 1/Type 2 ERQs with REW(\civ ) $\ge 100$ \AA\ as in Figures 9 and 10. The small blue dots are luminous blue quasars matched to the ERQs in $W3$ magnitude (as in Figures 8 and 16). The dashed vertical line at $W2-W3 = 3.08$ (5.0 Vega) is an approximate minimum color for HotDOGs (e.g., Figure~2 in Tsai et al. 2015). The two large green dots mark ERQs in the HotDOG sample of Tsai et al.}
\end{figure}

As expected from the SEDs, Figure~17 shows that ERQs are not generally as red as HotDOGs in $W2$$-$$W3$ and almost half of them fall below the threshold $W1-W2 \ga 0.8$ used to identify obscured quasars in other WISE-based studies \citep{Assef10,Assef13,Stern12,Banerji12,Banerji13}. The large $W2$$-$$W3$ threshold used for HotDOGs, in particular, finds a more heavily obscured population than ERQs. However, neither of these WISE color cuts is as effective as \imw\ for finding obscured quasars with exotic emission-line properties like the core ERQs. The small number of ERQs we find that are red like HotDOGs in $W2-W3$, including two that are in the HotDOG sample of \cite[][large green dots]{Tsai15}, exhibit a wide mix of REWs and other emission-line properties. The diverse emission-line properties of HotDOGs are also evident from the UV spectra presented by \cite{Wu12}. This is an important difference from the core ERQs. The core ERQs, with the added REW(\civ ) $>$ 100 \AA\ constraint, are a much more homogeneous red quasar population with a particular set of exotic line properties that point to exotic physical conditions in their line-forming regions (\S5.1, \S6.2, Hamann et al. 2016a, in prep.).

Another recent search for HotDOGs using WISE and SDSS photometry \citep{Toba16} places strong emphasis on $i-W4 > 7.5$ color. This is the same constraint used by \cite{Ross15} and, indeed, it appears to find a quasar population overlapping with the \cite{Ross15} study. \cite{Toba16} do not list individual sources, but of the four BOSS spectra shown in that paper (their Figure~8) all are quasars with no apparent stellar features in the rest UV. One is J000610+121501 in our core ERQ sample (shown above in Figure~12, also Figure 15 in \citealt{Ross15}) and another is J101362+611219 in our expanded sample (shown in Figure~18 below; note that the broad \nv\ and \ovi\ emission lines are mislabeled as \lya\ and \lyb , respectively, in the Toba \& Nagao plots). \cite{Toba16} refer to their sample as Hyper-Lumininous Infrared Galaxies (HyLIRGs), but it is not clear that starlight makes any significant contributions to the observed SDSS or WISE fluxes. The fits they show to the quasar SEDs attribute the flattening in the UV to hot stars; however, that seems unlikely due to the quasar-dominated BOSS spectra in the rest-frame UV (see our discussion of the SEDs in \S5.5 and \S6 for an alternative interpretation). The HyLIRG and ERQ populations probably do overlap, but, for sources discovered to have Type 1 quasar spectra in the UV, measuring the starlight to classify them as HyLIRGs or make comparisons to galaxy populations probably requires a detailed analysis of specific starlight spectral features or additional photometry in the far-IR \citep[e.g.,][]{Tsai15,Netzer16}. 

\subsection{Expanding the Sample}

The constraints we use to define the core ERQs, $i-W3 \ge 4.6$ and REW(\civ ) $\ge 100$ \AA\ in the redshift range 2.0--3.4, successfully find a nearly homogeneous sample of quasars with peculiar SEDs and emission-line properties (Figures 7--12, \S5.1). However, the strict selection thresholds also exclude quasars that appear closely related to the core ERQs. Here we perform additional searches to find more quasars with emission-line properties like the core ERQs. These searches are necessarily subjective as we seek to define ``ERQ-like'' properties using measured quantities in our emission-line catalog. However, they are helpful to understand the numbers of ERQ-like quasars in the BOSS catalog and the relationships of these exotic sources to the rest of the BOSS quasar population. 

One particular search in the redshift range 2.0--4.0 based on wingless \civ\ emission lines with $kt_{80} > 0.33$ and large peak heights relative to the continuum $>$ 4.0 (but with exceptions primarily to the peak height if \nv/\civ\ $>$ 1.5, \civ\ has an even larger kurtosis, or one of the core ERQ constraints is satisfied, e.g., $i-W3 \ge 4.6$ or REW(\civ ) $\ge 100$ \AA ) yields an additional 228 quasars not in the core ERQ sample. This automated search uses numerous conditional statements developed from many trials and visual inspection. We also perform visual inspections and additional line fits to all quasars with \imw\ $>$ 4.6 in DR12Q to find 7 more ERQ-like sources not in our emission-line catalog (as described for the core ERQs in \S5.1). Thus we define a total ERQ-like sample of 228+7 = 235 quasars (Table~1). These quasars are listed in Appendix~B. We note that other searches using different parameter thresholds or different criteria (such as \nv\ $>$ \lya\ line strengths) can find more ERQ-like quasars in DR12Q (also Alice Eltvedt, private communication). 

Figure~18 shows BOSS spectra of some Type 1s in the ERQ-like sample for comparison to the Type 1 core ERQs in Figure~12. In addition to the wingless line profiles, large REWs, and large \nv /\civ\ favored by the search, many of the ERQ-like sources also have unusually strong \aliii\ \lam 1860 like the core ERQs, although we do not quantify the strength of this line. A surprisingly large fraction of the ERQ-like quasars also have BALs based on the visual inspection flag in DR12Q (see Table B.1). It is important to note that BALs and other absorption lines are not considered in the ERQ-like quasar selection. We also confirm from inspection of the line fits that the BALs discovered in our search do not truncate the blue-side emission profiles in ways that produce spuriously large kurtosis values to favor their inclusion in the sample. Nonetheless, 107 (63\%) out of 169 ERQ-like quasars with FWHM(\civ ) $>$ 2000 \kms\ have \civ\ BALs according to the visual inspection flag in DR12Q. It is even more remarkable that 26 (39\%) of the 66 quasars with FWHM(\civ ) $<$ 2000 \kms\ (nominally Type 2s) also have \civ\ BALs flagged in DR12Q. Our own visual inspections support these numbers. In a few cases, the emission line widths might be artificially reduced by the broad absorption line eating into the blue side of the emission-line profiles. However, most of these quasars appear to have intrinsically narrow emission lines. 

These occurrences of BALs and BAL-like features in narrow-line quasars indicate that the Type 1/Type 2 boundary in this sample is even more blurry than the core ERQs (\S5.4). If we adopt the Type 1 definition from \S2, which requires FWHM(\civ ) $>$ 2000 \kms\ {\it or} a BAL by visual inspection, then this ERQ-like sample has 195 Type 1 quasars and 133 of them (68\%) have BALs. The high BAL fractions in this sample also indicate that ERQ-like emission-line properties are related somehow to the BAL outflow phenomenon (\S6.2). The common presence of BALs and BAL-like absorption might explain the observed weakness of \lya\ and the \nv\ $>$ \lya\ fluxes in many of core ERQs and ERQ-like quasars (Figures 12 and 18), e.g., if blueshifted \nv\ BALs cover the broad \lya\ emission lines (see Hamann et al. 2016a, in prep., for more discussion). 

\begin{figure*}
\includegraphics[scale=0.525,angle=0.0]{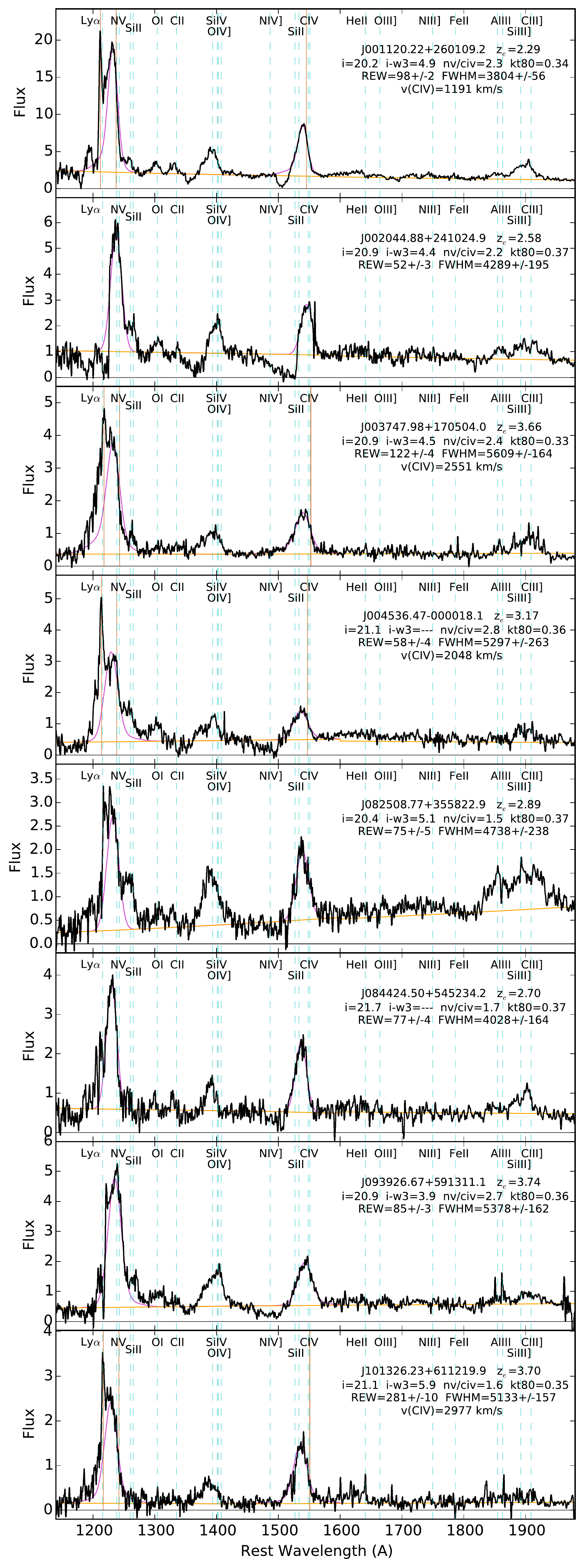}
\includegraphics[scale=0.525,angle=0.0]{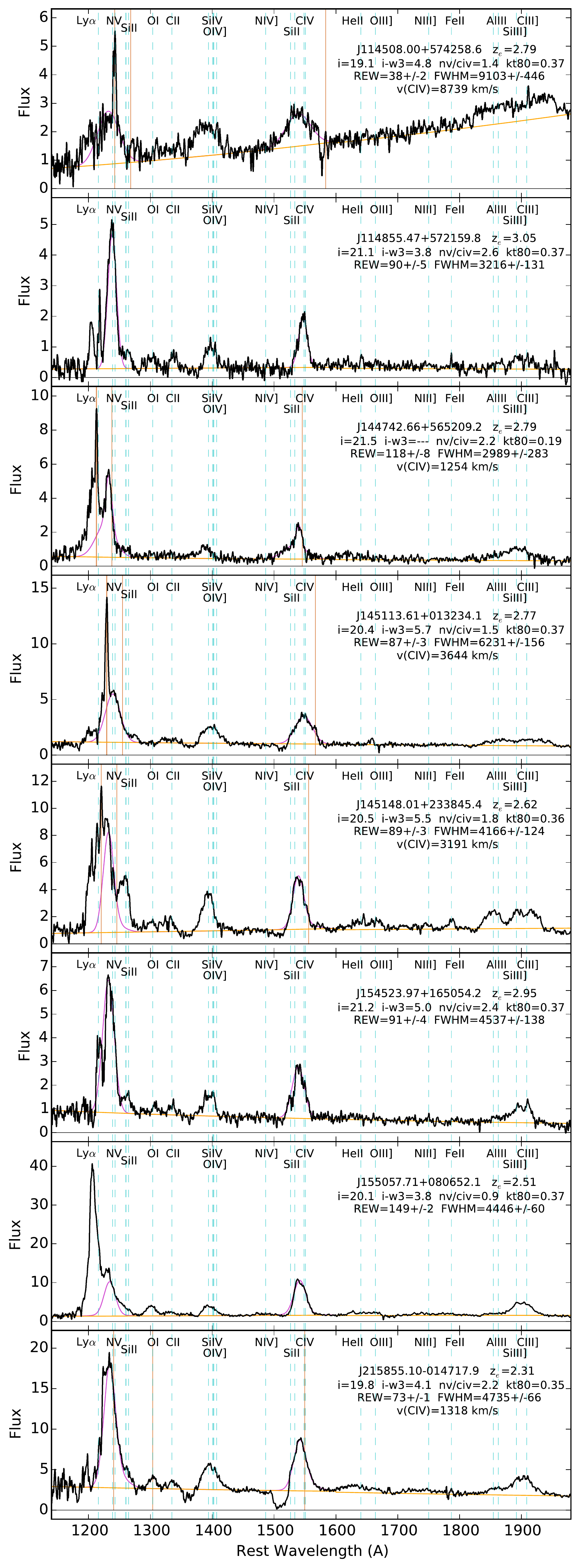}
\vspace{-10pt}
 \caption{BOSS spectra of select quasars in the ERQ-like sample, showing emission-line properties similar to the core ERQs. The brown vertical lines mark estimated systemic redshifts for some quasars that can reveal large \civ\ blueshifts listed as v(\civ ).
 See \S5.7, \S5.8, and Figure~12 for additional notes.}
\end{figure*} 

Perhaps the most important result from our additional searches is that quasars with ERQ-like emission-line properties strongly favor red \imw\ colors. For example, in a search through the $W3$-detected sample using only emission-line criteria as described above but no color constraints, the quasars selected have a median color, $\left< i-W3\right> \approx 4.66\pm 0.95$, that is $\sim$2 magnitudes redder than the median for all $W3$-detected quasars, $\left< i-W3\right> \approx 2.50\pm 0.57$ (\S4.1, Figure 4). This search finds many of the core ERQs. If we exclude the core ERQs as well as Type 2s based on FWHM $<$ 2000 \kms , the median color of the remaining 91 quasars is still $\left< i-W3\right> \approx 4.18\pm 0.74$. This reinforces our conclusion in \S4 that the peculiar emission-line properties of ERQs are closely related to red \imw\ color at these redshifts. 

\subsection{Extreme \civ\ Blueshifts}

Here we describe a surprising tendency for the core ERQs and ERQ-like quasars to have large blueshifts in their \civ\ and other high-ionization broad emission lines. Measuring these shifts from rest-frame UV spectra is a challenge because the strongest lines that dominate the redshift determinations are the same ones likely to be shifted. Low-ionization permitted lines such as \mgii\ \lam 2800, \oi\ \lam 1304 and \siii\ \lam1260 are generally regarded as good indicators of quasar systemic velocities because their shifts in well-studied sources are typically $\la$30 \kms\ \citep{Tytler92,Richards02,Hewett10,Shen16}. 

Another way to estimate the systemic redshifts is with narrow emission lines that are likely to form in the extended host galaxies or halos. Figures 12 and 18 include several core ERQs and ERQ-like sources with narrow emission spikes that appear to be \lya . See, for example, J000610+121501 and J104754+621300 in Figure~12 and J101326+611219, J114508+574258, and J145113+013234 in Figure~18. Our claim that these spikes are near the systemic redshifts is supported in most cases by good matches to the low-ionization lines mentioned above. In J145113+013234 and J145148+233845, the \lya\ spike identifications are also confirmed by narrow lines of \heii\ \lam 1640 and \feii\ \lam 1787, respectively, at the same redshift. In J114508+574258, there appears to be narrow associated absorption in \civ\ that is far to the red of the \civ\ broad emission line but consistent with a small blueshift ($\sim$1100 \kms ) relative to the \lya\ spike. There is also corresponding weak absorption in \nv . 

Brown vertical lines drawn for some quasars in Figures 12 and 18 mark redshift systems defined by well-measured \lya\ spikes or by approximate centers of the \oi\ \lam 1304 and \feii\ \lam 1787 emission lines (as available) that provides predicted systemic positions of \nv\ and \civ . The actual observed broad \nv\ and \civ\ emission lines can be very blueshifted from these positions. We measure these blueshifts relative to our preferred indicator of the \civ\ emission wavelength, which is the midpoint of the fitted profiles at their half maximum heights (Appendix A). The derived \civ\ velocity shifts relative to the \lya\ spikes are listed directly in Figures 12 and 18 when available. J102541+245424 is a BAL quasar where blueshifted \civ\ emission at v(\civ ) = 3215 \kms\ is accompanied by \civ\ absorption reaching almost 20,000 \kms . In J144742+565209 (Figure~18), the moderate blueshift of v(\civ ) = 1254 \kms\ is accompanied by a highly extended blue wing reaching velocities $>$9000 \kms . In the most extreme case, J114508+574258 (Figure~18), the blueshift is so large, v(\civ ) = 8739 \kms , that the broad \nv\ line appears on the {\it blue} side of the \lya\ spike. This quasar has unusually weak and broad lines compared to the core ERQs (cf., Figures 12 and 18); however, it is ``ERQ-like'' based on ERQ colors, \imw\ = 4.8, as well as \nv\ $>$ \lya , \nv /\civ\ = 1.4, $kt_{80} = 0.37$ and unusually strong \aliii\ \lam 1870 like many other core ERQs and ERQ-like sources. 

More work is needed to quantify the line shifts across the ERQ sample. However, two important results are evident already. First, the incidence of large blueshifts is much higher in ERQs than normal quasars. For example, the fraction of Type 1 quasars with blueshifts $\ge$2500 \kms\ in our combined core ERQ and ERQ-like samples is $\sim$8\% based on 7 out of 88 with systemic redshifts available from well-measured \lya\ spikes or low-ionization emission lines. In contrast to this, \cite{Richards11} find only 21 (0.13\%) out of 15,779 quasars in a well-measured SDSS sample have \civ\ blueshifts in this range\footnote{These numbers from Richards et al. might be slightly underestimated because they are derived from plots showing blueshifts up to only 3000 \kms\ \citep[but see also][for more discussion about these shifts]{Denney16}.}. Interestingly, we find very large blueshifts only in ERQs with large FWHMs compared to the average shown in Figure~2. 

It is also very interesting that the ERQs with large blueshifts have large \civ\ REWs. They to do not fit the strong trend in \cite{Richards11} for quasars with large \civ\ blueshifts to have exclusively {\it small} REWs. For example, in the Richards et al. SDSS sample, all 21 quasars with v$(\civ ) \ge 2500$ \kms\ have REW(\civ ) $\le 25$ \AA\ and the 3 most extreme cases with blueshifts $>$2800 \kms\ have REW(\civ ) $< 12$ \AA\ \citep[see also][]{Espey89,Corbin90}. Other well-studied quasars with large \civ\ blueshifts, e.g., PHL 1811 and its high-redshift analogs, have exclusively weak \civ\ lines with REW(\civ ) $\la 10$ \AA\ \citep{Leighly07b, Wu10,Wu12,Luo15}. In contrast to this, all but one of the core ERQs or ERQ-like sources with blueshifts $>$2500 \kms\ has REW(\civ ) $>$ 80 \AA\ and three have REW(\civ ) $>$ 200 \AA . The lone exception is J114508+574258, which still has a moderate REW(\civ ) = 38 \AA\ in spite of its unprecedented large blueshift v(\civ ) = 8739 \kms . We conclude that something different or more extreme is controlling the line strengths in large-blueshift ERQs compared to other large-blueshift quasar samples (see also \S6.2 below).

\subsection{Sky Densities}

The core ERQ sample described in \S5.1, which is drawn from our $W3$-detected sample (\S2), contains 97 quasars in the redshift range 2.0--3.4. Expanded searches to find more quasars with similar unusual emission-line properties indicate that the total number of ERQ-like quasars in BOSS is a few hundred. The particular search described in \S5.7 yields a total of 332 core ERQs plus ERQ-like quasars. The densities of these sources on the sky are just their numbers divided by the total effective area of the BOSS spectroscopic survey, 9376 square degrees \citep{Alam15}. Thus we find observed sky densities of 0.010 deg$^{-2}$ for the core ERQs and 0.035 deg$^{-2}$ for the expanded sample that includes ERQ-like quasars. 

These results are similar to other luminous red quasar studies. For example, \cite{Banerji15} find 21 luminous highly reddened Type 1 quasars in a search area of 1115 deg$^{2}$, indicating a sky density of 0.02 deg$^{-2}$, while \cite{Assef15} find an estimated 186 quasar-powered HotDOGs at redshifts $2\la z_e\la 4$ in a WISE search area of 32000 deg$^2$, corresponding to a sky density of 0.0058 deg$^{-2}$. ERQs and HotDOGs should have similar selection functions and completeness fractions based on WISE $W3$ detections alone \citep[see also][]{Eisenhardt12}. However, the ERQ detections are additionally limited by the rest-frame UV flux and UV color selection constraints of BOSS \citep{Ross12}. A thorough discussion of the selection effects is beyond the scope of this paper. We note simply that, owing to the additional BOSS constraints, the true source density of core ERQs plus ERQ-like quasars at redshifts $2\la z_e\la 4$ should be at least similar to and probably greater than the source density of HotDOGs. 

The space density of obscured quasars relative to blue/normal quasars can help to constrain models that connect these populations by evolution or orientation effects. \cite{Assef15} and \cite{Banerji15} claim that the completeness-corrected space densities of HotDOGs and highly reddened Type 1 quasars, respectively, are both similar to blue/normal quasars at the same redshifts and luminosities. However, that result requires large correction factors and seems highly uncertain. Here we note simply that the raw numbers of ERQs are small compared to blue quasars at similar luminosities in our study. In particular, the numbers of core ERQ and total core ERQs plus ERQ-like quasars listed above are 1.6\% and 5.1\%, respectively, compared to the 6119 luminous blue quasars with redshifts and $W3$ magnitudes similar to the ERQs (\S4.1). If these ERQs are part of an evolution sequence or a duty cycle with luminous blue quasars (\S1, \S6.2), then these results suggest that the core ERQ/ERQ-like phase lasts crudely a few percent of quasar lifetimes. 

\section{Discussion}

Our analysis in \S4 and \S5 shows that high-redshift ERQs defined by \imw\ $\ge$ 4.6 often have a suite of peculiar emission-line properties, a high incidence of broad outflow absorption lines, and unusual SEDs that are not consistent with simple reddening\citep[see also][]{Polletta08}. This ensemble of exotic properties identifies the ERQs as a unique new quasar population with unique physical characteristics. The additional requirement for REW(\civ ) $>$ 100 \AA\ in our core ERQ sample (\S5.1) helps to weed out interlopers that are just normal quasars reddened by dust to focus the sample more narrowly on quasars in this exotic new population. We present analyses of the physical conditions in the line-forming regions of the core ERQs in Hamann et al. (2016a, in prep.). Below we discuss possible causes for their red colors (\S6.1) and physical models that might explain the overall ERQ phenomenon (\S6.2). 

\subsection{On the Origins of Red \imw\ Colors}

The median color of the core ERQs is almost 3 magnitudes redder than the median overall in the $W3$-detected quasar sample (\S5.1). This could be caused by UV obscuration, enhanced emission in the mid-IR, or a combination of these factors. Given the extreme and peculiar nature of ERQs, we consider a number of possibilities. 

\subsubsection{UV Suppression}

Red \imw\ colors can be caused by dust obscuration in the rest-frame UV. This interpretation is not straightforward for the core ERQs because their relatively flat UV spectra are inconsistent with simple reddening (\S5.5, Figures~11 and 16). Most of the core ERQs have Type 1 emission lines, so we do appear to have direct views of the central engines in the UV. However, the flat UV spectral slopes resemble the high-redshift candidate Type 2 quasars described by \cite{Alexandroff13}. These authors attribute the Type 2 SEDs to moderate dust reddenings with typically $E(B-V) \sim 0.5$ plus scattered UV light or patchy obscuration that allows some direct UV/blue quasar emission to be viewed relatively unreddened \citep[see also][]{Greene14}. Patchy obscuration was also invoked by \cite{Veilleux13a, Veilleux16} to explain the spectrum of the low-redshift BAL quasar Mrk 231. This quasar's SED is very red across the visible but remarkably flat at rest wavelengths $\la$2400 \AA . Patchy obscuration is strongly favored over scattered light in Mrk 231 because the flat UV spectrum has negligible polarization \citep{Smith95}. 

Patchy obscuration could explain the unusual UV to mid-IR SEDs of the core ERQs, although scattered light contributions in the UV cannot be ruled out with existing data (see Alexandroff et al., in prep.). One important implication of patchy obscuration is that the dust patches/clouds cannot be much larger than the UV emission source in the accretion disk, which is only $\sim$0.01 pc across in luminous quasars \citep[e.g.,][and 2016c, in prep.]{Hamann11}. 

Patchy obscuration seems much more plausible than the explanations favored for HotDOG SEDs. In particular, \cite{Assef15} and \cite{Toba16} attribute HotDOG SEDs to very large dust extinctions in front of the quasars, typically with $E(B-V)\sim 7.8$ to explain the very red colors across the rest-frame visible to mid-IR, plus hot stars from high star formation rates that create the flat UV spectral slopes \citep[but cf.][]{Assef16}. This picture seems unlikely for the core ERQs because their UV spectra are quasar-dominated down to at least the wavelengths where we measure very strong and broad emission lines of \ovi\ \lam 1034 and \nv\ \lam 1240 \citep[Figures 8 and 12, also][]{Ross15}. It is not clear this picture can even apply generally to HotDOGs because many of them also have Type 1 quasar-dominated spectra in the rest-frame UV \citep[\S5.6][]{Wu12,Toba16}. 

Another way to suppress the UV flux might be viewing angle effects in a flattened accretion disk geometry. This could produce red \imw\ colors if the mid-IR emission is roughly isotropic while the observed UV flux is diminished by a $\sim$$\cos\theta$ factor that derives from the projected area of the UV-emitting disk \citep[with negligible limb darkening effects,][and $\theta$ measured from the disk axis]{Nemmen10}. The problem with this picture is that $\sim$3$^m$ of UV suppression in the core ERQs would require viewing angles very close to edge-on, with disk axis angles $\theta\sim 86^o$ from the line of sight. This seems highly implausible given that a dusty torus/wind must be present to intercept UV light and reprocess it into the observed strong mid-IR emission. The expected covering fractions of this dusty material ($\sim$50\%, see \S6.1.2 below) correspond to angular elevations $\sim$30$^o$ above and below the disk plane (ignoring clumpiness, which would lead to even higher elevations). Lines of sight that skim the edge of this torus/wind material, with $\theta\sim 60^o$, would produce only a factor of $\sim$2 UV flux suppression from projection effects, not the requisite $\sim$3$^m$. We do note however that, if the dusty torus/wind is clumpy, it might be possible to have some nearly-equatorial sight lines pass through the torus without being blocked by dusty clumps. 

A more exotic way to suppress the UV flux without UV reddening might be by electron scattering in a thick layer of highly-ionized gas above the UV-emitting accretion disk. Large amounts of this material are expected to develop in BAL winds due to the intense ionizating radiation from the inner accretion disk \citep{Murray95}. This gas fails to accelerate and can hover above the disk because it is too ionized and too transparent for radiative driving \citep[see also][and refs. therein]{Proga07, Sim10}. This highly-ionized failed wind material is believed to cause the X-ray weakness of BAL quasars and, in some cases, it might be Compton thick in front of the X-ray source even though the quasars are bright and visible to us in the UV \citep{Gallagher07}. The core ERQs might be extreme cases where a failed wind is spatially extended to cover larger portions of the UV-emitting disk. The amount of UV flux suppression and the resulting observed colors would depend on the geometry and optical depths of the obscuring material above different portions of the disk. However, a potential serious problem is that strong UV continuum fluxes are still needed to power the observed strong emission lines (Hamann et al. 2016a, in prep.). The failed wind material would need to be elevated above the disk enough to give the emission-line regions a clear view of the UV continuum source while our view is still substantially attenuated.

 Another exotic possibility is structural disruptions in the inner accretion disk. This should be a transient phenomenon with $\sim$3$^m$ drops in the UV emission followed by similar drops in the broad emission-line fluxes after time delays of $<$1 yr corresponding to the light travel time out to the broad emission-line regions \citep{Kaspi07}. This scenario appears to be ruled out by 12 ERQs in our sample with repeat spectroscopic observations in BOSS, SDSS-I/II or the VLT obtained $>$0.5 yrs apart in the rest frame (including 9 with spectra $>$1 year apart, Hamann et al. 2016a, in prep.). We do not find any instances of dramatic variability in the line shapes or the line strengths relative to the continuum. 

\subsubsection{Mid-IR Flux Enhancements}

Red \imw\ colors might also be caused by enhanced mid-IR emission. Hot dust in the inner torus is believed to dominate the SEDs of luminous Type 1 quasars from $\sim$2 \mum\ to $>$10 \mum\ \citep{Efstathiou95, Netzer07a, Mor09, Mor11, Deo11}. It overwhelms contributions from starlight and possible PAH emission at these wavelengths. The amount of mid-IR flux relative to the UV depends on our viewing angle and the dust covering factor as seen from the central source. 

Hot dust covering fractions in quasars are believed to be in the range 30--70\% \citep{Gaskell07,Mor09,Netzer16} with a nominal value around 50\%  \cite{Lawrence10}. Larger covering factors might occur in extreme cases if the dusty ``torus" is a dusty wind \citep{Konigl94,Gallagher15,Netzer15} extending farther than normal vertically above the accretion disk. This would allow the dust to reprocess more UV luminosity into mid-IR emission \citep[also][]{Wang11,Wang13}. Similarly, high-speed BAL winds might ablate dusty clumps off of a traditional torus to increase the dust covering fractions and enhance the mid-IR flux \citep{Wang13, Wagner13}. However, this scenario cannot explain the red colors of ERQs because, even with dust covering factors increased to unity (which is not realistic), the maximum mid-IR enhancement would be only $\sim$0.8$^m$ instead of the requisite $\sim$3$^m$.

Dust heated by star formation is another unlikely contributor to the mid-IR flux because the radiation temperatures needed for emission in $W3$ (rest-frame $\sim$3.4 \mum ) are very high, nominally $\sim$850 K. This is a natural temperature for dust near quasars \citep{Rowan-Robinson95,Efstathiou95, Nenkova08, Mor09}, but in star forming galaxies the dust temperatures are observed to be $\la$60 K with the dust emissions peaking at wavelengths $\ga$50 \mum\ \citep{Kirkpatrick12, Magnelli12,Melbourne12,Chapman05}. If dusty starbursts were somehow important to the mid-IR fluxes of ERQs, they should produce very red colors across the mid-IR (e.g., in \wtmwf ), which are not observed (Figure~10).  

Another possibility for mid-IR enhancements is strong PAH emissions in the bands at 3.2, 6.2 and 7.7 \mum . However, this can also be ruled out because i) PAHs are nominally destroyed by the hard UV radiation in quasar environments, and ii) the 3.2 \mum\ feature is weak compared to 6.2 and especially 7.7 \mum\ such that a PAH-dominated spectrum would produce red \wtmwf\ colors at the ERQ redshifts, which are not observed \citep[Figure 10, see][and refs. therein]{Sales10,Draine11}. 

\subsection{Toward a Physical Model}

We conclude from \S6.1 that the red \imw\ colors of the core ERQs are caused by dust obscuration, probably by a patchy medium that suppresses the observed UV fluxes without substantial UV reddening. A critical point for models of the core ERQs is that this obscuration is closely related to their peculiar line properties. Is this relationship defined by geometry and orientation effects or by unique physical conditions that are perhaps tied to a particular phase of quasar evolution? 

A good starting point is to ask where the obscuring dust is located. \cite{Banerji12,Banerji13} argue that the obscuration in highly reddened Type 1 quasars (HR1s) occurs on galactic scales. Recall that these quasars have SEDs consistent with a simple dust reddening screen (\S5.5, Figure~16). This is not the case for the core ERQs. If patchy obscuration is involved, then dust extinction on galactic scales might be ruled out by the requirement for dust patch/cloud sizes $\la$0.01 pc across (\S6.1.1), because clouds with these sizes are not expected in a galactic interstellar medium. Another problem is the close relationship between red \imw\ colors and the specific emission-line properties of ERQs. This seems to favor obscuration on small scales where it can be readily coupled to the orientation or physical conditions in quasar environments. 

It might be possible to couple the line properties of the core ERQs to obscuration on galactic scales if the quasars drive high-speed outflows that ablate and disperse dusty molecular clouds in the host galaxies. This is expected to occur when quasars provide ``feedback'' to their galactic environments \citep{Hopkins10, Faucher12, Wagner13}. It could, in principle, connect the physical conditions on small scales in quasar line-forming regions to the amounts of obscuration occurring on large scales in the host galaxies. Cloud shredding might also produce small dusty clumps capable of patchy obscuration far from the quasars. 

Alternatively, the obscuration might occur on small scales in a dusty torus/wind just outside the traditional BLR. Recent studies suggest that this material is clumpy \citep[e.g.,][]{Hoenig07, Nenkova08, Thompson09}, so it might also produce the purported patchy obscuration in the core ERQs. In clumpy torus models that posit self-gravitating clumps stable against tidal forces from the central black hole \citep{Hoenig07}, the predicted clump sizes satisfy a prerequisite for patchy obscuration in that they are smaller than the UV continuum source. For example, equation 5 in \cite{Hoenig07} predicts that dusty clumps 1 pc away from a black hole of mass $10^9$ M$_{\odot}$ should have maximum diameters of $\sim$0.001 pc.

One consequence of small-scale obscuration by a dusty torus/wind could be strong orientation effects in an axisymmetric geometry. However, it seems very difficult to explain the unique emission-line properties of the core ERQs if there are only orientation effects, e.g., if the core ERQs are like other quasars except for a particular viewing perspective that intersects $\sim$3$^m$ of dusty torus/wind material, Most problematic for an orientation-only model is the very broad and blueshifted [\oiii ] \lam 5007 lines that identify powerful outflows on galactic ($\sim$1 kpc) scales in the four core ERQs tested so far \citep[][Hamann et al. 2016a, in prep.]{Zakamska16}. It does not seem feasible to hide these features by orientation effects in a majority of quasars while revealing them only for specific viewing angles in ERQs. Given that the core ERQs are only about 2\% of the similarly-luminous quasar population in our $W3$-detected sample (\S5.9), the range of viewing angles that produces core ERQ properties would have to be only $\sim$2\%. Other aspects of the core ERQ emission lines are also not readily explained by orientation effects, including the peculiar flux ratios such as large \nv /\civ . We conclude that while orientation might play an important role in the ERQ phenomenon, their unique line properties overall appear to require unique physical conditions.

These physical conditions might be unusually powerful outflows that encompass large portions of line forming regions as well as the dusty ``torus.'' The [\oiii ] data mentioned above for a small subset of the core ERQs, along with high fractions of BALs and large \civ\ emission-line blueshifts (\S5.3 and \S5.8), indicate that outflows are pervasive in ERQs across a wide range of spatial scales. \civ\ blueshifts are often interpreted in terms of a two-component broad line region (BLR), with one component near the disk plane dominated by virial motions and another that is outflowing and vertically extended above the disk \citep[][and refs. therein]{Gaskell82, Collin88, Marziani96, Leighly04b, Leighly07b, Richards11}. Higher ionization lines like \civ\ and \nv\ are more likely to participate in the outflow while lower ions favor the denser and more radiatively shielded disk component. The outflow lines are blueshifted because the accretion disk bisects the flow and obscures receding material from our view. The ERQs might be extreme examples of outflow-dominated BLRs. 

The puzzle for ERQs in this BLR outflow picture is that they have dramatically larger REWs than other quasars with large \civ\ blueshifts (\S5.8). Well-studied quasars with large blueshifts \citep[e.g., PHL 1811 and its analogs,][]{Leighly04a,Leighly04b,Leighly07b,Wu11j,Wu12j,Luo15} are also X-ray weak. This is consistent with large-blueshift quasars having soft far-UV spectra favorable for radiative acceleration. The reasoning here is that soft far-UV spectra lead to moderate degrees of ionization in the outflow gas where ions like \civ\ and \ovi\ are then available for line driving in the near-UV \citep[see][and the Leighly et al. papers cited above]{Murray95}. Small \civ\ REWs are a natural consequence of this outflow picture because BLRs photoionized by a soft spectrum should produce less \civ\ emission relative to the near-UV continuum \citep[also][]{Korista97}. In contrast to this, all but one of the ERQs and ERQ-like quasars with blueshifts $>$2500 \kms\ have REW(\civ ) $\ge$ 87 \AA\ up to 281 \AA\ (in J101326+611219, Figure~18). 

ERQs might have unprecedented large REWs in this outflow picture if their BLRs are more vertically extended above the accretion disk than other quasars. This would lead to the line-forming regions intercepting more of the quasar continuum luminosity for reprocessing into line radiation (Hamann et al. 2016a, in prep.). Extended BLRs in an outflow might also blend smoothly with the high-speed low-density [\oiii ] gas much farther out. The idea of BLR outflows connecting to large-scale [OIII] outflows would be consistent with studies of less extreme lower-redshift quasars that show \civ\ blueshifts correlated with the blueshifts and outflow kinematics measured in [\oiii ] \citep{Zamanov02,Aoki05}. In the core ERQs, where the \civ\ and [\oiii ] lines can have similar widths, the distinction between ``broad'' and ``narrow'' line regions is particularly ambiguous \citep[\S5.4][Hamann et al. 2016a, in prep.]{Zakamska16}. 

Dusty winds \citep[e.g.,][]{Keating12,Gallagher15} in the ERQs might also be more vertically extended and participating in the same general outflow as the ionized gas. If the inner edge of the dusty wind is at the dust sublimation radius overlapping with the outer BLR \citep[as expected,][]{Gaskell09, Mor11, Goad12}, then portions of the BLR outflow would be dusty \citep[also][]{Wang13}. This outflow should be clumpy based on the evidence for a clumpy torus \citep{Nenkova08, Thompson09} and clumpy BLRs and BAL outflows \citep[e.g.,][]{deKool97, Arav97, Hamann13}. It could therefore produce patchy obscuration across the UV continuum source like we infer for the ERQs (\S6.1). Extended dusty outflows could also distribute small dusty clumps across larger fractions of sky as seen from the central quasar, thus avoiding the problem mentioned above that small-scale obscuration in a torus/wind would produce strong orientation effects. 

If the obscuration in the core ERQs does occur in dusty outflows, then all core ERQs should have considerable outflow gas accompanying the dust along our lines of sight and we might wonder why they do not all have strong outflow absorption lines in their spectra. The reason could be that the dusty clumps are opaque to UV radiation, so they suppress the observed UV flux without producing UV absorption lines \citep{Veilleux13a,Veilleux16}. BALs and BAL-like outflow features would form mainly on the periphery of these clumps, or in the spaces between them, where there is less dust and we do still see the UV continuum source. This could lead to a situation where BALs and BAL-like features are usually weak and sometimes absent from the observed spectra of ERQs even though powerful dusty outflows are present along our lines of sight. 

If outflows are the key ingredient to understanding the ERQ phenomenon, then we must explain why the outflows are more powerful or more spatially extended in ERQs compared to other quasars. Outflows driven by magneto-centrifugal forces \citep{Proga03b,Everett05,Fukumura10,Keating12} might be enhanced in ERQs if the quasars have unusually strong magnetic fields threaded vertically through their accretion disks. Flows driven by radiation pressure might be enhanced by higher accretion rates (relative to Eddington), higher metallicities (that can increase the opacities for radiative acceleration), or softer far-UV spectra (that favor line driving in the near-UV without over-ionization). There is some evidence for high metallicities in the ERQs based on the strong \nv\ \lam 1240 emission lines and large \siiv\ \lam 1400/\civ\ flux ratios \citep[][Hamann et al. 2016a, in prep.]{Polletta08}. Other studies have reported an observational link between large \civ\ blueshifts and high accretion rates \citep{Baskin05, Wang11, Wang13, Marziani12,Luo15}, which would be consistent with high accretion rates occurring in the ERQs.

The last critical question is whether the core ERQs are tied to a particular stage of quasar-galaxy evolution (\S1). Observations of ERQ host galaxies and extended environments are needed to address this question. The outflow scenario that we favor above does not directly connect the quasars to galaxy evolution because it emphasizes small-scale phenomena controlled by accretion physics. However, that connection could be established by ERQ outflows shredding dusty clouds in the galaxies to create patchy obscuration on galactic scales (\S6.1.1). The relationship of ERQs to galaxy evolution might also be more holistic, e.g., powerful outflows and high accretion rates are expected to occur generally in obscured quasars during the aftermath of a triggering event that funnels matter toward the central black hole in young gas-and-dust-rich galaxies \citep{Sanders88,Hopkins05, Hopkins08,Veilleux09b,Rupke11, Rupke13, Liu13}. ERQs might be interpreted within this paradigm like other red quasar populations -- caught in the transition between the initial triggering event and a more quiescent phase of galaxies hosting normal blue (unobscured) quasars (see refs. in \S1). If we want to place ERQs in a simple monotonic evolution sequence like this with HotDOGs \citep[see][and refs. therein]{Wu12,Fan16}, then the lesser obscurations in ERQs (Figure~16) suggest that they are in a slightly more advanced stage than HotDOGs, farther in time from the triggering event and closer to the blue quasar phase. We could also infer from the numbers of core ERQs and ERQ-like quasars compared to luminous blue quasars (\S5.9) that the lifetime of the core ERQ/ERQ-like phase is a few percent of total quasar lifetimes.  

\section{Summary}

This study follows up on the discovery by \cite{Ross15} of a population high-redshift extremely red quasars (ERQs) with rest-frame UV to mid-IR colors similar to dust obscured galaxies (DOGs) in the SDSS-III/BOSS and WISE surveys. We show that ERQs often have unusual SEDs that are surprisingly flat across the rest-frame UV given their red UV to mid-IR colors (Figures 11 and 16) as well as peculiar UV emission-line properties that include very large REWs, unusual wingless line profiles, and in many cases exotic line flux ratios such as \nv\ $>$ \lya , large \nv /\civ , and large \siiv /\civ\ (Figures 7, 8, and 12, also \citealt{Polletta08}). This ensemble of peculiar properties is fundamental to the nature of ERQs. 

We present a catalog of new measurements of the UV continua and \civ\ and \nv\ emission lines for 216,188 BOSS quasars in the redshift range $1.53 < z_e < 5.0$ (\S3 and Appendix A). We then focus on a subsample of 173,636 quasars at redshifts $2.0 < z_e < 3.4$ to i) characterize the exotic line properties of ERQs compared to the overall BOSS quasar population, ii) understand the relationships of these line properties to red quasar colors, and iii) revise the selection criteria for ERQs to find more of them with similar properties in the BOSS database. Our main results are the following:

{\it{\bf 1)} No Baldwin Effect:} The peculiar emission-line properties of ERQs have no relationship to luminosity in the Baldwin Effect. In fact, the ERQs behave opposite to this trend by favoring extremely large emission-line REWs (and smaller FWHMs) at high quasar luminosities (\S4.1).

{\it{\bf 2)} \imw\ Color Dependence:} The peculiar line properties of ERQs depend strongly on \imw\ color at the redshifts of our study (Figures 7--11). This dependence is stronger than \imwf\ and \rmwf\ due to scatter in \wtmwf\ that has no relationship to the ERQ phenomenon or to the UV line properties of quasars generally (\S4.4). ERQ line properties start to appear in a majority of quasars across a surprisingly sharp color boundary \imw\ $\ge 4.6 \pm 0.2$ (\S4.5).  In addition, quasars selected to have these peculiar line properties regardless of color are found to be red, with median \imw\ more than 2 magnitudes redder than $W3$-detected quasars overall (\S5.7). 

{\it{\bf 3)} Core ERQs:} The peculiar line properties and SEDs of ERQs are much more common in sources with large REW(\civ ), while ERQs with weak/normal line strengths are more often just normal blue quasars behind a dust reddening screen (\S4.2, \S4.5, Figures 7 and 11). Thus we combine our ERQ color selection, \imw\ $\ge 4.6$, with the constraint REW(\civ ) $\ge 100$ \AA\ to define a more homogeneous ``core" sample of 97 ERQs with median redshift $\left<z_e\right> = 2.50\pm 0.27$ and median luminosity $\left<\log L ({\rm ergs/s})\right> \sim 47.1\pm 0.3$ (\S5.1, Figures 8--12). 76\% of these core ERQs are Type 1 quasars based on FWHM(\civ ) $>$ 2000 \kms\ or BALs identified by visual inspection in DR12Q. However, the median line width of these Type 1 core ERQs, $\left<{\rm FWHM(\civ )}\right> \sim 3050\pm 990$ \kms , is only half that of blue quasars with similar luminosities (\S5.4). 

{\it{\bf 4)} Inconsistent with Simple Reddening:} The core ERQs have surprisingly flat rest-frame UV spectra given their extreme red \imw\ colors. This spectral shape and the peculiar emission-line properties are not consistent with normal quasars behind a simple dust reddening screen (\S5.5, Figures 15 and 16). 

{\it{\bf 5)} Radio Properties:} The fraction of radio-loud sources in the core ERQ sample, 6.7\%, is similar to other luminous quasar populations at these redshifts (\S5.2). 

{\it{\bf 6)} Sky Densities:} The 97 core ERQs in our study have an observed sky density of 0.010 deg$^{-2}$ (\S5.9). An expanded search for more BOSS quasars with ERQ-like emission-line properties (\S5.7) increases the sample to 332 quasars with sky density 0.035 deg$^{-2}$. 

{\it{\bf 7)} Outflow Signatures:} Outflow featrues are common in the core ERQs and the expanded sample of red ERQ-like quasars. In particular, they have i) high BAL fractions of 30--68\% compared to 14\% in our entire $W3$-detected sample (\S5.3, \S5.7), ii) a high incidence of large \civ\ emission-line blueshifts, e.g., with shifts $>$2500 \kms\ roughly fifty times more common in ERQs than normal quasars and reaching 8739 \kms\ in one quasar (\S5.8, Figure~18), and iii) high-velocity [\oiii ] \lam 5007 lines with FWHMs and blueshifts up to $\sim$5000 \kms\ that identify powerful outflows on galactic scales $\ga$1 kpc \citep[in 4 out of 4 core ERQs measured so far,][Hamann et al. 2016a, in prep.]{Zakamska16}. 

{\it{\bf 8)} A Unique Red Quasar Population:} The core ERQs have luminosities and sky densities similar to other high-redshift obscured quasar samples (\S5.8, \S5.9). They are less red than HotDOGs and the highly reddened Type 1 quasars found in other studies (Figures 16 and 17). However, the specific exotic characteristics of the core ERQs, involving both their line properties and SEDs, identify a unique new red quasar population. 

{\it{\bf 9)} Physical Models:} We argue that the red colors of ERQs are caused by dust obscuration, and that the exotic line properties in the core ERQs and other ERQ-like quasars are related to unusually powerful/extended outflows that encompass most of the line-forming regions and, perhaps, the dusty torus (\S6.1, \S6.2). The very large emission-line REWs, which are unprecedented in luminous quasars, might arise from outflow-dominated broad line regions that are vertically extended above the accretion disk. Patchy obscuration by small dusty clouds in the outflows (or in material ablated from dusty structures by these outflows) could produce the typical observed $\sim$3 magnitudes of UV extinction without substantial UV reddening and without strong BALs or other outflow absorption lines appearing in every spectrum.

\section*{Acknowledgements}

We are grateful to A. Baskin, D. M. Crenshaw, G. J. Ferland, S. B. Kraemer, and A. Laor for helpful conversations. FH acknowledges support from the USA National Science Foundation grant AST-1009628. KDD is supported by an NSF AAPF fellowship awarded under NSF grant AST-1302093. Funding for SDSS-III was provided by the Alfred P. Sloan Foundation, the Participating Institutions, the National Science Foundation, and the U.S. Department of Energy Office of Science. The SDSS-III Web site is http://www.sdss3.org/. SDSS-III is managed by the Astrophysical Research Consortium for the Participating Institutions of the SDSS-III Collaboration, including the University of Arizona, the Brazilian Participation Group, Brookhaven National Laboratory, University of Cambridge, Carnegie Mellon University, University of Florida, the French Participation Group, the German Participation Group, Harvard University, the Instituto de Astrofisica de Canarias, the Michigan State/Notre Dame/JINA Participation Group, Johns Hopkins University, Lawrence Berkeley National Laboratory, Max Planck Institute for Astrophysics, Max Planck Institute for Extraterrestrial Physics, New Mexico State University, New York University, Ohio State University, Pennsylvania State University, University of Portsmouth, Princeton University, the Spanish Participation Group, University of Tokyo, University of Utah, Vanderbilt University, University of Virginia, University of Washington, and Yale University.




\bibliographystyle{mnras}
\bibliography{bibliography}



\appendix

\section{Line \& Continuum Fits}

We fit the UV continuum and the \civ\ and \nv\ emission lines in the BOSS spectra of DR12 quasars at redshifts $1.53 < z_e < 5.0$. The redshift range is selected to ensure that the spectra cover the continuum on both sides of the \civ\ emission line. The general fitting procedure is outlined in \S3. The results are available in a FITS table described at the end of this appendix. All of the wavelengths mentioned here and listed in the catalog are in the quasar rest frames relative the best available DR12Q redshift (\S3). 

We begin with 226,984 quasars listed in DR12Q with signal-to-noise ratios per pixel $>$ 0.3 at 1700 \AA\ in the BOSS spectra. The signal-to-noise threshold excludes only a few quasars with very bad data. From this we reject 9963 more quasars that are reported in DR12Q to have strong \civ\ BALs at velocities v~$ > 15,000$ \kms\ (where they can interfere with a critical wavelength window used to define the continuum) or at v~$ < 2000$ \kms\ (where they can distort the emission-line profiles). These exclusion criteria are based on trial and error tests with different values of the balnicity index, the absorption index, and the BAL velocity limits listed in DR12Q. We reject fairly generously to have a problem-free catalog.

We then fit the remaining quasars using an iterative procedure as follows. First the fitting code searches the spectrum for narrow spikes above or below the spectra caused by cosmic rays or noise anomalies. These spikes are removed by interpolation from adjacent pixels. Then it fits a power law to the continuum locally beneath the \civ\ emission line. This fit is based on the median fluxes in several wavelength intervals between 1420 \AA\ and 1810 \AA\ that are free of emission lines and of broad absorption features (according to the data listed in DR12Q). Quasars are rejected if the fitted continuum is too steep to be realistic  (indicating a problem with the data), strong broad absorption is detected at the \civ\ emission line wavelengths, or there is no significant emission above the fitted continuum leading to failure in the fits. This rejects another 833 quasars. 

To fit the \civ\ emission lines, we fix the continuum calculated above, ignore the \civ\ doublet separation, and use either one or two Gaussian components depending on the results. We experimented with more Gaussian components and with Gauss-Hermite polynomials \citep[e.g.,][and refs. therein]{Denney16}, but they yield poor results in many cases because the data are not of sufficient quality to constrain the additional free parameters. We found that an iterative procedure that begins with a single Gaussian and adds a second component as needed provides the most robust and reliable fits across the full range of line properties and SNRs in the BOSS database. We also use the iterations to identify and mask from subsequent fits wavelength regions that have significant broad absorption lines.

The initial fit to \civ\ using a single Gaussian provides parameter guesses for subsequent iterations. It is sometimes adopted as the final best result in rare quasars with very noisy spectra or lines with little contrast above the continuum. Subsequent fits involving two Gaussian components introduce degrees of freedom (such as the relative widths and wavelength offsets of the Gaussian components) one at a time to obtain useful results even in bad data. The two Gaussian components are constrained in all cases so that one represents the peak/core of the line while the second broader component accounts for the wings and/or asymmetries in the base of the profiles. Specifically, the wing component must be broader than the core up to a maximum of several times broader (following a formulaic prescription based on visual inspections of many results), and the central wavelength of the core component cannot be outside of the FWHM range of the broader wing component. These constraints slightly limit the range of profiles that can be fitted accurately but they are necessary to avoid bad results in poor data.

Visual inspections of several thousand spectra indicate that the continuum and \civ\ line fits are generally excellent. We specifically examined results at the extremes of broad/narrow FWHMs, large/small REWs, strong BALs or other broad absorption that might overlap with the \civ\ emission-line profiles, and spectra with low SNRs in the continuum. The most common problem is fits that underestimate the line peak height and thus overestimate the FWHMs for observed lines that have a strong narrow core on top of much broader wings. However, this situation is extremely rare. 

Successful fits to the continuum and \civ\ line are obtained for 216,188 quasars in the redshift range $1.53 < z_e < 5.0$. The fitting software also measures the \civ\ REWs by direct integration above the fitted continuum and FWHMs by stepping across the line profiles in moderately smoothed spectra. These measurements interpolate across wavelengths identified by the fitting procedure to have significant absorption. The resulting REWs and FWHMs are generally in excellent agreement with values derived from the Gaussian fits. Occasional large disagreements can identify poor measurements or data that should be checked. We also include a quality flag, {\tt qflag}, in the catalog to identify possible problems. 

The continuum beneath the \nv\ line is fit by a second power law constrained initially by the same flux points used for the \civ\ continuum plus several shorter-wavelength intervals between 1134 and 1365 \AA . These fits are checked and flux points are discarded if significant absorption is detected (e.g., in the \lya\ forest) or if the flux points depart significantly from the fit defined by the other wavelengths. 

The final best fit to the \civ\ line profile is then used as a template that we shift and scale to measure the \nv\ emission line above this new \nv\ continuum. The profile width at \nv\ is increased to account for the different doublet separations between \civ\ (498 \kms ) and \nv\ (964 \kms ). The template scaling near \nv\ does not include contributions from the nearby \lya\ emission line. Instead, it avoids \lya\ by using the observed flux in a narrow wavelength window that encompasses the \nv\ line peak and most of the red side profile. This shift and scale technique also assumes implicitly that the \nv\ and \civ\ emitting regions have the same kinematics, which is reasonable given their similar ionizations and similar atomic physics in the lithium iso-electronic sequence. It is also supported by the generally excellent results revealed by inspection of the fits. Data entries of `$-$1' listed for \nv\ in the catalog indicate that a fit was not achieved or \nv\ is not within the spectral coverage. A quality flag, {\tt nvflag}, is also provided to help identify potential problems with the line and continuum fits at \nv . 

Finally, we fit a power law to the continuum across the entire range $\sim$1350 \AA\ to $\sim$2200 \AA , constrained by median fluxes in the wavelength intervals described above. At redshifts below $z_e\approx 1.86$, the shortest wavelength used to constrain this power law is 1420 \AA . Values of $\alpha_{\lambda}$ are not provided for quasars at $z_e > 3.4$ because the long-wavelength flux point is not available (or not reliable) in the BOSS spectra. This fit is performed and results are tabulated for the original BOSS spectra and for flux corrected versions using the prescription based on airmass in \cite{Harris15}. This continuum is not used for the line fitting.

The final catalog of 216,188 quasars is available as a binary FITS table from the University of California Digital Libraries with the digital object identifier DOI:10.6086/D1H59V at the URL https://dx.doi.org/10.6086/D1H59V. This archive also includes FITS tables that list data separately for the core ERQs and the ERQ-like quasar samples, as well as pdf plots that expand upon Figures 12 and 18 to show BOSS spectra for all quasars in these samples. All of the cataloged line data are measured from the Gaussian fits, except for {\tt rew\_di}, {\tt rewe\_di}, and {\tt fwhm\_di}, which are direct measurements of the observed lines above the fitted continuum. The 1 sigma uncertainties quoted for the fit parameters are based on pixel-to-pixel noise in the spectra returned by the fitting software. The uncertainties in the line parameters do not include uncertainties in the continuum fits. The catalog contents are summarized below:
\smallskip

\noindent {\tt SDSS\_Name} = from DR12Q.

\noindent{\tt Plate}, {\tt MJD}, {\tt FiberID}  = from DR12Q.

\noindent{\tt ThingID} = from DR12Q.

\noindent {\tt z\_dr12} = best available DR12Q redshift (PCA, or visual inspection of PCA is not available).

\noindent {\tt i} = $i$ magnitude from DR12Q corrected for Galactic extinction using offsets from \cite{Schlafly11}

\noindent {\tt i\_err} = error in above as listed in DR12Q.

\noindent {\tt W3} = $W3$ magnitude from DR12Q converted to AB using $W3$(AB)~--~$W3$(Vega) = 5.24.

\noindent {\tt W3\_snr} = signal-to-noise ratio in above as listed in DR12Q.

\noindent {\tt cc\_flags} = WISE contamination and confusion flag as listed in DR12Q.

\noindent {\tt bal\_flag\_vi} = BAL visual inspection flag from DR12Q (1 indicates that a BAL is present).

\noindent {\tt f1450} = flux in the uncorrected BOSS spectrum at 1450 \AA\ rest (10$^{-17}$ ergs s$^{-1}$ cm$^{-2}$ \AA$^{-1}$) used to anchor the power law continuum fits beneath \civ\ and \nv , e.g., $f_{\lambda} = f_{1450}\, (\lambda /1450{\rm \AA})^{\alpha}$.

\noindent {\tt alpha\_civ} = power law continuum slope ($f_{\lambda}\propto \lambda^{\alpha}$) measured from the BOSS spectrum on either side of \civ .

\noindent {\tt alpha\_nv} = as above but for \nv .

\noindent {\tt alpha\_all} = power law continuum slope ($f_{\lambda}\propto \lambda^{\alpha}$) between 1350 \AA\ and 2200 \AA\ in the uncorrected BOSS spectrum. 

\noindent {\tt alphae\_all} = 1 sigma uncertainty in above.

\noindent {\tt alpha\_allc} = power law continuum slope ($f_{\lambda}\propto \lambda^{\alpha}$) between 1350 \AA\ and 2200 \AA\ in the flux corrected BOSS spectrum, listed only for quasars with $z_e \le 3.4$.

\noindent {\tt alphae\_allc} = 1 sigma uncertainty in above.

\noindent {\tt rew} = \civ\ REW (\AA ) from the line profile fit.

\noindent {\tt rewe} = 1 sigma uncertainty in above.

\noindent {\tt rewc} = \civ\ REW (\AA ) for the core Gaussian component.

\noindent {\tt reww} = \civ\ REW (\AA ) for the wing Gaussian component.

\noindent {\tt fhwm} = \civ\ FWHM (km/s) from the line profile fit.

\noindent {\tt fwhme} = 1 sigma uncertainty in above.

\noindent {\tt fwhmc} = \civ\ FWHM (km/s) for the core Gaussian only.

\noindent {\tt fwhmw} = \civ\ FWHM (km/s) for the wing Gaussian only.

\noindent {\tt sigma} = \civ\ velocity dispersion (km/s) measured from the profile fit \citep{Peterson04}.

\noindent {\tt peak} = peak height of the \civ\ line profile fit relative to the fitted continuum at the \civ\ wavelength.

\noindent {\tt peaksnr} = SNR in above.

\noindent {\tt rat} = ratio of peak heights of the core/wing Gaussian components.

\noindent {\tt wcent} = wavelength of \civ\ line profile centroid.

\noindent {\tt wciv0} = wavelength of the \civ\ profile measured as the midpoint of the fitted profile at half the profile peak height.

\noindent {\tt wcore} = central wavelength of core Gaussian in \civ\ fit.

\noindent {\tt wcoree} = 1 sigma uncertainty in above.

\noindent {\tt shift} = wavelength shift (\AA ) of the wing Gaussian center from the core Gaussian center.

\noindent {\tt asy} = \civ\ profile asymmetry index \citep{Marziani96}.

\noindent {\tt kt75} = \civ\ profile kurtosis 75 index \citep{Marziani96}.

\noindent {\tt kt80} = \civ\ profile kurtosis 80 index \citep[$kt_{80}$, modified from][see \S3]{Marziani96}.

\noindent {\tt rew\_di} = \civ\ REW (\AA ) measured by direct integration of the data above the fitted continuum.

\noindent {\tt rewe\_di} = 1 sigma uncertainty in above.

\noindent {\tt fwhm\_di} = \civ\ FWHM (km/s) measured by stepping across the line in a smoothed spectrum.

\noindent {\tt qflag} = quality flag for the \civ\ fit: 0 = no problems, 1~= line peak height $<$ 80\% of the SNR per pixel at the \civ\ wavelength OR derived FHWM(\civ ) $<$ 400 \kms , 2~= failure in an intermediate step of the profile fitting that might indicate bad data or strong absorption on top of the \civ\ emission line OR a median flux point that constrains the power law continuum has SNR $<$ 0.5 OR derived continuum slope is very steep with $\vert\alpha_{\lambda}\vert > 9$ suggesting a problem in the BOSS data or contamination by a BAL, 3~= significant mismatch between the data and the continuum fit on the red side of the \civ\ emission line usually indicating a BAL or a data anomaly corrupting the continuum fit (e.g., a difference of $>$20\% measured at SNR $>$ 4 significance in the median fluxes in select wavelength intervals 35 to 60 \AA\ wide between 1620 and 1730 \AA ). 

\noindent {\tt rew\_nv} = \nv\ REW (\AA ) from the scaled \civ\ template fit.

\noindent {\tt frat\_nv\_civ} = line flux ratio \nv /\civ .

\noindent {\tt nvflag} = quality flag for the \nv\ fit: 0 = no problems, 1 = failed or unreliable fit to the line or the local continuum.

\section{ERQ-like Quasars}

The table below lists 235 quasars with ERQ-like emission-line properties from the expanded search described in \S5.7. Some of these quasars have ERQ colors, e.g., \imw\ $>$ 4.6, but none are in the core ERQ sample because they do not satisfy both \imw\ $>$ 4.6 and REW(\civ ) $>$ 100 \AA .  
\begin{table*}
\begin{center}
\begin{minipage}{176mm}
\caption{ERQ-like quasars. See \S5.1 and Table 2 for descriptions of the table contents.}
  \begin{tabular}{@{}lccccccccccc@{}}
  \hline
 Quasar Name& $z_{e}$& $i$ & \imw & REW & FWHM& $kt_{80}$& \nv /\civ& BAL & $\alpha_{\lambda}$& $E(B$$-$$V)$& FIRST\\
 & & (mag) & (mag) & (\AA ) & (km/s)& & & & & & (mJy)\\
\hline
J000315.85+061331.5&  2.25&  20.6&   4.0&   59$\pm$2&   3978$\pm$109&   0.35&   1.98&   1&   $-$0.02&   0.19$\pm$0.01&  0.0\cr
J001120.22+260109.2&  2.29&  20.2&   4.9&   98$\pm$2&   3804$\pm$56&   0.34&   2.26&   1&   $-$0.72&   0.21$\pm$0.01&  ---\cr
J001809.42+042527.7&  2.55&  20.4&   3.3&   48$\pm$3&   4095$\pm$185&   0.35&   2.47&   1&   -1.11&   0.06$\pm$0.01&  0.0\cr
J001911.87+263515.3&  2.24&  20.9&   ---&   123$\pm$7&   1654$\pm$92&   0.33&   0.86&   1&   $-$2.02&   ---&  ---\cr
J002044.88+241024.9&  2.58&  20.9&   4.4&   52$\pm$3&   4289$\pm$195&   0.37&   2.22&   1&   -0.45&   0.09$\pm$0.02&  -1.0\cr
J003747.98+170504.0&  3.66&  20.9&   4.5&   122$\pm$4&   5609$\pm$164&   0.33&   2.38&   0&   ---&   ---&  ---\cr
J004015.89+263902.3&  2.02&  21.1&   ---&   72$\pm$3&   4317$\pm$136&   0.36&   1.72&   1&   $-$0.47&   ---&  ---\cr
J004106.71+035201.4&  3.19&  21.7&   ---&   147$\pm$9&   4210$\pm$181&   0.37&   0.86&   0&   0.62&   ---&  0.0\cr
J004536.47-000018.1&  3.17&  21.1&   ---&   58$\pm$4&   5297$\pm$263&   0.36&   2.83&   1&   -0.21&   0.08$\pm$0.02&  0.0\cr
J005503.19+025135.1&  3.55&  20.5&   ---&   73$\pm$2&   2731$\pm$66&   0.34&   1.42&   1&   ---&   ---&  0.0\cr
J010129.16+331349.8&  2.63&  20.7&   4.0&   50$\pm$2&   2769$\pm$122&   0.34&   2.64&   1&   $-$1.87&   0.12$\pm$0.01&  ---\cr
J010312.59+252657.8&  2.11&  20.7&   4.1&   51$\pm$2&   2716$\pm$105&   0.35&   2.40&   1&   $-$1.88&   0.29$\pm$0.01&  ---\cr
J010725.36$-$003602.1&  2.16&  20.0&   4.1&   94$\pm$2&   4396$\pm$104&   0.34&   2.16&   1&   $-$2.09&   0.15$\pm$0.01&  0.0\cr
J010928.11+045759.5&  2.19&  21.3&   ---&   98$\pm$6&   4721$\pm$227&   0.35&   1.50&   0&   $-$1.96&   0.11$\pm$0.02&  0.0\cr
J011557.25-015842.6&  2.21&  20.7&   4.2&   63$\pm$2&   5637$\pm$140&   0.35&   2.25&   1&   -0.58&   0.19$\pm$0.01&  0.0\cr
J011601.43$-$050503.9&  3.18&  21.8&   6.2&   94$\pm$10&   2291$\pm$253&   0.19&   1.56&   0&   $-$2.01&   0.20$\pm$0.02&  0.0\cr
J012552.08$-$015218.3&  3.18&  22.2&   ---&   127$\pm$6&   1720$\pm$68&   0.35&   0.74&   0&   $-$0.63&   ---&  0.0\cr
J013343.24+153015.7&  2.54&  20.9&   4.0&   65$\pm$3&   4418$\pm$158&   0.35&   2.08&   1&   -0.02&   0.16$\pm$0.01&  -1.0\cr
J013413.22-023409.7&  2.39&  19.4&   3.3&   57$\pm$2&   4534$\pm$114&   0.35&   2.77&   1&   -1.25&   0.07$\pm$0.01&  0.0\cr
J013459.91+152027.8&  2.06&  22.2&   ---&   150$\pm$6&   3097$\pm$115&   0.33&   1.26&   1&   $-$1.49&   ---&  ---\cr
J014145.76$-$010135.4&  2.82&  21.3&   ---&   51$\pm$4&   2983$\pm$195&   0.37&   2.25&   1&   $-$0.34&   ---&  0.0\cr
J014415.19+052603.0&  3.32&  21.0&   ---&   61$\pm$5&   3264$\pm$182&   0.35&   1.71&   0&   $-$0.35&   0.11$\pm$0.02&  0.0\cr
J014632.04+154309.5&  3.17&  20.9&   4.1&   61$\pm$5&   3375$\pm$202&   0.37&   1.39&   1&   0.75&   0.21$\pm$0.01&  ---\cr
J014818.11+070952.6$^a$&  2.69&  21.2&   4.7&   87$\pm$3&   3469$\pm$92&   0.37&   1.86&   1&   $-$0.03&   0.20$\pm$0.01&  0.0\cr
J015341.97+094551.5&  2.49&  21.6&   ---&   165$\pm$5&   3452$\pm$111&   0.33&   0.72&   0&   $-$1.17&   ---&  0.0\cr
J015556.70+174943.0&  3.34&  20.8&   ---&   86$\pm$3&   1187$\pm$48&   0.33&   1.07&   1&   0.45&   0.19$\pm$0.01&  ---\cr
J020005.87$-$053052.9&  2.32&  21.1&   ---&   60$\pm$3&   3107$\pm$136&   0.35&   1.51&   1&   0.24&   ---&  0.0\cr
J020006.77$-$031126.8$^a$&  2.80&  20.0&   5.1&   58$\pm$2&   2797$\pm$84&   0.36&   1.40&   1&   $-$0.48&   0.22$\pm$0.01&  0.0\cr
J020728.19+033833.5&  2.75&  21.6&   ---&   138$\pm$5&   1647$\pm$49&   0.37&   0.94&   1&   $-$0.90&   ---&  0.0\cr
J020822.78+250639.8&  2.58&  20.5&   ---&   63$\pm$3&   4102$\pm$150&   0.37&   1.67&   0&   $-$0.34&   0.06$\pm$0.01&  ---\cr
J022702.10$-$065829.1&  2.73&  20.4&   4.5&   76$\pm$4&   3435$\pm$135&   0.37&   3.07&   1&   $-$2.12&   0.14$\pm$0.01&  0.0\cr
J023728.11+002702.1&  2.87&  21.7&   4.1&   76$\pm$5&   1906$\pm$87&   0.37&   1.19&   1&   $-$1.38&   0.20$\pm$0.02&  0.0\cr
J024002.49+000711.7&  3.02&  23.3&   ---&   80$\pm$5&   2231$\pm$115&   0.37&   1.08&   0&   0.00&   ---&  0.0\cr
J025020.87$-$032610.4&  2.74&  21.2&   4.7&   75$\pm$3&   1785$\pm$74&   0.34&   1.96&   1&   $-$1.23&   ---&  0.0\cr
J025422.35$-$020652.5&  2.27&  20.4&   ---&   52$\pm$2&   2396$\pm$89&   0.36&   2.27&   1&   0.82&   ---&  0.0\cr
J074614.86+421526.2&  2.62&  20.5&   ---&   55$\pm$4&   3719$\pm$185&   0.37&   1.75&   0&   0.32&   0.00$\pm$0.02&  0.0\cr
J075252.99+095513.1&  2.28&  21.8&   ---&   85$\pm$7&   2256$\pm$162&   0.33&   1.08&   0&   $-$1.81&   ---&  ---\cr
J075948.24+381639.7&  3.05&  21.5&   ---&   40$\pm$3&   2288$\pm$147&   0.37&   1.36&   1&   $-$0.61&   ---&  0.0\cr
J080420.36+302546.8&  2.27&  21.7&   ---&   153$\pm$5&   2273$\pm$58&   0.36&   0.88&   0&   $-$1.53&   ---&  0.0\cr
J080926.63+234534.6&  2.79&  21.0&   ---&   105$\pm$10&   1651$\pm$317&   0.14&   1.90&   0&   $-$0.84&   0.26$\pm$0.01&  0.0\cr
J081011.82+064020.6&  2.09&  21.7&   5.9&   62$\pm$5&   6798$\pm$465&   0.33&   2.25&   1&   $-$0.83&   ---&  0.0\cr
J082224.01+583932.8&  2.55&  20.2&   4.8&   65$\pm$2&   5474$\pm$167&   0.36&   1.85&   0&   1.07&   0.32$\pm$0.00&  0.0\cr
J082349.75+100439.1&  2.52&  22.2&   5.0&   68$\pm$4&   2908$\pm$140&   0.37&   1.97&   1&   $-$0.75&   0.20$\pm$0.02&  0.0\cr
J082418.56+395423.6&  2.50&  20.9&   ---&   114$\pm$3&   1880$\pm$40&   0.35&   0.76&   1&   0.51&   0.15$\pm$0.01&  0.0\cr
J082508.77+355822.9&  2.89&  20.4&   5.1&   75$\pm$5&   4738$\pm$238&   0.37&   1.55&   0&   2.07&   0.27$\pm$0.01&  0.0\cr
J082618.04+565345.9&  2.32&  21.5&   4.6&   82$\pm$4&   3509$\pm$135&   0.37&   2.52&   1&   $-$1.16&   0.28$\pm$0.01&  0.0\cr
J083306.26+273845.2&  3.25&  22.0&   ---&   107$\pm$5&   1843$\pm$67&   0.36&   1.34&   0&   $-$0.40&   ---&  0.0\cr
J083402.67+252421.5&  2.63&  21.8&   ---&   176$\pm$8&   2906$\pm$123&   0.33&   1.48&   0&   $-$0.34&   ---&  0.0\cr
J083534.58+090717.2&  2.60&  20.4&   4.3&   91$\pm$2&   3723$\pm$93&   0.35&   1.67&   1&   $-$0.66&   0.18$\pm$0.01&  0.0\cr
J084151.84+313821.0&  2.56&  20.9&   4.9&   99$\pm$4&   6126$\pm$213&   0.36&   1.41&   0&   $-$1.46&   0.24$\pm$0.01&  0.0\cr
J084424.50+545234.2&  2.70&  21.7&   ---&   77$\pm$4&   4028$\pm$164&   0.37&   1.69&   1&   0.35&   ---&  0.0\cr
J084600.05+151031.2&  2.44&  21.7&   ---&   137$\pm$7&   1940$\pm$100&   0.34&   1.23&   0&   -4.64&   0.23$\pm$0.01&  0.0\cr
J084808.48+080223.7&  2.34&  21.9&   ---&   60$\pm$3&   1648$\pm$67&   0.37&   1.10&   1&   $-$1.27&   ---&  0.0\cr
J085229.65+524730.8&  2.27&  21.2&   4.7&   64$\pm$4&   1291$\pm$70&   0.25&   1.71&   0&   2.49&   0.34$\pm$0.01&  0.0\cr
J085825.63+262540.2&  3.47&  21.1&   4.5&   71$\pm$2&   1156$\pm$42&   0.34&   1.08&   1&   ---&   ---&  4.0\cr
J090053.50+293819.2&  2.53&  21.2&   4.7&   59$\pm$4&   2271$\pm$142&   0.36&   1.67&   1&   $-$1.44&   0.17$\pm$0.01&  0.0\cr
J090152.11+272344.9&  2.57&  20.9&   ---&   51$\pm$3&   3128$\pm$163&   0.35&   1.66&   0&   $-$3.29&   ---&  0.0\cr
J090257.60+192701.2&  3.71&  20.4&   ---&   54$\pm$5&   3162$\pm$200&   0.35&   1.36&   1&   ---&   ---&  0.0\cr
J090502.28+401239.5&  3.10&  21.5&   ---&   83$\pm$3&   1970$\pm$93&   0.23&   1.83&   0&   $-$0.08&   ---&  1.1\cr
J091025.50+042944.3&  3.78&  21.6&   ---&   186$\pm$6&   1327$\pm$37&   0.37&   1.00&   0&   ---&   ---&  0.0\cr
J091159.79+240938.3&  3.04&  20.6&   3.6&   65$\pm$4&   3865$\pm$206&   0.35&   1.50&   0&   0.55&   0.19$\pm$0.01&  0.0\cr
J091500.76+103619.8&  2.12&  21.1&   ---&   90$\pm$4&   3749$\pm$153&   0.34&   1.79&   0&   $-$0.47&   0.24$\pm$0.01&  0.0\cr
\hline
\end{tabular}
\end{minipage}
\end{center}
\end{table*}
\setcounter{table}{0}
\begin{table*}
\begin{center}
\begin{minipage}{176mm}
\caption{ERQ-like Quasars (Continued)}
  \begin{tabular}{@{}lccccccccccc@{}}
  \hline
 Quasar Name& $z_{e}$& $i$ & \imw & REW & FWHM& $kt_{80}$& \nv /\civ& BAL & $\alpha_{\lambda}$& $E(B$$-$$V)$& FIRST\\
 & & (mag) & (mag) & (\AA ) & (km/s)& & & & & & (mJy)\\
\hline
J091541.56+541234.2&  3.17&  20.9&   4.9&   77$\pm$5&   4743$\pm$217&   0.37&   3.30&   1&   $-$2.02&   0.26$\pm$0.01&  0.0\cr
J091751.98+431233.1&  2.86&  20.0&   2.6&   51$\pm$2&   3308$\pm$140&   0.35&   1.44&   1&   $-$0.69&   0.08$\pm$0.01&  0.0\cr
J092419.57+314338.1&  2.20&  20.8&   4.1&   79$\pm$3&   1788$\pm$75&   0.29&   1.77&   1&   $-$1.19&   0.14$\pm$0.01&  0.0\cr
J092600.13+271322.6&  2.02&  20.1&   4.0&   112$\pm$2&   3135$\pm$70&   0.33&   1.74&   1&   $-$0.60&   ---&  0.0\cr
J092604.08+524652.9&  2.35&  21.8&   4.7&   86$\pm$5&   3053$\pm$156&   0.36&   3.03&   0&   $-$2.05&   ---&  0.0\cr
J092910.47+044033.5&  2.45&  21.4&   4.6&   54$\pm$3&   1979$\pm$100&   0.36&   1.18&   0&   $-$0.59&   0.33$\pm$0.01&  0.0\cr
J093138.12+263945.2&  2.31&  21.4&   4.8&   62$\pm$4&   3330$\pm$182&   0.37&   1.84&   1&   $-$1.82&   0.19$\pm$0.01&  0.0\cr
J093141.25+300300.0&  2.92&  21.8&   ---&   76$\pm$7&   1207$\pm$87&   0.37&   1.21&   0&   $-$2.08&   ---&  0.0\cr
J093659.08+484048.5&  2.37&  21.4&   ---&   59$\pm$2&   1764$\pm$72&   0.34&   1.23&   0&   $-$2.46&   ---&  0.0\cr
J093926.67+591311.1&  3.74&  20.9&   3.9&   85$\pm$3&   5378$\pm$162&   0.36&   2.72&   1&   ---&   ---&  0.0\cr
J094054.29+302316.0&  2.57&  20.8&   5.1&   72$\pm$3&   2371$\pm$90&   0.35&   1.69&   1&   0.97&   0.33$\pm$0.01&  0.0\cr
J094405.66+211200.8&  3.39&  20.9&   ---&   126$\pm$4&   3875$\pm$89&   0.36&   0.67&   0&   0.58&   ---&  0.0\cr
J094613.19+315521.6&  2.67&  21.4&   ---&   366$\pm$23&   1967$\pm$111&   0.34&   1.16&   0&   $-$1.81&   ---&  0.0\cr
J094728.10+363033.1&  3.05&  21.3&   ---&   149$\pm$5&   1451$\pm$47&   0.35&   0.90&   0&   $-$0.50&   0.12$\pm$0.02&  0.0\cr
J094831.90+480111.6&  2.00&  21.3&   4.5&   41$\pm$3&   1488$\pm$78&   0.37&   1.28&   1&   $-$0.16&   ---&  0.0\cr
J095259.90+094916.0$^a$&  2.76&  20.6&   5.2&   56$\pm$3&   1244$\pm$55&   0.35&   1.78&   0&   1.33&   0.39$\pm$0.01&  0.0\cr
J095652.51+390234.7&  3.16&  22.2&   ---&   195$\pm$13&   3421$\pm$212&   0.33&   1.93&   0&   $-$0.80&   ---&  0.0\cr
J095710.31+572902.2&  2.63&  21.2&   5.3&   50$\pm$2&   2423$\pm$78&   0.35&   1.52&   1&   $-$1.62&   0.23$\pm$0.01&  0.0\cr
J100113.40+120951.3&  2.81&  21.7&   ---&   148$\pm$6&   1289$\pm$45&   0.36&   1.10&   0&   $-$1.05&   0.18$\pm$0.02&  0.0\cr
J100901.15+345937.6&  2.40&  22.3&   ---&   311$\pm$23&   2315$\pm$145&   0.35&   1.87&   0&   $-$1.21&   ---&  0.0\cr
J100916.92+031128.8&  2.68&  21.9&   ---&   155$\pm$5&   1324$\pm$42&   0.34&   0.89&   0&   $-$0.46&   ---&  0.0\cr
J100917.93+532500.5&  2.26&  20.7&   ---&   50$\pm$2&   3351$\pm$126&   0.36&   1.40&   1&   $-$1.48&   0.02$\pm$0.02&  0.0\cr
J101204.96+220802.3&  2.70&  20.5&   3.1&   48$\pm$2&   2599$\pm$92&   0.35&   1.64&   1&   $-$0.16&   0.07$\pm$0.01&  0.0\cr
J101254.73+033548.6&  3.19&  21.3&   4.5&   127$\pm$9&   3224$\pm$173&   0.37&   1.05&   0&   $-$0.19&   0.19$\pm$0.02&  1.4\cr
J101326.23+611219.9&  3.70&  21.1&   5.9&   281$\pm$10&   5133$\pm$157&   0.35&   1.56&   0&   ---&   ---&  0.0\cr
J101909.57+074600.6&  3.00&  20.0&   4.2&   55$\pm$2&   3107$\pm$115&   0.35&   2.17&   1&   $-$0.77&   0.12$\pm$0.01&  0.0\cr
J102049.71+305956.7&  2.61&  21.0&   4.5&   64$\pm$4&   3164$\pm$155&   0.36&   2.20&   1&   $-$0.01&   0.14$\pm$0.01&  0.0\cr
J102250.16+112500.8&  2.19&  20.3&   3.5&   48$\pm$3&   3068$\pm$150&   0.36&   1.55&   1&   0.06&   0.17$\pm$0.01&  0.0\cr
J102315.66+070452.7&  3.63&  19.4&   ---&   80$\pm$1&   4042$\pm$58&   0.35&   1.54&   1&   ---&   ---&  0.0\cr
J102322.10+083337.0&  2.75&  20.5&   3.5&   70$\pm$3&   4063$\pm$130&   0.36&   1.45&   1&   0.67&   0.14$\pm$0.01&  0.0\cr
J102919.89+183053.9&  3.58&  20.2&   3.7&   68$\pm$4&   1924$\pm$116&   0.30&   2.02&   1&   ---&   ---&  0.0\cr
J102928.52+352352.0&  3.96&  20.9&   4.6&   45$\pm$3&   3802$\pm$185&   0.36&   2.23&   1&   ---&   ---&  0.0\cr
J103558.15$-$023321.1&  2.27&  21.0&   ---&   68$\pm$3&   1545$\pm$67&   0.34&   1.51&   0&   $-$0.40&   ---&  0.0\cr
J103748.63+371407.4&  2.85&  21.0&   ---&   76$\pm$4&   1859$\pm$113&   0.27&   1.64&   1&   $-$1.52&   0.14$\pm$0.02&  0.0\cr
J103807.71+325515.9&  2.41&  20.9&   4.6&   67$\pm$2&   4295$\pm$79&   0.37&   1.46&   1&   $-$0.51&   0.20$\pm$0.01&  0.0\cr
J103919.17+040130.3&  3.03&  21.2&   ---&   89$\pm$4&   2085$\pm$85&   0.20&   1.36&   0&   $-$0.03&   ---&  0.0\cr
J104617.99+060952.7&  2.73&  19.2&   4.0&   73$\pm$1&   4306$\pm$61&   0.36&   1.59&   1&   0.13&   0.14$\pm$0.01&  0.0\cr
J105018.31+230505.3&  2.58&  20.7&   4.2&   68$\pm$3&   3697$\pm$111&   0.37&   1.47&   0&   0.56&   0.07$\pm$0.01&  0.0\cr
J105057.90+545432.1&  2.33&  20.8&   3.2&   114$\pm$8&   2814$\pm$244&   0.23&   1.43&   0&   $-$1.76&   0.02$\pm$0.02&  0.0\cr
J110442.94+650451.7$^a$&  2.24&  20.6&   4.6&   70$\pm$2&   2347$\pm$58&   0.34&   1.94&   1&   0.06&   0.23$\pm$0.01&  ---\cr
J110455.29+603004.0&  2.80&  20.4&   ---&   86$\pm$5&   3397$\pm$178&   0.36&   2.17&   1&   $-$2.17&   0.03$\pm$0.02&  0.0\cr
J110735.97+072155.5&  2.17&  21.2&   ---&   100$\pm$5&   2174$\pm$126&   0.29&   1.38&   1&   $-$0.97&   ---&  0.0\cr
J111922.98+422311.6&  3.01&  20.3&   4.2&   84$\pm$4&   5235$\pm$187&   0.36&   1.95&   0&   0.55&   0.19$\pm$0.01&  0.0\cr
J112124.55+570529.6&  2.38&  20.0&   5.1&   28$\pm$1&   1780$\pm$74&   0.36&   1.35&   0&   2.05&   0.42$\pm$0.01&  0.0\cr
J112444.11+623303.8&  2.44&  21.1&   ---&   112$\pm$5&   1027$\pm$44&   0.33&   0.78&   1&   $-$1.42&   0.19$\pm$0.01&  0.0\cr
J112827.10+131406.4&  3.26&  21.0&   ---&   56$\pm$2&   3004$\pm$77&   0.37&   1.37&   1&   $-$0.06&   0.11$\pm$0.02&  0.0\cr
J113707.89+311127.7&  2.98&  19.5&   3.9&   61$\pm$2&   4883$\pm$110&   0.36&   2.52&   1&   -0.73&   0.13$\pm$0.01&  0.0\cr
J113721.46+142728.8$^a$&  2.30&  20.0&   4.9&   98$\pm$2&   4734$\pm$99&   0.36&   2.01&   1&   $-$0.27&   0.23$\pm$0.01&  0.0\cr
J114208.45+131706.7&  2.29&  21.2&   ---&   75$\pm$3&   3503$\pm$131&   0.36&   1.75&   0&   $-$2.32&   ---&  0.0\cr
J114325.84+465901.8&  2.27&  22.4&   ---&   129$\pm$4&   1208$\pm$34&   0.36&   0.78&   0&   $-$0.87&   ---&  0.0\cr
J114508.00+574258.6&  2.79&  19.1&   4.8&   38$\pm$2&   9103$\pm$446&   0.37&   1.41&   0&   2.35&   0.28$\pm$0.00&  0.0\cr
J114542.07+401318.3&  3.30&  21.9&   ---&   209$\pm$4&   1275$\pm$19&   0.37&   0.74&   0&   $-$1.71&   ---&  0.0\cr
J114855.47+572159.8&  3.05&  21.1&   3.8&   90$\pm$5&   3216$\pm$131&   0.37&   2.62&   0&   $-$0.26&   0.16$\pm$0.01&  0.0\cr
J120025.75+102314.1&  2.64&  21.1&   4.3&   121$\pm$4&   2502$\pm$72&   0.36&   0.89&   1&   $-$0.85&   0.16$\pm$0.01&  0.0\cr
J120150.13+191632.0&  2.90&  22.3&   ---&   156$\pm$6&   1768$\pm$63&   0.35&   1.40&   0&   $-$1.67&   ---&  0.0\cr
J120346.36+220846.9&  2.34&  20.5&   4.6&   89$\pm$3&   3460$\pm$84&   0.37&   1.94&   1&   1.04&   0.34$\pm$0.01&  0.0\cr
J121429.97+641125.6&  3.01&  20.3&   ---&   47$\pm$2&   3333$\pm$95&   0.36&   2.05&   1&   -0.63&   ---&  0.0\cr
J121634.84+025714.3&  3.11&  20.4&   4.8&   70$\pm$4&   6147$\pm$280&   0.33&   1.71&   0&   1.47&   0.24$\pm$0.01&  3.1\cr
J122232.46$-$014435.4&  3.90&  21.7&   ---&   92$\pm$5&   4060$\pm$156&   0.37&   1.66&   1&   ---&   ---&  0.0\cr
J122519.31+565733.0&  2.24&  20.4&   ---&   80$\pm$3&   1328$\pm$42&   0.34&   1.14&   1&   0.65&   ---&  1.2\cr
J122835.79+105949.4&  2.82&  21.5&   ---&   82$\pm$3&   1523$\pm$55&   0.36&   1.42&   0&   $-$0.58&   ---&  0.0\cr
J122900.89+374934.3&  3.15&  21.8&   5.2&   82$\pm$4&   2202$\pm$98&   0.34&   1.08&   0&   $-$0.80&   0.26$\pm$0.02&  0.0\cr
\hline
\end{tabular}
\end{minipage}
\end{center}
\end{table*}
\setcounter{table}{0}
\begin{table*}
\begin{center}
\begin{minipage}{176mm}
\caption{ERQ-like Quasars (Continued)}
  \begin{tabular}{@{}lccccccccccc@{}}
  \hline
 Quasar Name& $z_{e}$& $i$ & \imw & REW & FWHM& $kt_{80}$& \nv /\civ& BAL & $\alpha_{\lambda}$& $E(B$$-$$V)$& FIRST\\
 & & (mag) & (mag) & (\AA ) & (km/s)& & & & & & (mJy)\\
\hline
J123009.78+451039.7&  2.18&  21.5&   4.6&   112$\pm$3&   1344$\pm$31&   0.36&   0.83&   0&   $-$1.17&   0.29$\pm$0.01&  0.0\cr
J123506.05+503040.2&  2.91&  21.5&   4.3&   198$\pm$5&   1302$\pm$29&   0.36&   0.68&   0&   0.73&   0.19$\pm$0.02&  0.0\cr
J123643.41+010908.4&  2.91&  21.9&   ---&   68$\pm$5&   2039$\pm$149&   0.26&   1.56&   1&   $-$1.60&   0.21$\pm$0.03&  0.0\cr
J123713.55+010029.3&  3.12&  21.5&   ---&   67$\pm$5&   1726$\pm$115&   0.34&   1.29&   0&   0.35&   0.21$\pm$0.02&  0.0\cr
J123843.19+465120.4&  2.00&  19.1&   3.6&   54$\pm$2&   1832$\pm$62&   0.36&   3.09&   1&   3.47&   ---&  5.4\cr
J124025.83+614830.6&  2.70&  20.0&   2.6&   39$\pm$2&   3647$\pm$125&   0.36&   2.92&   1&   -1.33&   0.05$\pm$0.01&  0.0\cr
J124543.46+362901.3&  3.12&  20.9&   ---&   46$\pm$3&   2638$\pm$122&   0.36&   2.21&   1&   $-$0.32&   0.07$\pm$0.02&  0.0\cr
J125009.73+170934.5&  2.57&  21.4&   ---&   63$\pm$4&   3102$\pm$165&   0.34&   1.50&   1&   $-$0.85&   ---&  0.0\cr
J125737.98+165828.7&  2.32&  20.2&   3.8&   59$\pm$2&   2742$\pm$80&   0.35&   1.79&   1&   $-$0.17&   0.09$\pm$0.01&  0.0\cr
J130003.84+580628.5&  3.05&  20.7&   4.4&   91$\pm$3&   3467$\pm$90&   0.37&   1.01&   0&   $-$0.60&   0.15$\pm$0.01&  0.0\cr
J130544.31+121609.0&  2.36&  21.7&   ---&   138$\pm$4&   2350$\pm$69&   0.34&   1.11&   0&   1.00&   ---&  84.1\cr
J130711.76+280029.6&  3.87&  20.2&   2.9&   69$\pm$2&   3020$\pm$93&   0.35&   2.02&   1&   ---&   ---&  0.0\cr
J130736.35+183337.2&  3.10&  19.7&   3.1&   58$\pm$2&   4269$\pm$117&   0.35&   2.05&   1&   0.14&   0.08$\pm$0.01&  0.0\cr
J130741.38+364843.0&  2.35&  20.8&   3.3&   124$\pm$4&   3633$\pm$97&   0.34&   1.08&   0&   $-$0.69&   0.13$\pm$0.01&  0.0\cr
J131133.17+052459.4&  2.03&  20.4&   4.0&   42$\pm$2&   3506$\pm$115&   0.37&   2.12&   1&   -0.06&   ---&  0.0\cr
J131330.67+625957.2&  2.37&  22.2&   4.7&   82$\pm$5&   1589$\pm$112&   0.20&   1.56&   1&   $-$1.20&   0.15$\pm$0.02&  0.0\cr
J131628.32+045316.2&  2.14&  21.3&   5.7&   63$\pm$2&   3010$\pm$66&   0.37&   1.99&   1&   $-$0.97&   0.45$\pm$0.01&  0.0\cr
J131639.71+342617.3&  2.53&  20.8&   5.4&   55$\pm$4&   1149$\pm$121&   0.17&   1.11&   1&   1.08&   0.25$\pm$0.01&  0.0\cr
J131853.46+295736.4&  3.02&  21.6&   4.2&   81$\pm$6&   4138$\pm$195&   0.37&   1.42&   1&   0.23&   0.15$\pm$0.03&  0.0\cr
J132020.40+012316.3&  2.16&  21.5&   ---&   108$\pm$5&   2739$\pm$119&   0.34&   0.86&   0&   $-$1.94&   ---&  0.0\cr
J132026.05+074310.7&  2.61&  20.4&   3.1&   53$\pm$4&   4887$\pm$250&   0.37&   2.02&   1&   -2.05&   0.11$\pm$0.01&  0.0\cr
J132210.44+640225.1&  2.71&  20.5&   4.9&   33$\pm$3&   3813$\pm$310&   0.37&   2.63&   1&   $-$0.65&   0.23$\pm$0.01&  0.0\cr
J132654.95$-$000530.1&  3.32&  19.9&   4.7&   77$\pm$3&   1607$\pm$88&   0.16&   1.12&   0&   1.11&   0.32$\pm$0.00&  0.0\cr
J133340.76+084753.9&  3.12&  21.7&   4.6&   69$\pm$5&   2028$\pm$157&   0.30&   2.13&   0&   0.14&   0.31$\pm$0.02&  0.0\cr
J133345.80+383631.4&  2.66&  21.4&   4.6&   76$\pm$3&   3919$\pm$115&   0.35&   2.14&   1&   $-$2.00&   0.16$\pm$0.01&  0.0\cr
J133401.30+123337.7&  2.87&  20.7&   3.8&   40$\pm$4&   3512$\pm$227&   0.37&   3.00&   1&   0.34&   0.15$\pm$0.01&  0.0\cr
J133634.42+131637.9&  2.56&  21.8&   ---&   101$\pm$8&   1806$\pm$106&   0.36&   0.80&   1&   $-$0.98&   0.22$\pm$0.01&  0.0\cr
J133651.11+280544.1&  2.41&  21.1&   4.7&   32$\pm$4&   3146$\pm$274&   0.37&   1.95&   0&   0.63&   0.14$\pm$0.01&  0.0\cr
J133922.09+315440.7&  2.64&  21.1&   ---&   54$\pm$2&   2171$\pm$80&   0.37&   1.51&   1&   $-$0.90&   0.12$\pm$0.01&  0.0\cr
J134015.04+585304.8&  2.71&  20.1&   ---&   80$\pm$2&   3617$\pm$105&   0.33&   1.36&   0&   $-$0.58&   0.07$\pm$0.01&  0.0\cr
J134248.86+605641.1&  2.41&  21.5&   ---&   66$\pm$3&   2652$\pm$102&   0.34&   1.55&   0&   $-$1.42&   0.13$\pm$0.01&  0.0\cr
J134254.45+093059.3&  2.34&  20.8&   4.9&   66$\pm$1&   3246$\pm$57&   0.36&   2.73&   1&   $-$1.19&   0.19$\pm$0.01&  0.0\cr
J134614.93+124619.7&  2.84&  21.3&   ---&   90$\pm$6&   3777$\pm$234&   0.34&   1.45&   1&   $-$1.28&   ---&  0.0\cr
J134800.13$-$025006.4&  2.25&  21.6&   5.7&   87$\pm$5&   3654$\pm$151&   0.37&   2.79&   0&   $-$1.26&   0.33$\pm$0.01&  0.0\cr
J135118.82+270023.2&  2.17&  20.2&   4.8&   62$\pm$2&   3326$\pm$78&   0.34&   1.65&   1&   $-$0.62&   0.21$\pm$0.01&  0.0\cr
J135736.60+184459.5&  3.24&  21.0&   3.9&   78$\pm$2&   4439$\pm$111&   0.37&   2.45&   1&   $-$0.70&   0.10$\pm$0.02&  0.0\cr
J135815.74+472846.4&  2.59&  21.0&   ---&   56$\pm$4&   1682$\pm$102&   0.35&   1.76&   1&   $-$1.41&   0.10$\pm$0.01&  0.0\cr
J135837.89+112855.8&  2.56&  21.2&   ---&   44$\pm$4&   2630$\pm$202&   0.34&   1.71&   1&   $-$0.97&   ---&  0.0\cr
J140347.33+454649.6&  3.21&  21.1&   ---&   75$\pm$4&   3164$\pm$140&   0.36&   1.73&   1&   $-$0.86&   ---&  0.0\cr
J141002.95+455455.8&  2.98&  21.9&   ---&   119$\pm$6&   1737$\pm$72&   0.36&   1.19&   0&   $-$0.88&   0.17$\pm$0.02&  0.0\cr
J141538.29+035410.2&  3.32&  20.1&   4.6&   108$\pm$2&   3058$\pm$44&   0.34&   0.86&   0&   0.64&   0.34$\pm$0.01&  0.0\cr
J142121.54$-$001138.9&  2.05&  21.3&   4.8&   51$\pm$5&   1330$\pm$208&   0.19&   2.43&   1&   $-$0.86&   ---&  0.0\cr
J143050.54+524430.0&  3.02&  20.8&   ---&   55$\pm$3&   3737$\pm$169&   0.37&   1.36&   1&   $-$0.62&   ---&  0.0\cr
J144111.65+130837.4&  2.85&  20.9&   ---&   55$\pm$3&   2279$\pm$118&   0.34&   1.41&   0&   $-$0.99&   ---&  0.0\cr
J144742.66+565209.2&  2.79&  21.5&   ---&   118$\pm$8&   2989$\pm$283&   0.19&   2.23&   0&   $-$0.60&   ---&  0.0\cr
J144833.43+112525.0&  2.87&  21.2&   ---&   58$\pm$4&   1924$\pm$119&   0.35&   1.06&   1&   $-$1.43&   ---&  0.0\cr
J144855.62+105728.2&  3.70&  21.2&   4.2&   78$\pm$3&   2749$\pm$91&   0.36&   1.42&   1&   ---&   ---&  0.0\cr
J144908.48+540215.7&  2.03&  20.5&   ---&   51$\pm$2&   1629$\pm$62&   0.35&   1.09&   1&   $-$1.01&   ---&  0.0\cr
J144932.66+235437.2$^a$&  2.34&  21.5&   4.7&   98$\pm$3&   1352$\pm$47&   0.34&   0.79&   1&   $-$0.66&   0.26$\pm$0.01&  0.0\cr
J145113.61+013234.1&  2.77&  20.4&   5.7&   87$\pm$3&   6231$\pm$156&   0.37&   1.54&   1&   $-$0.19&   0.16$\pm$0.01&  0.0\cr
J145148.01+233845.4&  2.62&  20.5&   5.5&   89$\pm$3&   4166$\pm$124&   0.36&   1.77&   1&   0.96&   0.30$\pm$0.01&  0.0\cr
J145457.52+250824.4&  2.25&  19.7&   2.8&   70$\pm$2&   3212$\pm$71&   0.34&   1.55&   1&   $-$0.09&   0.06$\pm$0.01&  0.0\cr
J145756.84+364606.3&  3.28&  21.1&   3.8&   60$\pm$4&   5664$\pm$266&   0.37&   2.16&   0&   -0.83&   0.08$\pm$0.02&  0.0\cr
J145926.48+273022.1&  2.88&  20.7&   3.2&   81$\pm$3&   3979$\pm$122&   0.33&   1.98&   1&   $-$1.25&   0.13$\pm$0.01&  0.0\cr
J150549.72+074308.9&  3.32&  21.2&   4.3&   199$\pm$4&   1414$\pm$23&   0.35&   0.77&   0&   $-$1.33&   0.21$\pm$0.01&  0.0\cr
J151113.83+275233.0&  2.15&  21.1&   4.4&   42$\pm$1&   3709$\pm$104&   0.36&   2.18&   1&   -1.21&   0.16$\pm$0.01&  0.0\cr
J152209.18+201137.1&  2.74&  20.4&   2.7&   73$\pm$3&   3629$\pm$143&   0.33&   1.50&   1&   $-$0.16&   0.01$\pm$0.02&  0.0\cr
J152838.91+560938.9&  2.07&  21.6&   ---&   88$\pm$3&   2224$\pm$71&   0.35&   1.24&   0&   $-$1.17&   ---&  0.0\cr
J153441.83+542541.1&  3.46&  20.2&   4.6&   112$\pm$6&   1455$\pm$141&   0.17&   2.90&   0&   ---&   ---&  0.0\cr
J154517.30+451920.8&  2.46&  21.8&   ---&   86$\pm$4&   3864$\pm$147&   0.33&   1.82&   1&   $-$1.26&   ---&  0.0\cr
J154523.97+165054.2&  2.95&  21.2&   5.0&   91$\pm$4&   4537$\pm$138&   0.37&   2.40&   0&   $-$1.37&   0.17$\pm$0.02&  0.0\cr
J154709.04+131708.4&  2.83&  21.5&   ---&   51$\pm$3&   2047$\pm$121&   0.35&   1.89&   0&   $-$1.09&   0.14$\pm$0.02&  0.0\cr
\hline
\end{tabular}
\end{minipage}
\end{center}
\end{table*}
\setcounter{table}{0}
\begin{table*}
\begin{center}
\begin{minipage}{176mm}
\caption{ERQ-like Quasars (Continued)}
  \begin{tabular}{@{}lccccccccccc@{}}
  \hline
 Quasar Name& $z_{e}$& $i$ & \imw & REW & FWHM& $kt_{80}$& \nv /\civ& BAL & $\alpha_{\lambda}$& $E(B$$-$$V)$& FIRST\\
 & & (mag) & (mag) & (\AA ) & (km/s)& & & & & & (mJy)\\
\hline
J155057.71+080652.1&  2.51&  20.1&   3.8&   149$\pm$2&   4446$\pm$60&   0.37&   0.91&   0&   0.59&   0.12$\pm$0.01&  1.3\cr
J155459.46+555330.9&  2.57&  19.9&   3.9&   57$\pm$2&   3823$\pm$98&   0.37&   2.53&   0&   0.74&   0.12$\pm$0.00&  0.0\cr
J155520.62+551906.6&  2.69&  20.4&   3.6&   44$\pm$2&   3743$\pm$151&   0.37&   2.04&   0&   -0.03&   0.01$\pm$0.01&  0.0\cr
J155725.27+260252.7&  2.82&  21.8&   4.9&   56$\pm$4&   1182$\pm$79&   0.34&   1.08&   0&   $-$0.36&   0.32$\pm$0.02&  0.0\cr
J160733.82+192817.3&  3.02&  20.5&   ---&   53$\pm$3&   2425$\pm$99&   0.36&   1.37&   1&   $-$1.38&   0.08$\pm$0.01&  0.0\cr
J161033.30+352204.7&  2.56&  20.8&   3.8&   98$\pm$4&   4499$\pm$159&   0.34&   1.63&   0&   0.66&   0.15$\pm$0.01&  0.0\cr
J161305.49+104055.2&  3.24&  20.7&   ---&   62$\pm$3&   4098$\pm$124&   0.36&   2.49&   1&   $-$0.63&   ---&  0.0\cr
J161622.14+384559.2&  2.92&  20.4&   ---&   75$\pm$3&   4063$\pm$143&   0.37&   2.22&   1&   0.24&   ---&  0.0\cr
J161800.49+435538.6&  2.51&  20.6&   3.6&   76$\pm$3&   3907$\pm$124&   0.35&   1.37&   0&   0.38&   0.19$\pm$0.01&  0.0\cr
J162244.14+284044.0&  2.33&  21.6&   ---&   74$\pm$3&   3862$\pm$149&   0.35&   1.61&   1&   $-$0.45&   0.18$\pm$0.01&  0.0\cr
J162327.66+312204.2&  2.34&  21.4&   4.5&   164$\pm$7&   1572$\pm$68&   0.34&   1.54&   0&   $-$1.24&   0.15$\pm$0.01&  0.0\cr
J162518.66+144509.9&  2.39&  20.4&   4.7&   79$\pm$2&   4209$\pm$73&   0.35&   2.51&   1&   $-$1.56&   0.19$\pm$0.01&  0.0\cr
J162650.28+371344.9&  2.45&  19.7&   ---&   50$\pm$1&   2572$\pm$78&   0.34&   1.90&   1&   $-$1.72&   ---&  0.0\cr
J162920.36+495705.3&  2.76&  20.9&   ---&   74$\pm$3&   2575$\pm$72&   0.37&   1.85&   0&   0.27&   ---&  0.0\cr
J162943.03+395844.4&  2.20&  20.4&   4.5&   63$\pm$4&   1707$\pm$128&   0.21&   2.02&   0&   0.51&   0.27$\pm$0.01&  0.0\cr
J164325.87+401119.8&  2.29&  20.9&   3.6&   100$\pm$3&   2247$\pm$75&   0.33&   0.73&   0&   $-$0.58&   0.13$\pm$0.01&  0.0\cr
J164745.64+161443.8&  2.62&  21.7&   ---&   105$\pm$5&   2010$\pm$95&   0.34&   0.86&   0&   $-$3.01&   ---&  0.0\cr
J165053.78+250755.4&  3.32&  18.5&   3.6&   56$\pm$1&   3348$\pm$55&   0.36&   2.16&   1&   $-$0.71&   0.13$\pm$0.00&  0.0\cr
J170047.07+400238.7&  2.90&  21.6&   4.4&   135$\pm$5&   1336$\pm$50&   0.34&   0.89&   1&   0.54&   0.21$\pm$0.02&  0.0\cr
J170110.12+301502.8&  3.27&  20.7&   3.6&   93$\pm$3&   1572$\pm$48&   0.36&   1.37&   0&   0.20&   0.17$\pm$0.01&  0.0\cr
J170327.66+305148.9&  2.70&  20.9&   3.8&   55$\pm$4&   3464$\pm$188&   0.36&   1.99&   1&   0.78&   0.15$\pm$0.01&  0.0\cr
J171728.91+394102.5&  2.80&  19.7&   4.2&   94$\pm$3&   4308$\pm$88&   0.36&   1.38&   1&   0.10&   0.13$\pm$0.01&  2.2\cr
J211329.61+001841.7&  2.00&  23.1&   7.1&   171$\pm$6&   1565$\pm$46&   0.35&   0.66&   0&   $-$1.77&   ---&  0.0\cr
J212951.40$-$001804.3&  3.21&  21.8&   ---&   125$\pm$6&   1904$\pm$71&   0.37&   1.99&   0&   $-$0.22&   ---&  0.0\cr
J214236.60+000534.4&  2.51&  21.7&   ---&   64$\pm$3&   1864$\pm$87&   0.34&   1.66&   1&   $-$1.27&   ---&  0.0\cr
J214437.72$-$001553.9&  3.32&  20.8&   4.6&   41$\pm$2&   1137$\pm$80&   0.26&   2.08&   0&   0.46&   0.27$\pm$0.01&  0.0\cr
J214621.58+022528.1&  2.04&  22.2&   5.4&   79$\pm$3&   1456$\pm$46&   0.36&   1.21&   0&   $-$0.78&   ---&  0.0\cr
J215855.10$-$014717.9&  2.31&  19.8&   4.1&   73$\pm$1&   4735$\pm$66&   0.35&   2.23&   1&   $-$0.45&   0.20$\pm$0.00&  0.0\cr
J221225.14+220122.7&  2.42&  21.9&   4.3&   234$\pm$7&   1723$\pm$53&   0.35&   0.71&   0&   $-$0.13&   ---&  ---\cr
J221322.84+091611.9&  2.23&  21.0&   ---&   101$\pm$4&   4102$\pm$134&   0.35&   1.58&   0&   $-$0.61&   0.17$\pm$0.01&  0.0\cr
J222128.44+001322.4&  3.29&  20.9&   ---&   90$\pm$2&   2937$\pm$49&   0.35&   1.91&   1&   $-$1.03&   ---&  0.0\cr
J222307.12+085701.7&  2.29&  21.3&   5.6&   77$\pm$2&   3661$\pm$78&   0.37&   2.52&   1&   $-$0.57&   0.23$\pm$0.01&  0.0\cr
J223348.09+024932.8&  2.58&  22.1&   ---&   94$\pm$4&   1089$\pm$53&   0.33&   1.46&   0&   0.47&   ---&  0.0\cr
J223808.50+192541.6&  2.75&  20.8&   4.6&   50$\pm$10&   3586$\pm$360&   0.37&   1.78&   1&   $-$0.22&   0.33$\pm$0.02&  ---\cr
J224010.95+183818.2&  2.13&  21.3&   4.3&   67$\pm$2&   2422$\pm$73&   0.34&   1.76&   1&   $-$2.09&   0.20$\pm$0.01&  ---\cr
J224014.48+084518.0&  2.30&  21.2&   ---&   43$\pm$2&   1781$\pm$81&   0.34&   1.38&   0&   $-$0.10&   ---&  0.0\cr
J224758.69+212826.4&  3.93&  19.0&   2.9&   49$\pm$1&   2435$\pm$62&   0.36&   1.92&   1&   ---&   ---&  ---\cr
J224926.63$-$024703.5&  2.20&  21.2&   ---&   81$\pm$4&   2022$\pm$95&   0.35&   1.16&   0&   $-$1.42&   ---&  0.0\cr
J230114.14+040421.7$^a$&  2.19&  20.7&   4.6&   80$\pm$2&   3266$\pm$72&   0.36&   1.31&   1&   $-$0.53&   0.24$\pm$0.01&  0.0\cr
J231804.15+111632.4&  2.90&  21.1&   4.0&   157$\pm$6&   1858$\pm$72&   0.34&   0.77&   0&   1.04&   ---&  0.0\cr
J232007.21+084847.4&  2.09&  20.4&   ---&   41$\pm$3&   3827$\pm$233&   0.37&   2.41&   1&   -0.95&   ---&  0.0\cr
J232231.64+010813.5&  2.56&  21.6&   ---&   48$\pm$5&   2704$\pm$284&   0.33&   1.63&   1&   $-$0.16&   ---&  0.0\cr
J232348.64+013528.1&  3.12&  21.1&   ---&   71$\pm$5&   1423$\pm$89&   0.33&   1.40&   1&   0.02&   0.20$\pm$0.01&  0.0\cr
J232611.97+244905.7&  2.37&  21.2&   4.5&   131$\pm$4&   2402$\pm$52&   0.36&   2.40&   1&   $-$1.61&   0.25$\pm$0.01&  ---\cr
J233507.23+074335.5&  2.76&  20.8&   4.7&   75$\pm$2&   2074$\pm$52&   0.35&   1.91&   0&   $-$0.63&   0.21$\pm$0.01&  0.0\cr
J233720.66$-$015438.9&  3.16&  21.2&   ---&   61$\pm$3&   2029$\pm$80&   0.34&   1.02&   1&   $-$0.88&   ---&  0.0\cr
J234933.29+322628.9&  2.33&  22.1&   ---&   112$\pm$6&   2366$\pm$113&   0.35&   1.86&   0&   $-$1.33&   ---&  ---\cr
J235010.00+185021.0&  2.50&  20.7&   3.6&   46$\pm$3&   3641$\pm$184&   0.36&   2.18&   1&   -1.30&   0.09$\pm$0.01&  -1.0\cr
J235556.69+014747.7&  2.25&  20.2&   3.1&   49$\pm$2&   3295$\pm$104&   0.37&   2.07&   0&   $-$0.10&   0.12$\pm$0.01&  0.0\cr
\hline
\multicolumn{12}{l}{$^a$These quasars are in DR12Q but not in our emission line catalog.}
\end{tabular}
\end{minipage}
\end{center}
\end{table*}


\bsp	
\label{lastpage}
\end{document}